%% file: AVSQS_review.tex
\documentclass[aip,showpacs,amsmath,amssymb,amsfonts,lengthcheck,longbibliography,superscriptaddress]{revtex4-2}

\usepackage{mathrsfs}  
\usepackage[usenames,dvipsnames]{color}
\usepackage{soul}
\usepackage{cancel}
\usepackage[
verbose,
colorlinks=true,
naturalnames=true,
linkcolor=blue,
]{hyperref}

\usepackage{graphicx}
\usepackage{subfigure}
\usepackage{amsthm}
\usepackage{verbatim}
\usepackage{dcolumn}
\usepackage{bm}
\usepackage{epsf}

\hypersetup{colorlinks,linkcolor={blue},citecolor={blue},urlcolor={blue}} 

\newcommand{\bra}[1]{\left\langle #1\right|}
\newcommand{\ket}[1]{\left|#1\right\rangle}
\newcommand{\braket}[2]{\left\langle #1|#2\right\rangle}

\newcommand{\tr}[1]{\mathrm{tr}\left\{#1\right\}}

\newcommand{\la}{\left\langle}
\newcommand{\ra}{\right\rangle}
\newcommand{\pd}{\partial}

\newcommand{\etal}{\textit{et al. }}
\newcommand{\e}[1]{\exp{\left(#1\right)}}
\newcommand{\lo}[1]{\ln{\left(#1\right)}}
\newcommand{\id}{\mathbb{I}}

\newcommand{\co}[1]{\cos{\left(#1\right)}}

\newcommand{\ct}[1]{\coth{\left(#1\right)}}
\newcommand{\bla}{bla\\bla\\bla\\bla\\bla}
\newcommand{\mb}[1]{\mbox{\boldmath$#1$}}
\newcommand{\mc}[1]{\mathcal{#1}}

\newcommand{\mrm}[1]{\mathrm{#1}}

\def\dbar{{\mathchar'26\mkern-12mu {\rm d}}}

\newcommand{\draftmode}{1}    
\newcommand{\notetoself}[1]{\ifnum \draftmode=1 {\color[rgb]{0,0,0.8} [#1]} \fi}  
\newcommand{\cuttext}[1]{\ifnum \draftmode=1 {\color[rgb]{0,0.5,0} [#1]} \fi}  
\newcommand{\warntext}[1]{\ifnum \draftmode=1 {\color[rgb]{0.9,0.6,0} #1} \else {#1} \color{black} \fi}

\begin{document} 

\title{Quantum thermodynamic devices: from theoretical proposals to experimental reality} 

	\author{Nathan M. Myers}
	\email{myersn1@umbc.edu}
	\affiliation{Department of Physics, University of Maryland, Baltimore County, Baltimore, Maryland 21250, USA}
	\affiliation{Computer, Computational and Statistical Sciences Division, Los Alamos National Laboratory, Los Alamos, New Mexico 87545, USA}

	\author{Obinna Abah}
	\email{o.abah@qub.ac.uk}
	\affiliation{Joint Quantum Centre (JQC) Durham-Newcastle, School of Mathematics, Statistics, and Physics, Newcastle University, Newcastle upon Tyne, NE1 7RU, United Kingdom}
	
	\author{Sebastian Deffner}
	\email{deffner@umbc.edu}
	\affiliation{Department of Physics, University of Maryland, Baltimore County, Baltimore, Maryland 21250, USA}
	\affiliation{Instituto de F\'isica `Gleb Wataghin', Universidade Estadual de Campinas, 13083-859, Campinas, S\~{a}o Paulo, Brazil}
	
\date{\today}

\begin{abstract}
Thermodynamics originated in the need to understand novel technologies developed by the Industrial Revolution.  However, over the centuries the description of engines, refrigerators, thermal accelerators, and heaters has become so abstract that a direct application of the universal statements to real-life devices is everything but straight forward.  The recent, rapid development of quantum thermodynamics has taken a similar trajectory, and, e.g., ``quantum engines'' have become a widely studied concept in theoretical research.  However, if the newly unveiled laws of nature are to be useful, we need to write the dictionary that allows us to translate abstract statements of theoretical quantum thermodynamics,  to physical platforms and working mediums of experimentally realistic scenarios.  To assist in this endeavor, this review is dedicated to providing an overview over the proposed and realized quantum thermodynamic devices, and to highlight the commonalities and differences of the various physical situations. 
\end{abstract}

\maketitle

\tableofcontents

\input{sections/intro.tex}
%
\input{sections/history.tex}
%
\input{sections/theory_cycles.tex}
\input{sections/theory_endoreversible.tex}
\input{sections/theory_many.tex}
\input{sections/theory_STA.tex}

%
%
%
\input{sections/experiment_ion.tex}

\input{sections/experiment_measurement.tex}
\input{sections/experiment_optomech.tex}

\input{sections/experiment_mems.tex}

\input{sections/experiment_NMR.tex}
\input{sections/experiment_NV.tex}
\input{sections/experiment_BEC.tex}
\input{sections/experiment_JJ.tex}
\input{sections/experiment_photo.tex}
\input{sections/experiment_dirac.tex}
%
%
\input{sections/conclude.tex}
%

\acknowledgements{We would like to thank Akram Touil and Maxwell Aifer for comments on the manuscript. N. M. gratefully acknowledges support from Harry Shaw of NASA Goddard Space Flight Center and Kenneth Cohen of Peraton. This material is based upon work supported by the U.S. Department of Energy, Office of Science, Office of Workforce Development for Teachers and Scientists, Office of Science Graduate Student Research (SCGSR) program. The SCGSR program is administered by the Oak Ridge Institute for Science and Education for the DOE under contract number DE‐SC0014664. O.A. acknowledges support from the UK EPSRC EP/S02994X/1. S.D. acknowledges support from the U.S. National Science Foundation under Grant No. DMR-2010127.  }

\bibliography{AVSQS}

\end{document}

%% file: sections/intro.tex
\section{Introduction}

For many students,  grasping the concepts of \emph{thermodynamics} poses significant challenges.  Arguably, the reason is that, as a phenomenological theory, thermodynamics relies much more on abstraction of complex problems than many other areas of theoretical physics.  The most prominent topic of the theory are heat engine cycles, which are abstract, idealized,  and cyclic processes that describe the universal working principles of converting heat to work.

It is often useful to remind oneself that thermodynamics was invented concurrently with the Industrial Revolution \cite{Kondepudi1998}, and that its original purpose was to understand and optimize the operation of steam engines. Rather remarkably, we are in a similar situation which occasionally is referred to as the Second Quantum Revolution \cite{MacFarlane2003}.  Recent years have witnessed significant domestic and international efforts  \cite{Raymer2019QST,Riedel2019QST,Yamamoto2019QST,Sussman2019QST,Roberson2019QST} to realize market-ready quantum technologies. However, to fully exploit the capabilities of these new technologies it appears obvious that we will need to train a ``quantum literate workforce'' \cite{Aiello2021QST}. However, very similar to the situation the public faced at the invention of steam engines,  new quantum technologies are complex \cite{Sanders2017} and markedly different from what societies are used to \cite{Roberson2021QST}. While many fundamental as well as quite practical questions remain to be addressed \cite{Auffeves2021}, it does appear obvious that a thermodynamic approach to quantum technologies may provide an effective and pedagogical entry point for a large audience.  

\begin{figure}
\includegraphics[width=.48\textwidth]{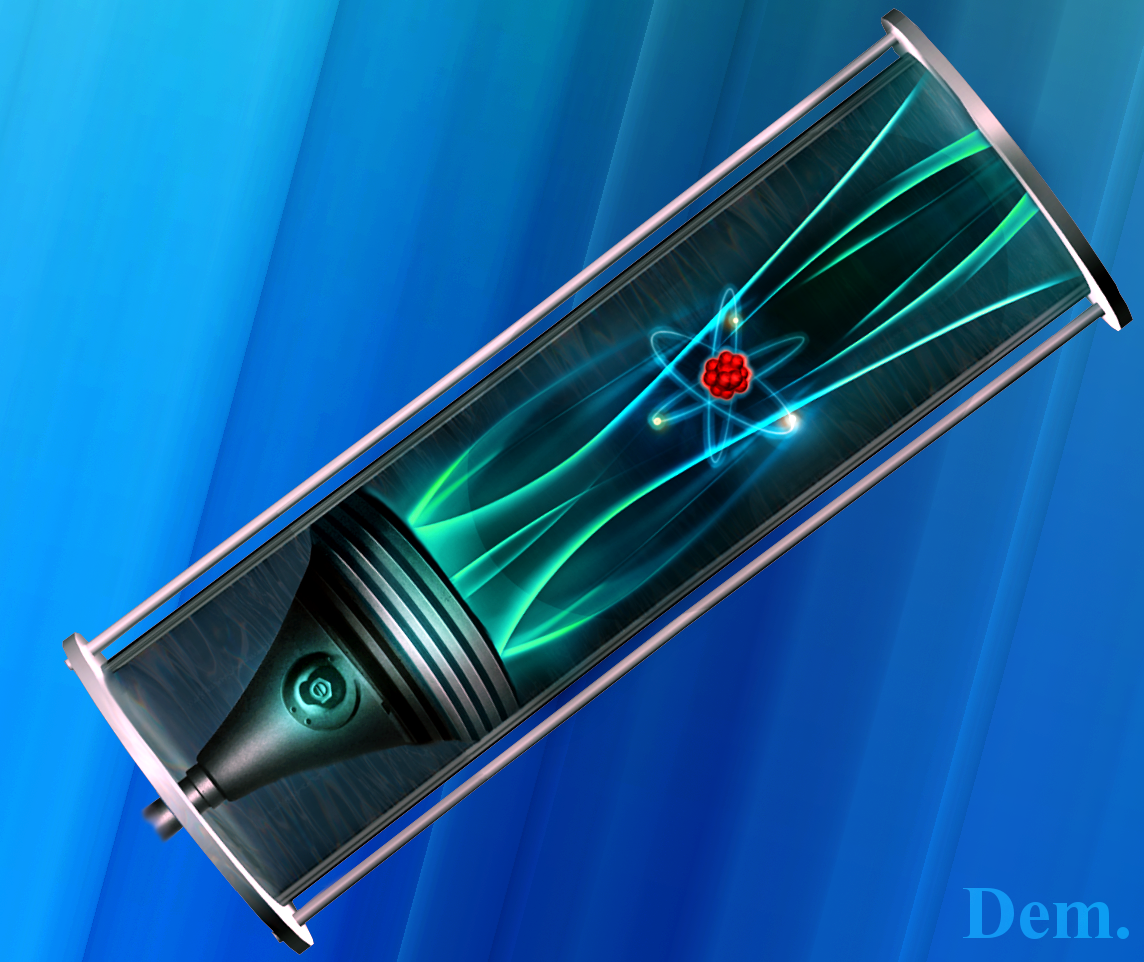}
\caption{\label{fig:qm_pist} Schematic representation of a quantum heat engine, here a quantum particle in a piston.  \emph{Figure adopted from Ref.~\cite{Gardas2015}.}}
\end{figure}

\emph{Quantum thermodynamics} \cite{Deffner2019book} is a relatively young field, which only rather recently has grown into a vibrant branch of modern research.  One of its main objectives is the study and description of converting heat, work, and information in quantum systems. In complete analogy to conventional thermodynamics \cite{Callen1985}, quantum thermodynamics relies on idealized processes that universally describe a wide range of scenarios. Remarkably, the first \emph{quantum heat engine} was proposed already in the late 1950s  \cite{Scovil1959}.  However, only over the last decade (or so) has the literature on quantum thermal devices really grown.  \cite{Binder2018}.  This may be explained by the fact that quantum heat engines (cf. Fig.~\ref{fig:qm_pist}) provide simple, pedagogical descriptions of otherwise much more complex quantum scenarios, as well as by the fact that the first experimental realizations of genuinely quantum devices have been reported.

The present review paper seeks to provide an overview of the current state of the art, and to categorize the plethora of different proposals according to their distinct platforms. To this end, we outline the beginnings of quantum heat engines in Sec.~\ref{sec:hist}, before we summarize the main concepts and theoretical tools in Sec.~\ref{sec:theory}. The majority of this article, however, is dedicated to the various physical platforms and selected implementations of quantum thermal devices in Sec.~\ref{sec:exp}. The review is concluded with a few remarks in Sec.~\ref{sec:con}.

When writing this review, we strove for a comprehensive,  objective, and pedagogically valuable account of the current state of the art. However, given the enormous number of publications on the topic we had to make some hard choices and not everything could be covered in detail. However, we do hope that we have done justice to the field, and that this review may be useful for newcomers as well as seasoned practitioners of quantum thermodynamics. 

%% file: sections/history.tex
\section{\label{sec:hist} From masers to genuine quantum devices}

We begin with a brief historical account of the beginnings of the field.  Interestingly, the concept of a \emph{quantum heat engine} is intimately related to the theoretical description of masers and lasers.

\subsection{\label{sec:maser} The maser: a first quantum heat engine}

Already more than six decades ago,  Scovil and Schultz-DuBois \cite{Scovil1959} suggested that the performance of a maser (microwave amplification by stimulated emission of radiation) \cite{Gordon1955} can be assessed akin to a continuous engine.  The maser was the direct predecessor of the laser (light amplification by stimulated emission of radiation) and it operates by the same working principle \cite{Schawlow1958}.

\begin{figure}
\includegraphics[width=.48\textwidth]{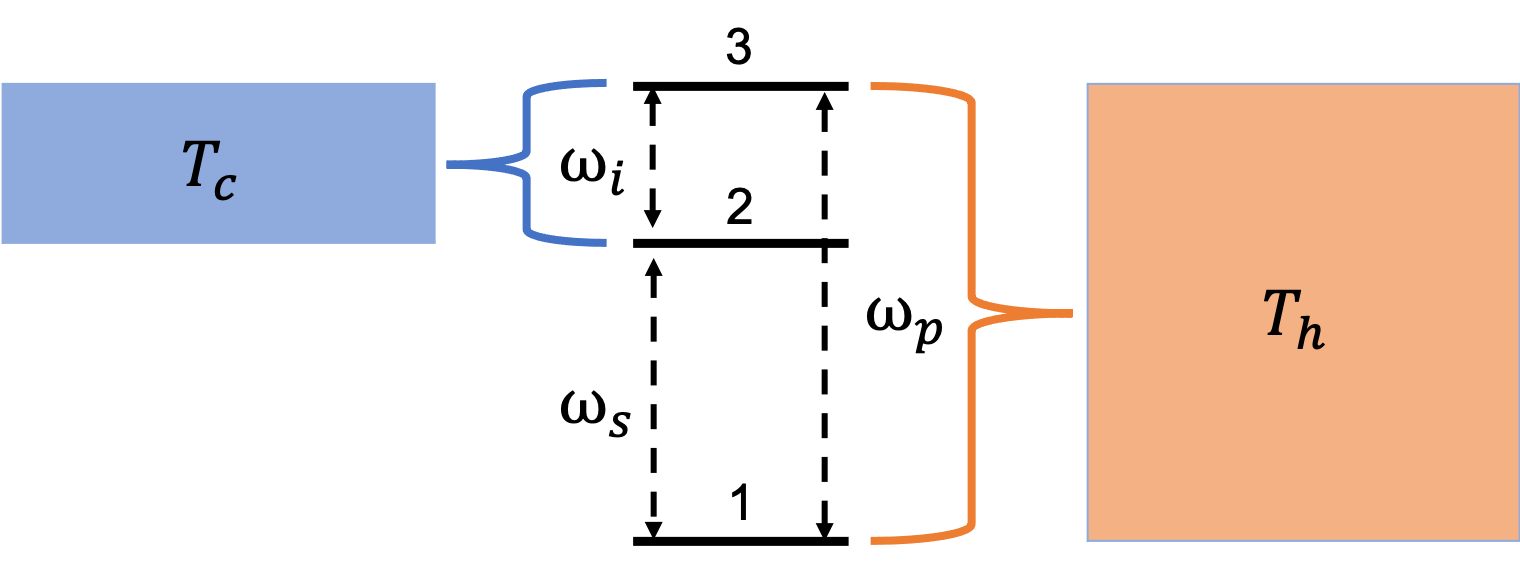}
\caption{\label{fig:maser} Schematic representation of a maser as a quantum heat engine.  Idler mode with frequency $\omega_i$ is coupled to a cold reservoir, and pumping mode is in contact with a hot reservoir. The signal with frequency $\omega_s$ is interpreted as the work extraction.}
\end{figure}

In its simplest form, a maser consists of a three level system which amplifies a signal with frequency $\omega_s$, by being pumped at frequency $\omega_p$. The excess energy is dumped into an idler mode with frequency $\omega_i=\omega_p-\omega_s$.  Scovil and Schultz-DuBois then realized that the maser becomes a continuous heat engine if the idler mode is coupled to a cold heat reservoir with temperature $T_c$, and the pumping is facilitated by a hot reservoir at $T_h$, see Fig.~\ref{fig:maser} for an illustration of the set-up.
Amplification is successful if the occupation number of the first excited state is larger than of the ground state, $n_2\geq n_1$.  This is commonly known as ``population inversion''. 

From a thermodynamic perspective, for any quantum of heat $Q_h=\hbar \omega_p$ that is absorbed from the hot reservoir, the three-level engine exhausts $Q_c=\hbar\omega_i$ to the cold reservoir, and work $W=\hbar\omega_s$ can be extracted. Therefore, the thermodynamic efficiency simply becomes
\begin{equation}
\label{eq:eff_maser}
\eta_\mrm{M}=\frac{W}{Q_h}=\frac{\omega_s}{\omega_p}\,.
\end{equation}
This can be further elaborated on by considering the thermally driven transitions between the levels 2 and 1.  In particular, we have
\begin{equation}
\label{eq:ratio}
\frac{n_2}{n_1}=\frac{n_2}{n_3}\,\frac{n_3}{n_1}=\e{\beta_c\,\hbar\omega_i}\,\e{-\beta_h\,\hbar\omega_p}
\end{equation}
where $n_3$ is the occupation number of the second excited state, and $\beta=1/k_B T$ is the inverse temperature.

Equation~\eqref{eq:ratio} can be re-written to read
\begin{equation}
\frac{n_2}{n_1}=\e{\beta_c\,\hbar \omega_s\,\left(\frac{\eta_\mrm{M}}{\eta_\mrm{C}}-1\right)}
\end{equation}
where we used Eq.~\eqref{eq:eff_maser} and where we introduced the Carnot efficiency,
\begin{equation}
\eta_\mrm{C}=1-\frac{T_c}{T_h}\,.
\end{equation}
It is then interesting to note that for the maser population inversion is identical to the positive working condition. Namely,  work can be extracted if and only if $n_2\geq n_1$. Thus, we also immediately conclude that the Carnot efficiency is the natural upper bound on the maser efficiency \eqref{eq:eff_maser},
\begin{equation}
\eta_\mrm{M}\leq \eta_\mrm{C}\,.
\end{equation}
The inequality becomes tight in the limit of vanishing inversion, i.e., $n_2\simeq n_1$.

Scovil and Schultz-DuBois \cite{Scovil1959} concluded their analysis by noting that for $n_2\leq n_1$ the device would operate as a refrigerator.  Thus, the maser is a simple and pedagogically valuable system to study the whole range of thermodynamic devices. Remarkably, this three level system is, however, not only a toy model, but actually a good description of experimentally realistic masers \cite{Gordon1955,Kleppner1962,Thomas2020}.  

However, the original analysis of the maser as a ``quantum'' heat engine \cite{Scovil1959} is still somewhat rudimentary. In particular, describing the transitions between the quantum levels by classical, thermal fluctuations does not leave room for genuine quantum effects. Nevertheless, the maser has become the prototypical example and the foundation for the study of continuous heat engines \cite{Geusic1967,Geva1994,Tonner2005PRE,Opatrny2005,Boukobza2006,Boukobza2007,Youssef2009PRE,Goswami2013,Abah2014EPL,Correa2014,Kosloff2014review,Gonzales2016,Silva2016,Ghosh2018,Niedenzu2019,Mitchison2019,Feldmann2000,Kalaee2021}.

\subsection{\label{sec:afterburner}Laser-maser quantum afterburner}

\begin{figure*}
\includegraphics[width=.48\textwidth]{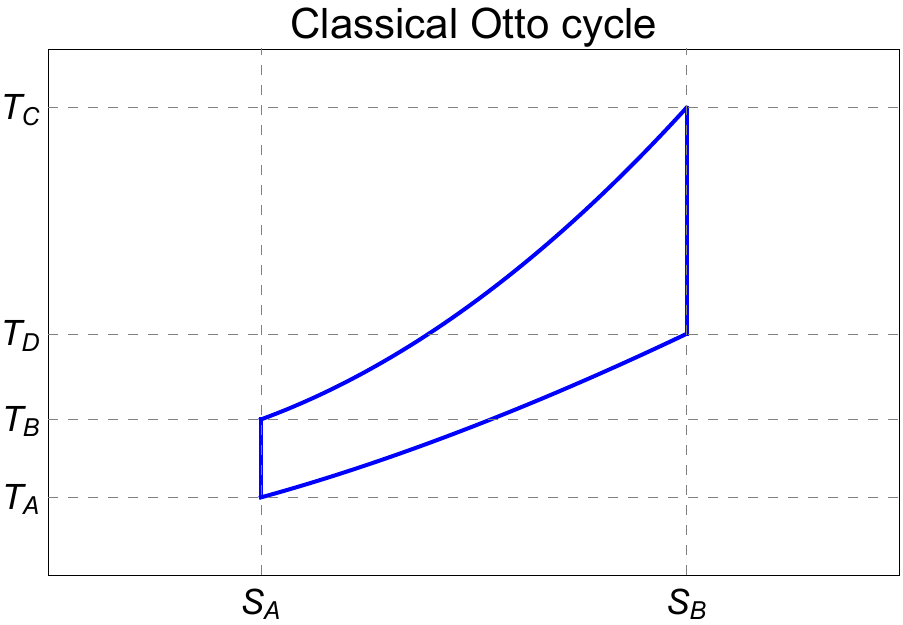}
\hfill
\includegraphics[width=.48\textwidth]{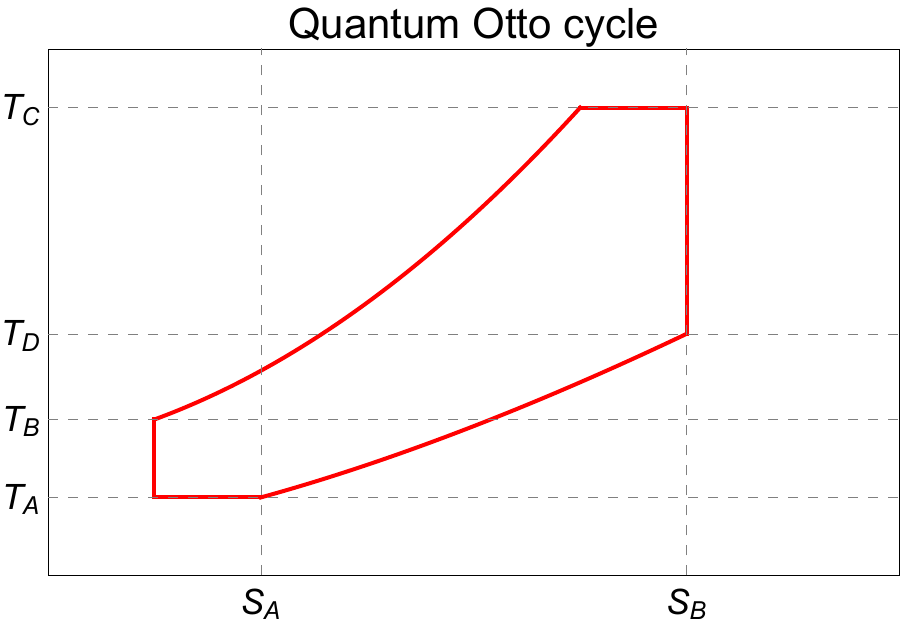}
\caption{\label{fig:otto} TS-diagram of the classical (left panel) and quantum (right panel) Otto cycle as suggest by Scully \cite{Scully2002}. Additional work can be extracted in the quantum case if the Otto cycle operates in an optical cavity,  and a resonant laser is coupled at low entropy state $A$. }
\end{figure*}

The next major step towards truly quantum engines was taken by Scully \cite{Scully2002} with the proposal of a ``quantum afterburner''.  This device consists of a laser-maser system that undergoes a joint Otto cycle.  The standard Otto cycle consists of four strokes,  ($A\rightarrow B$) isentropic compression,  ($B\rightarrow C$) isochoric heating, ($C\rightarrow D$) isentropic expansion,  and ($D \rightarrow A$) ischoric cooling \cite{Callen1985}.  A typical sketch  of resulting TS-diagram is depicted in the left panel of Fig.~\ref{fig:otto}. 

Scully realized that if the the working medium is an optical cavity, then a laser on resonance with the cavity mode can extract further work from the low entropy state $A$.  This is possible, since the thermal radiation of the cavity will excite the laser into population inversion, and the total energy of cavity plus laser will decrease through the spontaneous emission of the laser. The correspondingly modified TS-diagram is sketched in the right panel of Fig.~\ref{fig:otto}.

Thus, this ``quantum afterburner'' can extract additional energy from an ideal Otto cycle by exploiting the quantum states of the laser. The explicit expressions for extracted work and efficiency depend on the set-up and the actual design of the system \cite{Scully2002,Hill2004SPIE}. However, the idea of modifying ideal, classical cycles to harness quantum resources has proven to be versatile and powerful, see,  for instance,  an engine operating with transmon qubits \cite{Cherubim2019entropy}. More importantly,  we will see in the following that heat engines operating with nonequilibrium or squeezed reservoirs \cite{Scully2001PRL,Huang2012PRE,Zagoskin2012PRB,Rossnagel2014,Altintas2014PRE,Long2015PRE,Agarwalla2017PRB,Xiao2018PLA,Niedenzu2018,Wang2019PRE,Zhang2020QIP,Zhang2020PhysA,Assis2020PRE} also follow essentially the same design principles.

\subsection{Exploiting quantum coherence}

Arguably, the most prominent example among the early analyses of a heat engine that exploits genuine quantum effects is the photo-Carnot engine proposed by Scully \etal \cite{Scully2003} Inspired by the discovery of \emph{lasing without inversion}\cite{Kocharovskaya1992},  Scully \etal \cite{Scully2003} considered a scenario in which the hot heat bath supports a small amount of quantum coherence.

More specifically, in the photo-Carnot engine the working medium is simply the radiation pressure,  and the actual ``engine'' is a microlaser cavity \cite{Meschede1985,An1994}. Its mechanical equation of state can be written as
\begin{equation}
P V=\hbar\omega\,\bar{n}\,,
\end{equation}
where $P$ is the radiation pressure,  $V$ is the cavity volume,  $\omega$ its frequency, and $\bar{n}$ is the average number of photons in the mode in thermal equilibrium.

\begin{figure}
\centering
\includegraphics[width=.4\textwidth]{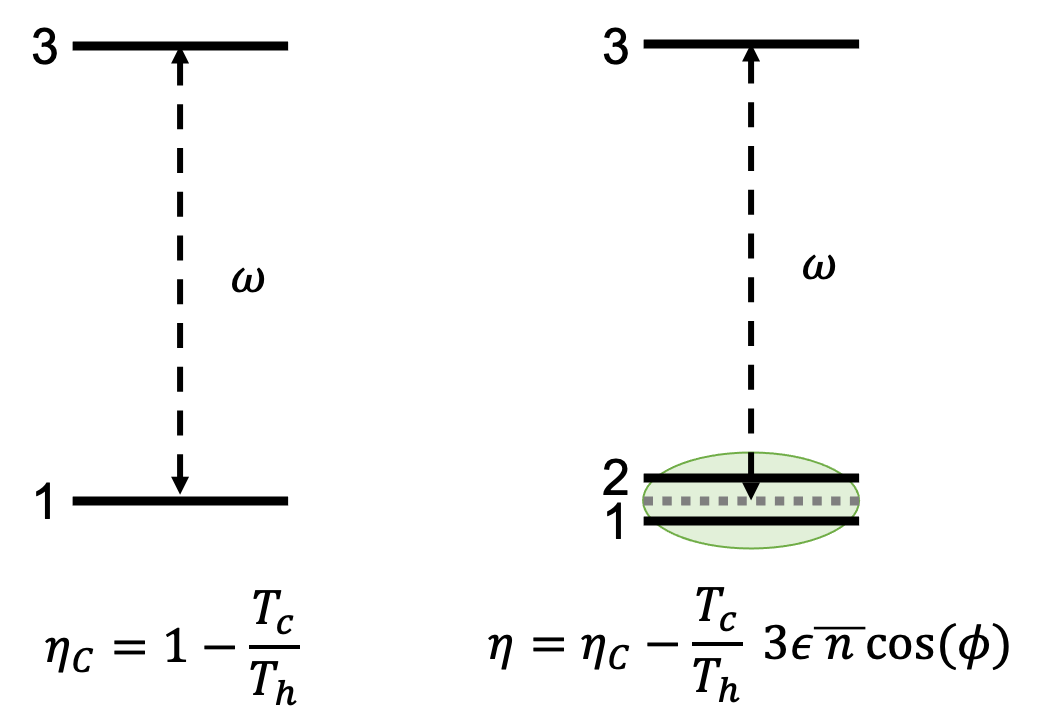}
\caption{\label{fig:phaseonium} Two-level atoms (left panel) vs. phaseonium (right panel). The quantum coherence between energy levels 1 and 2 can be exploited to outperform the classical Carnot efficiency.}
\end{figure}

To exploit quantum effects, the hot reservoir is then  assumed to consist of phaseonium,  a three level atom whose the lower two states are nearly energetically degenerate.  These nearly degenerate states are prepared so that they support a small amount of quantum coherence, see Fig.~\ref{fig:phaseonium} for a sketch.  Scully \etal \cite{Scully2003} showed that the average number of photons in the cavity can be written as
\begin{equation}
\bar{n}_\phi=\frac{1}{\beta_h\hbar \omega}\,\left(1-\epsilon\,\bar{n}\co{\phi}\right)\,,
\end{equation}
where $\epsilon$ measures the magnitude of the coherence, and $\phi$ is the phase.

Since the radiation field generated by the phaseonium is still thermal, we can further identify
\begin{equation}
T_\phi=\hbar \omega\,\bar{n}_\phi/k_B=T_h \left(1-\epsilon\,\bar{n}\co{\phi}\right)\,.
\end{equation}
The photo-Carnot engine otherwise operates similarly to a standard Carnot engine.  Therefore, it is a simple exercise to show that its efficiency becomes
\begin{equation}
\label{eq:photo_C}
\eta=\eta_C-\frac{T_c}{T_h}\,3\epsilon \,\bar{n}\co{\phi}\,.
\end{equation}
We immediately observe that in the case of vanishing coherence, $\epsilon=0$, the standard Carnot efficiency is recovered. However, we also see that there are particular phase values $\phi$ for which this photo-Carnot engine outperforms classical devices.

In their interpretation of this result, Scully \etal \cite{Scully2003} are very clear. While Eq.~\eqref{eq:photo_C} is in full agreement with the second law of thermodynamics, it does demonstrate that there are genuine quantum resources that can be exploited through judicious design of heat engines. Thus,  it may not be too bold to claim that Scully \etal \cite{Scully2003} put the ``quantum'' into quantum heat engines, and it inspired significant work on thermodynamic devices with quantum coherences \cite{Quan2006PRE,Scully2011,Rahav2012PRA,Harbola2012EPL,Killoran2015JCP,Chen2016PRE,Uzdin2016,Chen2017entropy,Peng2018IJTP,Dorfman2018PRE,Camati2019PRA,Shi2020JPA,Hammam2021NJP}, and entanglement \cite{Zhang2007PRA,Zhang2008EPJD,Wang2009PRE,He2012CPB,He2012PhysA,Huang2013PS,Albayrak2013,Sun2017EPJD,Brunner2014PRE,Zhao2017QIP,Yon2020PhysA}. 
 
In the following sections, we will be following a similar logic. Motivated and inspired by experimentally realistic scenarios we survey existing proposals and realizations of heat engines that exploit quantum resources. However, it is worth emphasizing that none of the following scenarios constitute a violation of physical principles of thermodynamics.

%% file: sections/theory_cycles.tex
\section{\label{sec:theory} Theoretical preliminaries and concepts}

The principal component of any cyclic heat engine analysis is the identification of the heat exchanged and work done during each stroke. Let us consider the general case of a quantum system, described by density operator $\rho$, coupled to a thermal environment. We take the Hamiltonian for the system to be $H(\lambda)$, where $\lambda$ is an external control parameter. The system's dynamics can then be described by $\dot{\rho} = \mathcal{L}_{\lambda}(\rho)$ where the superoperator $\mathcal{L}_{\lambda}$ accounts for the unitary dynamics generated by $H(\lambda)$ as well as the nonunitary evolution that arises from the interaction of the system with the thermal environment. In the limit of ultraweak coupling the equilibrium state of the system is the Gibbs state \cite{Deffner2019book},
\begin{equation}
	\rho^{\mathrm{eq}} = \frac{1}{Z} \exp\left(-\beta H\right),
\end{equation}
where $Z = \tr{\exp\left(-\beta H\right)}$ is the partition function. The internal energy of the system can be found from, 
\begin{equation}
	\label{eq:internalE}
	E = \langle H \rangle = \tr{\rho^{\mathrm{eq}} H}.
\end{equation}
For the Gibbs state, the thermodynamic entropy is given by the Gibbs entropy, $S = -\tr{\rho^{\mathrm{eq}} \ln \rho^{\mathrm{eq}}}$. For an isothermal, quasistatic process the change in entropy is then \cite{Deffner2019book},  
\begin{equation}
	\label{eq:entropyChange}
	dS = \beta \, \tr{d\rho^{\mathrm{eq}}H}.
\end{equation}
Using Eq. \eqref{eq:internalE} we can separate the change in internal energy, $dE$ into two contributions, one associated with a change in entropy and the other from a change in the Hamiltonian \cite{Deffner2019book},
\begin{equation}
	\label{eq:FirstLaw}
	dE = \tr{d\rho^{\mathrm{eq}} H}+\tr{\rho^{\mathrm{eq}} dH} \equiv \dbar Q + \dbar W. 
\end{equation}
In complete analogy to classical thermodynamics we can define the first as heat and the second as work. 

It is important to note that these definitions of heat and work are valid \textit{if and only if} $\rho^{\mathrm{eq}}$ is a Gibbs state \cite{Gardas2015}. In the case that the system-environment coupling is not ultraweak,  the energetic back-action due to correlations between the system and environment must be accounted for. For a non-Gibbsian equilibrium state $\rho^{\mathrm{ss}}$ the change in entropy can be expressed as \cite{Gardas2015},
\begin{equation}
	d\mathcal{H} = \beta \left(\dbar Q_{\mathrm{tot}} - \dbar Q_{\mathrm{c}}\right),
\end{equation}     
where $\dbar Q_{\mathrm{tot}} \equiv \tr{d\rho^{\mathrm{ss}}H}$ is the total heat and $\dbar Q_{\mathrm{c}} \equiv d\mathcal{F} - \tr{\rho^{\mathrm{ss}}dH}$ is the energetic price to maintain quantum coherence and correlations. Here $\mathcal{F}$ is the ``information free energy" \cite{Sagawa2015}, which is determined from the Helmholtz free energy and the quantum relative entropy between $\rho^{\mathrm{ss}}$ and $\rho^{\mathrm{eq}}$.     

\subsection{Reversible quantum cycles}

With heat and work identified, quantum analogues of the classical isothermal, isochoric, adiabatic, and isobaric processes can be found \cite{Bender2000JPA,Feldmann2003PRE,Quan2007PRE, Quan2009PRE}, allowing for the study of quantum implementations of heat engine cycles including Carnot, Otto, Stirling, Brayton and Diesel.  

For instance,  Ref. \cite{Allahverdyan2008PRE} provides a detailed examination of the performance optimization of a quantum heat engine consisting of two finite-level quantum systems both coupled to a work source and separately to thermal baths at two different temperatures. Notably this analysis is not limited to the equilibrium regime, with the intermediate states of the engine allowed to be arbitrarily far from equilibrium. For finite-time performance the Curzon-Ahlborn efficiency \cite{Curzon1975} is found to be the \textit{lower} bound on the efficiency when the work output is maximized, achieved in the macroscopic limit. Furthermore, it is shown that as the efficiency approaches the Carnot efficiency, the work output of the finite-time engine vanishes. However, if the system-bath interaction is optimized, finite work output can be achieved at close to Carnot efficiency, provided that the cycle duration is long.         

Outside of the realm of equilibrium systems, the notion of temperature is no longer straightforward to define. Ref. \cite{Johal2009PRE} examines this question using a quantum engine consisting of two qubits prepared in different thermal states that then undergo a thermally isolated unitary work extraction process. The final state of the total system is a nonequilibrium one, with a different local temperature for each subsystem. The behavior of three different definitions for the nonequilibrium temperature of the composite system are examined, and it is shown that, while all three definitions agree at mutual equilibrium, in general they show radically different behavior.

A motivating factor in the study of quantum heat engines is the idea that quantum resources can be exploited to enhance the engine performance. This prompts the immediate question of which parameter regimes the engine must operate in in order to maintain its quantum nature. This is the primary question of Ref. \cite{Friedenberger2017EPL} which uses violations of the Leggett-Garg inequality to quantify the regimes in which a  Otto cycle with a two-level working medium displays nonclassical properties. The cycle is found to operate in three distinct regimes, one in which the dynamics are entirely classical, one in which they are entirely quantum, and a third transition regime in which the dynamics are quantum over certain temperature ranges and classical in others. Furthermore, it is shown that decreasing the cycle duration to avoid decoherence can actually lead to incoherent dynamics arising from thermodynamic constraints.         

Finally, it is important to highlight the distinction between cyclic and continuous heat engines. This distinction is the central focus of Ref. \cite{Humphrey2005PhysE}. In a cyclic engine the working medium interacts alternately between a hot and cold reservoir in a cyclic process, while for continuous engines heat is exchanged between the reservoirs via a flow of particles that do work against an external field during this process. For this reason Ref. \cite{Humphrey2005PhysE} refers to continuous thermal machines as ``particle-exchange" engines. Common examples of continuous heat engines include thermionic, thermoelectric, and photovoltaic devices. Notably the conditions for reversibility are distinct for each implementation, with cyclic engines achieving reversibility when the heat transfer is isothermal, while continuous engines achieve reversibility when the particle transfer is isentropic \cite{Humphrey2005PhysE}. Ref. \cite{Humphrey2005PhysE} notes that the prototypical quantum heat engine, the three level maser (see Sec. \ref{sec:maser}), should be properly classified as a continuous engine.

In the following sections we will elaborate on the various heat engine and refrigerator cycles, as well as continuous thermodynamic devices in the context of realistically available physical platforms. Before discussing specific systems, however, it is instructive to outline the main concepts and notions.

\subsubsection{Quantum Carnot engines}

In Ref. \cite{Alicki1979} Eq. \eqref{eq:FirstLaw} was applied to derive bounds on the efficiency of a heat engine consisting of an open quantum system weakly coupled to $N$ independent thermal reservoirs. For slowly varying external conditions, the evolution of the system can be described by a Markovian master equation \cite{Alicki1979}. Over the course of a full cycle of period $\tau$ the periodicity conditions \cite{Alicki1979},
\begin{equation}
	H_0 = H_{\tau}, \,\, \tr{\rho_0 H_0} = \tr{\rho_{\tau} H_{\tau}}, \,\, \text{and} \,\, S(\rho_0) = S(\rho_{\tau}) 
\end{equation}
must be followed. The total work performed by the system per cycle is,
\begin{equation}
	\label{eq:Alickiwork}
	-W = - \int_{0}^{\tau} dt \, \tr{\rho_t \dot{H_t}} = \int_{0}^{\tau} dt \, \tr{\dot{\rho_t} H_t} = Q_1 + Q_2, 
\end{equation}
where $Q_1$ and $Q_2$ are the total heat exchanged with the hot and cold reservoirs, respectively, over a full cycle. Taking into account the periodicity condition for the entropy leads to,
\begin{equation}
	\label{eq:EntropyCond}
	\int_{0}^{\tau} dt \, \dot{S}(\rho) = \int_{0}^{\tau} dt \, \sigma(t) + \beta_1 Q_1 + \beta_2 Q_2 = 0,
\end{equation}          
where $\sigma(t)$ is the entropy production. Noting that $\sigma(t) \geq 0$ Eq. \eqref{eq:EntropyCond} yields \cite{Alicki1979} 
\begin{equation}
	\label{eq:AlickiHeat}
	\beta_1 Q_1 + \beta_2 Q_2 \leq 0.
\end{equation} 
By combining Eqs. \eqref{eq:Alickiwork} and \eqref{eq:AlickiHeat} it can immediately be seen that the efficiency of this cyclic engine is bounded by the Carnot efficiency \cite{Alicki1979},
\begin{equation}
	\eta = - \frac{W}{Q_1} \leq 1 - \frac{T_2}{T_1}.
\end{equation}  

Outside of the assumption of weak coupling between the system and the thermal environments, the energetic cost of the system-environment interaction must be carefully accounted for. Such an analysis is carried out in Ref. \cite{Newman2017PRE}, however it is important to note that doing so requires specifying the microscopic model of the heat engine.  

The availability of quantum resources, such as coherence and entanglement, opens up a natural question as to whether these resources can be leveraged to design an engine that can outperform the Carnot efficiency. Reference \cite{Gardas2015} demonstrates that this is not the case, and that no quantum heat engine operating in a quasistatic Carnot cycle can harness quantum correlations. More specifically, the energetic back action arising from the correlation of system and environment must be accounted for. During any quasistatic process a portion of the energy exchanged with the environment is paid as an energetic price to maintain the necessarily non-Gibbsian state arising from the presence of coherences and correlations. When this energetic cost is properly accounted for the Carnot efficiency is maintained as the fundamental bound on cycle efficiency. In Ref. \cite{Gardas2016} it has even been shown that the Carnot bound holds beyond the typical framework of Hermitian quantum mechanics and can be extended to pseudo-hermitian systems. 

The quantum Carnot cycle has been studied for a range of working mediums including single spins \cite{Tonner2006FP}, and two level systems, where it has been shown that the Carnot efficiency can be achieved in experimentally relevant scenarios by cycling over a range of values for the energy gap \cite{Beretta2012EPL}.

\subsubsection{Quantum Otto engines}

As the basis for the internal combustion engine, the Otto cycle is perhaps the most widely implemented thermodynamic cycle in practice. This lends particular practical importance to understanding how the Otto cycle can be implemented for a quantum working medium. The classical Otto cycle consists of four strokes: (1) adiabatic compression, (2) isochoric heating (3), adiabatic expansion, and (4) isochoric cooling \cite{Callen1985}. 

Let us consider the adiabatic strokes first. By definition, an adiabatic process is one in which no heat transfer between the working medium and thermal environment occurs. Considering Eq. \eqref{eq:FirstLaw} such a process corresponds to one in which the eigenenergies are varied while the populations remain fixed, thus ensuring that any change in the internal energy is associated with work. As the populations of each energy level do not change, we can immediately see that the von Neumann entropy of the system remains constant. In the framework of classical thermodynamics an adiabatic process is typically accomplished by performing the compression or expansion process rapidly enough that no heat exchange can occur. However, for a quantum working medium the quantum adiabatic theorem guarantees that any rapid perturbation will generate excitations that will lead to changes in the energy level populations. Thus, to remain in accordance with the adiabatic theorem, the adiabatic strokes of the quantum Otto cycle must be performed \textit{quasistatically}. This is a crucial difference between the classical and quantum implementations of the Otto cycle \cite{Kosloff2017entropy,Deffner2019book}.      

The fact that each stroke of the Otto cycle corresponds to an exchange of either work or heat, but never both simultaneously, makes it particularly attractive for implementation in quantum systems. Work and heat can be immediately identified from the change in internal energy during each stroke, without the need to differentiate each contribution, as must be done in the case of, e.g., isothermal processes.

Let us now consider the isochoric strokes. During the heating (cooling) stroke the working medium is placed in contact with the hot (cold) bath until it achieves thermal equilibrium. The external control parameter of the Hamiltonian is held fixed during this process. Considering Eq. \eqref{eq:FirstLaw}, we see that such a process corresponds to one in which the eigenenergies remain fixed while the populations of each energy level change, thus ensuring that any change in internal energy is associated with heat. As the populations of each energy level are varying, we note that this change in internal energy is associated with a change in the von Neumann entropy.

It is important to note that the quantum Otto cycle is inherently irreversible. For most of the cycle the working medium is \textit{not} in a state of thermal equilibrium, only achieving such a state at the ends of the heating and cooling strokes. The process of thermalization that takes place during these strokes is fundamentally irreversible and leads to the overall irreversibility of the cycle.  

In Ref. \cite{Quan2007PRE} the quantum Otto cycle is analyzed for a multilevel working medium with energy eigenvalues $E_n$ (with $n = 0,1,2,...$). During the adiabatic processes it is assumed that the energy gaps between each level change by the same ratio $\alpha$, such that,
\begin{equation}
	E_n^h - E_m^h = \alpha \left( E_n^l - E_m^l \right),
\end{equation}
where $E_n^h$ ($E_n^l$) is the $n$th eigenenergy of the system during the isochoric heating (cooling) process. The heat absorbed by the working medium during the isochoric heating process is,
\begin{equation}
	\label{eq:OttoHeatIn}
	Q_{\textrm{in}} = \sum_n \int_{A}^{B}d\wp_n\, E_n  = \sum_n E_n^h \left[\wp_n(B) - \wp_n(A)\right]\,,
\end{equation}      
where $\wp_n$ is the occupation probability of the $n$th energy eigenstate. Similarly, the heat released to the low temperature bath during the isochoric cooling process is, 
\begin{equation}
	\label{eq:OttoHeatOut}
	Q_{\textrm{out}} = -\sum_n \int_{C}^{D}d\wp_n\, E_n  = \sum_n E_n^l \left[\wp_n(C) - \wp_n(D)\right].
\end{equation}
The adiabatic nature of the expansion and compression strokes means that the occupation probabilities must fulfill the conditions
\begin{equation}
	\label{eq:OttoCond}
	\wp_n(B) = \wp_n(C) \quad \text{and} \quad \wp_n(A) = \wp_n(D).
\end{equation} 
Using Eqs. \eqref{eq:OttoHeatIn}, \eqref{eq:OttoHeatOut}, and \eqref{eq:OttoCond} we can determine the net work done during the cycle,
\begin{equation}
	W = Q_{\textrm{in}} - Q_{\textrm{out}} = \sum_n \left(E_n^h - E_n^l\right) \left[\wp_n(B) - \wp_n(A)\right].
\end{equation}
The efficiency can now be found as the ratio of the net work to the absorbed heat,
\begin{equation}
	\eta = \frac{W}{Q_{\textrm{in}}} = 1 - \frac{E_n^l - E_m^l}{E_n^h - E_m^h} = 1 - \frac{1}{\alpha}.
\end{equation} 
It is worth noting that this efficiency is completely analogous to the classical Otto efficiency, with the parameter $\alpha$ playing the role of an inverse ``compression ratio."   

The quantum Otto cycle has been examined in a huge variety of contexts including, but not limited to, implementations with working mediums of single spin systems \cite{Geva1992JCP,Henrich2007EPJST, Feldmann2000}, coupled spin systems \cite{Huang2013PS, Huang2014EPJP, Feldmann2004PRE}, harmonic oscillators \cite{Abah2016EPL, Abah2012, Myers2020PRE}, relativistic oscillators \cite{Myers2021NJP}, an ideal Bose gas \cite{Wang2009JAP}, a Bose-Einstein condensate \cite{MyersBECArXiv}, anyons \cite{Myers2021PRXQ, Myers2021Symm}, a two-level atom \cite{Wang2012PRE}, coupled spin-3/2 biquartits \cite{Ivanchenko2015PRE}, and an $\mathrm{NI}_2$ dimer \cite{Hubner2014PRB}. Furthermore, it has been shown that the cycle performance can be enhanced with a ``quantum afterburner" \cite{Rostovtsev2003PRE} (as elaborated on in Sec.~\ref{sec:afterburner}), non-Markovian reservoirs \cite{Zhang2014JPA}, and nonequilibrium effects \cite{Li2013EPJD}. 

\subsubsection{Quantum Stirling and beyond}

While Carnot and Otto are perhaps the most ubiquitously studied, there exist a range of other heat engine cycles including the Stirling, Diesel, and Brayton cycles. As with Carnot and Otto, these cycles can also be generalized to quantum working mediums.

\begin{figure}
	\includegraphics[width=.48\textwidth]{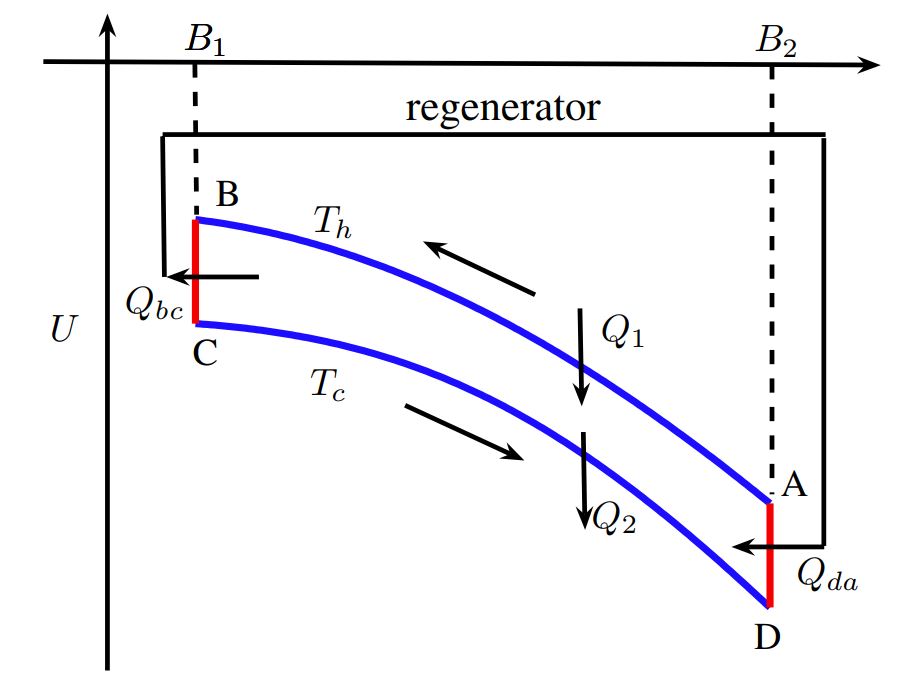}
	\caption{\label{fig:StirlingDiagram} Magnetic field - energy diagram of a quantum Stirling cycle. \textit{Figure adopted from Ref.~\cite{Huang2014EPJD}}.}
\end{figure}

Reference \cite{Huang2014EPJD} examines the performance of a quantum Stirling engine with a single spin or pair of coupled spins under an external magnetic field as the working medium. Like its classical counterpart, the quantum Stirling cycle consists of four strokes, illustrated in Fig. \ref{fig:StirlingDiagram}:

(1) \textit{Isothermal expansion:} While the working medium remains in contact with the hot bath the external magnetic field is decreased slowly from $B_2$ to $B_1$ such that the working medium remains in thermal equilibrium with the hot bath at temperature $T_h$. Work is done by the working medium and heat is absorbed from the hot bath during this process. 

(2) \textit{Isochoric cooling:} The working medium is placed in contact with the cold bath and allowed to cool to temperature $T_c$ while the magnetic field is held constant at $B_1$. No work is done and heat is absorbed by the cold bath during this process. 

(3) \textit{Isothermal compression:} While the working medium remains in contact with the cold bath the external magnetic field is increased slowly from $B_1$ to $B_2$ such that the working medium remains in thermal equilibrium with the cold bath at temperature $T_c$. Work is done on the working medium and heat is absorbed by the cold bath during this process. 

(4) \textit{Isochoric heating:} The working medium is placed in contact with the hot bath and allowed to heat to temperature $T_h$ while the magnetic field is held constant at $B_2$. No work is done and heat is absorbed from the hot bath during this process.

The use of a regenerator is often considered when analyzing the performance of the Stirling cycle. The regenerator improves the cycle efficiency by capturing the heat released by the working medium to the hot bath during the isochoric cooling stroke and allowing it to be absorbed back into the working medium during the isochoric heating stroke.         

For a single spin the Hamiltonian of the working medium is simply, $H = B \sigma^z/2$,  where $B$ is the strength of the external magnetic field and $\sigma^z$ is the Pauli spin operator. The heat exchanged during the isothermal processes can be found by using $\dbar Q = T dS$ where $S$ is, as before, the von Neumann entropy. Thus,
\begin{equation}
	\begin{split}
		Q_1 = &\int_{A}^{B} dS \, T = T_h \left(S_B - S_A \right) \\
		& = \frac{1}{\beta_h} \ln\left[\cosh \left(\frac{\beta_h B_1}{2}\right)\right] - \frac{1}{\beta_h} \ln\left[\cosh \left(\frac{\beta_h B_2}{2}\right)\right]\\ 
		& \qquad + \frac{B_2}{2}\tanh\left(\frac{\beta_h B_2}{2}\right) - \frac{B_1}{2}\tanh\left(\frac{\beta_h B_1}{2}\right).
	\end{split}
\end{equation}   
Similarly,
\begin{equation}
	\begin{split}
		Q_2 = &\int_{C}^{D} dS \, T = T_h \left(S_D - S_C \right) \\
		& = \frac{1}{\beta_c} \ln\left[\cosh \left(\frac{\beta_c B_2}{2}\right)\right] - \frac{1}{\beta_c} \ln\left[\cosh \left(\frac{\beta_c B_1}{2}\right)\right]\\ 
		& \qquad + \frac{B_1}{2}\tanh\left(\frac{\beta_c B_1}{2}\right) - \frac{B_2}{2}\tanh\left(\frac{\beta_c B_2}{2}\right).
	\end{split}
\end{equation}
During the isochoric processes, as no work is done, the change in internal energy of the working medium can be entirely attributed to heat. As such,
\begin{equation}
	Q_{bc} = U_C - U_B = \frac{B_1}{2}\left[\tanh\left(\frac{\beta_h B_1}{2}\right) - \tanh\left(\frac{\beta_c B_1}{2}\right)\right].
\end{equation}
Similarly,  
\begin{equation}
	Q_{da} = U_A - U_D = \frac{B_2}{2}\left[\tanh\left(\frac{\beta_c B_2}{2}\right) - \tanh\left(\frac{\beta_h B_2}{2}\right)\right].
\end{equation}
The net work per cycle can then be found using the first law, $W = \sum Q$. Thus,
\begin{equation}
	W = \frac{1}{\beta_h} \ln\left[\frac{\cosh \left(\beta_h B_1/2\right)}{\cosh \left(\beta_h B_2/2\right)}\right] + \frac{1}{\beta_c} \ln\left[\frac{\cosh \left(\beta_c B_2/2\right)}{\cosh \left(\beta_c B_1/2\right)}\right]. 
\end{equation}

The heat transfer between the system and the regenerator is given by the sum of the heats exchanged during the isochoric strokes,
\begin{equation}
	\Delta Q = Q_{bc} + Q_{da}. 
\end{equation}   
For a classical ideal gas working medium the Stirling cycle is perfectly regenerative, with $\Delta Q = 0$ always. Reference~\cite{Huang2014EPJD} highlights that this is not always the case for the quantum Stirling cycle, allowing for three possibilities. If $\Delta Q = 0$ perfect regeneration is achieved, however this only occurs for a restrictive set of values of the bath temperatures and magnetic field strengths. If $\Delta Q < 0$ the regenerator absorbs more heat than it releases. This redundant heat must be dissipated to the cold bath to keep it from building up in the regenerator. If $\Delta Q > 0$ the regenerator absorbs less heat than it releases. In this case additional heat from the hot bath must be absorbed by the working medium in compensation.

The efficiency can be found in the typical manner, by taking the ratio of the net work and the heat absorbed from the hot bath. For the classical Stirling cycle the efficiency is identical to the Carnot efficiency. This is no longer true for the quantum Stirling cycle, with the efficiency depending not just on the bath temperatures, but also on the initial and final magnetic field strengths. The efficiency is maximized for initial and final magnetic field strengths that meet the perfect regeneration of $\Delta Q = 0$. For some parameter regimes the maximum efficiency can even exceed the Carnot efficiency, however Ref. \cite{Huang2014EPJD} notes that this is not in violation of the second law as in these regimes the regenerator consumes additional energy.      

The authors extend the analysis to a working medium of two coupled spins, resulting in an additional tunable parameter in the coupling strength. Notably for coupled spins the condition of perfect regeneration can more easily be met for arbitrary bath temperatures and magnetic field strengths by adjusting the coupling strength.  The analysis of the quantum Stirling cycle has been extended to the finite time regime, using a two-level system as the working medium \cite{Raja2021NJP}. 

The Diesel cycle, consisting of one isobaric, one isochoric, and two isentropic strokes, has also been examined for ideal Bose and Fermi gas working mediums \cite{He2009ECM}. Similarly, the quantum Brayton cycle, consisting of two isobaric and two isentropic strokes, has been analyzed for a variety of working mediums, including a harmonic oscillator \cite{Wang2013PS}, coupled spins \cite{Huang2013PRE}, noninteracting spins \cite{Lin2004JPA}, and an ideal Bose gas \cite{Wang2008PhysB}. Quantum working mediums also allow for the implementation of non-conventional cycles,. This includes cycles that extract work from a single heat bath whose energy input arises from nonselective measurement of the working medium \cite{Ding2018PRE}, cycles that incorporate isoenergetic strokes during which the expectation value of the Hamiltonian is held constant \cite{Wang2012JAP, Munoz2012PRE, Ou2016EPL, Liu2016entropy, Pena2016PRE, Gabriel2018entropy, WangR2012PRE}, and cycles that utilize non-thermal baths \cite{Cherubim2019entropy}.           

%% file: sections/theory_endoreversible.tex
\subsection{Endoreversiblity and finite power}

\label{sec:endo}

In equilibrium thermodynamics the optimal efficiency of any heat engine cycle is bounded by the Carnot efficiency, regardless of the properties of the working medium \cite{Callen1985}. However, this efficiency is obtained in the limit of infinitely slow, quasistatic strokes, resulting in zero power output. A figure of merit of more practical use, the \textit{efficiency at maximum power} (EMP), was introduced by Curzon and Ahlborn using the framework of \textit{endoreversible thermodynamics} \cite{Curzon1975, Rubin1979, Hoffmann1997}. In endoreversible thermodynamics the system is assumed to be in a state of \textit{local equilibrium} at all times, but with dynamics that occur quickly enough that \textit{global equilibrium} with the environment is not achieved. This results in a process that is locally reversible, but globally irreversible \cite{Hoffmann1997}.  In the context of heat engines, this means that while the working medium may be in a local equilibrium state at temperature $T$, there is a thermal gradient at the boundaries where the working medium comes into contact with the bath at temperature $T_{\mathrm{bath}}$. 

Curzon and Ahlborn \cite{Curzon1975} found the EMP of a endoreversible Carnot engine to be,
\begin{equation}
	\eta_{\mathrm{CA}} = 1 - \sqrt{\frac{T_c}{T_h}}\,,
\end{equation}  
where $T_c$ ($T_h$) is the cold (hot) reservoir temperature. Remarkably, this result, now known as the \textit{Curzon-Ahlborn} (CA) efficiency, has emerged in the analysis of heat engines in a wide variety of contexts outside the original case of a classical working medium undergoing a finite-time Carnot cycle. The EMP of endoreversible Otto \cite{Leff1987}, Brayton \cite{Leff1987}, and Stirling \cite{Erbay1997} cycles have all been shown to be equivalent to the CA efficiency. Even quantum working mediums, including an open system of harmonic oscillators undergoing a quasistatic Otto cycle \cite{Rezek2006}, a single harmonically trapped particle undergoing an Otto cycle \cite{Abah2012, Rossnagel2014}, and an endoreversible Otto cycle for a relativistic quantum particle in a Dirac oscillator potential \cite{Myers2021NJP}, have been found to assume the CA efficiency.               

This wide applicability seems to suggest that the CA efficiency may be a universal characterization of finite-time performance, akin to the bound on pure efficiency given by the Carnot efficiency. However, despite its ubiquitous appearances, the CA efficiency is not universal, and can be exceeded under the right circumstances. For the case of the finite-time Carnot cycle whether or not CA efficiency is achieved at maximum power is determined by the symmetry of the dissipation during the cold and hot isothermal strokes, with the CA efficiency being achieved in the limit of symmetric dissipation and exceeded when the dissipation is significantly larger during the hot isotherm \cite{Esposito2010}. Furthermore, within the regime of linear response, it has been shown that an Otto cycle with a working medium of a classical harmonic oscillator can come arbitrarily close to the Carnot efficiency at finite power with a specific choice for the parameterization of the potential \cite{Bonanca2019}.     

In Ref. \cite{Deffner2018Entropy} it was shown that the EMP of an endoreversible Otto engine with a single harmonically-trapped quantum particle as the working medium can exceed the Curzon-Ahlborn efficiency.  For an equilibrium thermal state of such a particle, $\rho \propto \text{exp}(-\beta H)$, the corresponding internal energy reads,
\begin{equation}
	\label{eq:IntEnergy}
	E =\la H\ra= \frac{\hbar \omega}{2}\coth \left(\frac{\beta \hbar \omega}{2}\right)\,,
\end{equation}
where as always $\beta=1/k_B T$. The four strokes of the Otto cycle are:    

(1) \textit{Isentropic compression}:  During this stroke the working medium remains in a state of constant entropy, exchanging no heat with the environment. Using the first law $\Delta E = Q + W$ the change in internal energy can be identified completely with work,
\begin{equation}
	\label{eq:Wcomp}
	W_{\mathrm{comp}} = E(T_B, \omega_2) - E(T_A, \omega_1).
\end{equation}  

(2) \textit{Isochoric Heating}: During this stroke the externally-controlled work parameter (the trap frequency in the case of the harmonic engine) is held constant, resulting in zero work. By the first law, the change in internal energy can be completely identified with heat,      
\begin{equation}
	\label{eq:Qh}
	Q_{h} = E(T_C, \omega_2) - E(T_B, \omega_2).
\end{equation}
In the endoreversible regime the working medium does not fully thermalize with the hot reservoir during this stroke. This yields the condition $T_B \le T_C \leq T_h$. The change in temperature during the stroke depends on the properties of the working medium and can be determined using Fourier's law \cite{Callen1985},    
\begin{equation}
	\label{eq:fh}
	\frac{dT}{dt} = -\alpha_h (T(t)-T_h),
\end{equation}
where $\alpha_h$ is a constant determined by the heat capacity and thermal conductivity of the working medium.

(3) \textit{Isentropic expansion}: In exactly the same manner as the compression stroke, the change in internal energy during the expansion stroke can be identified with work,
\begin{equation}
	\label{eq:Wexp}
	W_{\mathrm{exp}} = E(T_D, \omega_1) - E(T_C, \omega_2).
\end{equation} 

(4) \textit{Isochoric Cooling}: As in the heating stroke, the change in internal energy during the cooling stroke is identified with heat,
\begin{equation}
	Q_{c} = E(T_A, \omega_1) - E(T_D, \omega_1).
\end{equation}
The temperature change can again be determined from Fourier's law,
\begin{equation}
	\label{eq:fc}
	\frac{dT}{dt} = -\alpha_c (T(t)-T_c),
\end{equation}
where $T_D > T_A \geq T_c$.

The efficiency of the engine is given by the ratio of the total work and the heat exchanged with the hot reservoir,
\begin{equation}
	\label{eq:eff}
	\eta = -\frac{W_{\mathrm{comp}}+W_{\mathrm{exp}}}{Q_h},
\end{equation}
and the power output by the ratio of the total work to the cycle duration, $\tau_\mrm{cyc}$,,
\begin{equation}
	P = -\frac{W_{\mathrm{comp}}+W_{\mathrm{exp}}}{\gamma (\tau_h + \tau_c)}.
\end{equation} 
Note that only the durations of the heating and cooling strokes are accounted for explicitly, with $\gamma$ serving as a multiplicative factor that implicitly incorporates the duration of the isentropic strokes,  $\tau_\mrm{cyc}\equiv \gamma (\tau_h + \tau_c)$.

In order to ensure that the compression and expansion strokes are isentropic we have the conditions,
\begin{equation}
	T_A\, \omega_2 = T_B\, \omega_1 \quad \text{and} \quad T_C\, \omega_1 = T_D\, \omega_2.
\end{equation} 
Applying these conditions, along with the solutions to Eqs. \eqref{eq:fh} and \eqref{eq:fc}, the efficiency is determined to be,
\begin{equation}
	\eta = 1 - \kappa,
\end{equation} 
where $\kappa = \omega_1/\omega_2$. Similarly, the expression for power can be found in terms of the hot and cold bath temperatures, the thermal conductivities, the initial and final frequencies, and the stroke times. 

\begin{figure}
	\includegraphics[width=.48\textwidth]{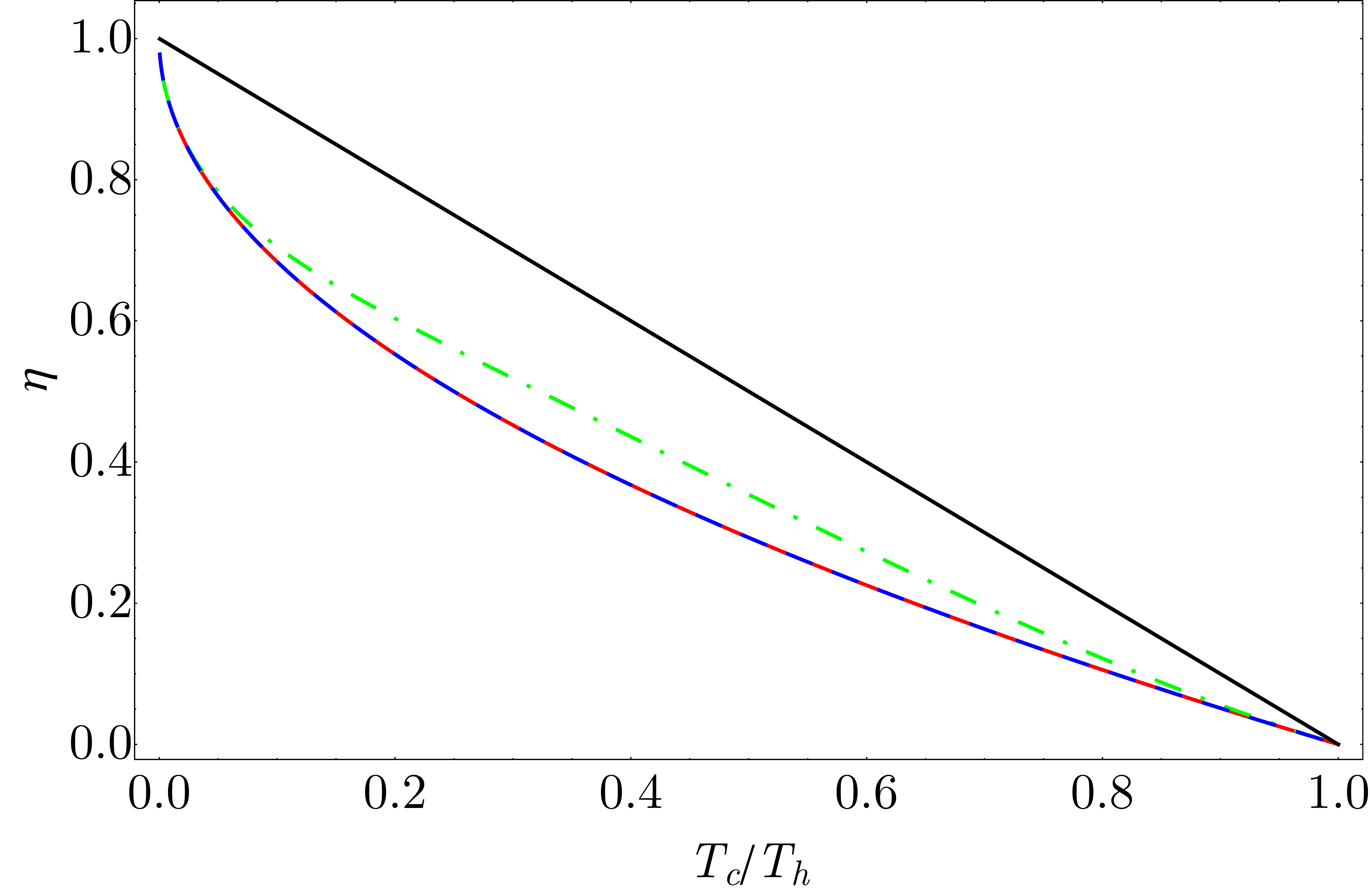}
	\caption{\label{fig:EMPClass} EMP as a function of the ratio of bath temperatures for an endoreversible single particle harmonic Otto engine in the classical regime  corresponding to $\hbar \omega_2/k_{\mathrm{B}} T_c = 0.1$ (blue, dashed) and in the quantum regime corresponding to $\hbar \omega_2/k_{\mathrm{B}} T_c = 10$ (green, dot dashed). The Curzon-Ahlborn efficiency (bottom solid, red) and the Carnot efficiency (top solid, black) are given in comparison. Parameters are $\alpha_c= \alpha_h=\gamma = 1$, and $\tau_c=\tau_h=0.5$. \textit{This figure is original work produced following the methods established in Ref. \cite{Deffner2018Entropy}.}}
\end{figure}

The EMP can then be determined by maximizing the power with respect to the compression ratio $\kappa$. The EMP for the engine operating in the classical regime, corresponding to $\hbar \omega_2/k_{\mathrm{B}} T_c = 0.1$, and in the quantum regime, corresponding to $\hbar \omega_2/k_{\mathrm{B}} T_c = 10$, is shown in Fig. \ref{fig:EMPClass}, with the Carnot and CA efficiencies given for comparison. It is clear that the engine significantly exceeds the CA efficiency when operating in the quantum regime.     

As demonstrated by these results, the EMP of the Otto cycle is not universally given by the CA efficiency, but determined by the nature of the working medium. Further work has investigated the power and EMP of Otto cycles using working mediums with polynomial fundamental relations (e.g. photonic gases) \cite{Smith2020}, two-level working mediums \cite{Feldmann2000, Yan2012PRE}, degenerate quantum gas working mediums \cite{Wang2009ATE},  quantum statistics \cite{Myers2020PRE, Myers2021Symm, Myers2021PRXQ}, and Bose-Einstein condensate working mediums \cite{MyersBECArXiv}. The power and EMP of other cycle types has also been examined, including Carnot cycles with two-level quantum systems as a working medium \cite{Guo2013JAP}, and a uniquely quantum cycle consisting of isoenergetic, isothermal, and adiabatic strokes \cite{Ou2016EPL, Liu2016entropy}. 

%% file: sections/theory_many.tex
\subsection{Single particles vs. collective performance}
\label{sec:many}

As a field, thermodynamics has always been aware of the principle that ``more is different" \cite{Anderson1972S}. Many-body systems display collective behavior beyond the sum of the individual constituent behaviors, with phase transitions serving as a prime example. Quantum many-body systems introduce a range of new collective behaviors such as wave function symmetrization, superradiance, quantum phase transitions, and many-body localization. For quantum thermal machines with multi-particle working mediums, many-body effects can have significant impacts on performance.  Thus, we continue with a particularly instructive situation, for which single vs. many-particle performance has been thoroughly investigated.

\subsubsection{Single particle in harmonic trap}

\begin{figure}
\includegraphics[width=.48\textwidth]{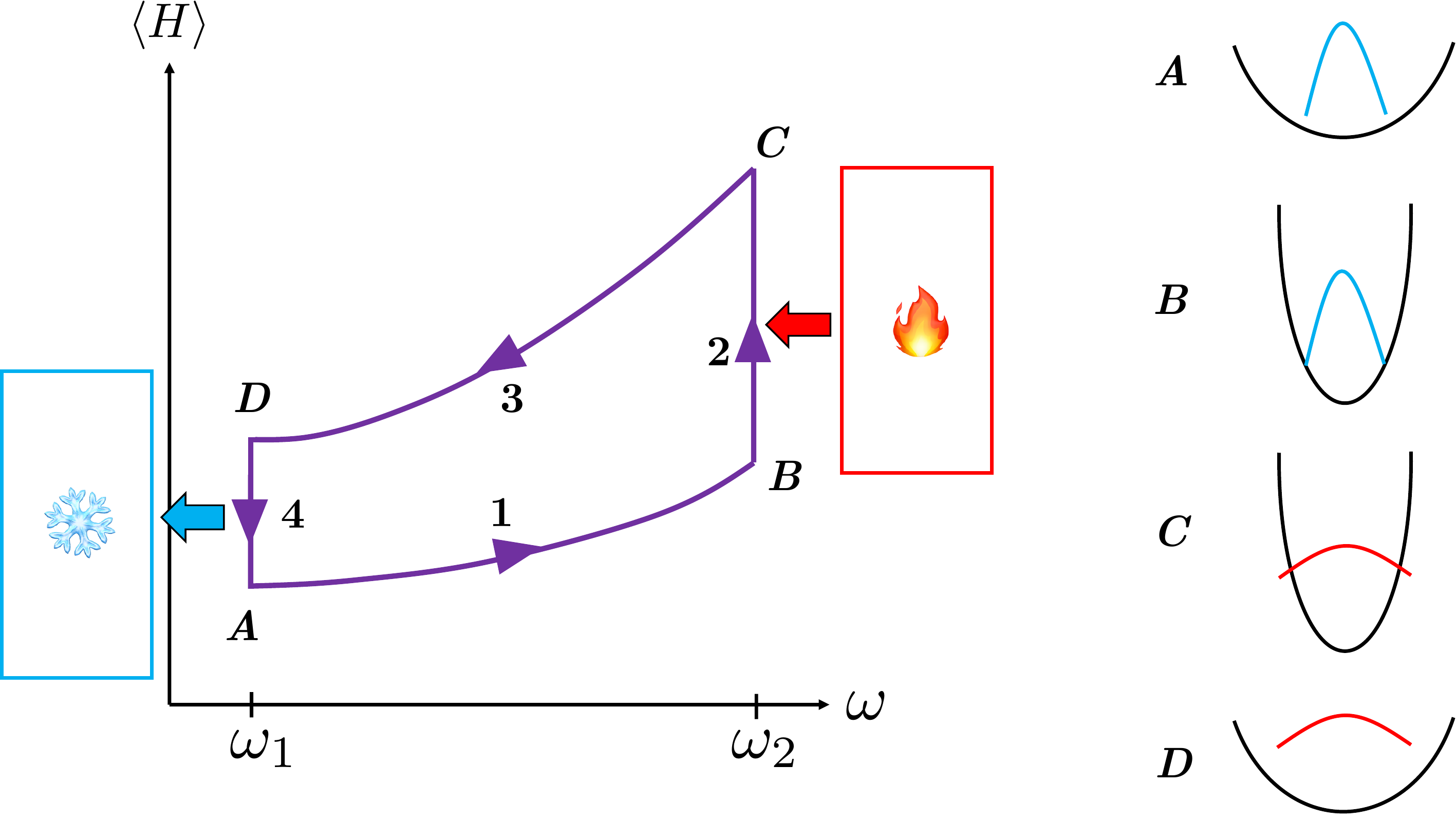}
\caption{\label{fig:STAcycle}  Schematic representation of the four-stroke  quantum Otto  cycle for a harmonic trap with time-dependent frequency.  \textit{Figure adopted from Ref.~\cite{Myers2020PRE}}.}
\end{figure}

We begin with a single particle trapped in a harmonic oscillator, which has probably become the most studied setup for finite-time quantum Otto engines, see for instance  Refs.~\cite{Abah2012,Kosloff2014review,Abah2016EPL,Wiedmann2020NJP}.  Its Hamiltonian reads
\begin{equation}
\label{eq:harm_osc}
H(t)=\frac{p^2}{2 m} + \frac{1}{2}\,m\omega^2(t)\, x^2\,,
\end{equation}
where $x$ and $p$ are the position and momentum operators of an oscillator of mass $m$.  The angular frequency plays the role of ``inverse volume'', and we consider situations in which $\omega(t)$ can be varied between $\omega_1$ and $\omega_2$.  In addition, the particle is alternatingly coupled to two heat baths at inverse temperatures, $\beta_1$ and $\beta_2$, see Fig.~~\ref{fig:STAcycle} for an illustration of the scenario.

The quantum Otto cycle is then implemented through the following four strokes:

(1) \textit{Isentropic compression} $A(\omega_1,\beta_1)\rightarrow B(\omega_2,\beta_1)$: the frequency is varied during time $\tau_1$ while the system is isolated. The evolution is unitary and the von Neumann entropy of the oscillator is thus constant. Note that state $B$ is non-thermal even for slow (adiabatic) processes. In this step, work $W_1=\la H\ra_B-\la H\ra_A$ is performed on the medium, and the average thermal energy is again given by Eq.~\eqref{eq:IntEnergy}.  At the end of the unitary stroke we have  \cite{Husimi53,Deffner2008PRE,Deffner2010CP,Deffner2013PRE},
\begin{equation}
\la H\ra_B=\frac{\hbar \omega_2}{2} Q^\ast_\mrm{comp} \coth\left(\frac{\beta_1\hbar\omega_1}{2}\right)\,.
\end{equation}
Here $Q^*_\mathrm{comp}$ is a dimensionless parameter that measures the degree of adiabaticity of the isentropic compression and expansion strokes, respectively \cite{Husimi53,Deffner2008PRE,Deffner2010CP,Deffner2013PRE}.  Its exact form is determined by the protocol under which the trapping frequency is modulated, but in general $Q^* \geq 1$ with $Q^* = 1$ corresponding to a completely adiabatic stroke.

(2) \textit{Hot isochore} $B(\omega_2,\beta_1)\rightarrow C(\omega_2,\beta_2)$: the oscillator is weakly coupled to a reservoir at inverse temperature $\beta_2$ at fixed frequency and allowed to relax during time $\tau_2$ to the thermal state $C$. This equilibration is much shorter than the expansion/compression strokes and only an amount of heat $Q_2=\la H\ra_C-\la H\ra_B$ is transferred.  Note that at $C$ the system is again in equilibrium, and its energy is accordingly given by Eq.~\eqref{eq:IntEnergy}.

(3) \textit{Isentropic expansion} $C(\omega_2,\beta_2)\rightarrow D(\omega_1,\beta_2)$: the frequency is changed back to its initial value during time $\tau_3$. The  isolated oscillator evolves unitarily into the non-thermal state $D$ at constant entropy. An amount of work $W_3=\la H\ra_D-\la H\ra_C$ is extracted from the medium during this stroke, which we can compute with  \cite{Husimi53,Deffner2008PRE,Deffner2010CP,Deffner2013PRE},
\begin{equation}
\la H\ra_B=\frac{\hbar \omega_1}{2} Q^\ast_\mrm{exp} \coth\left(\frac{\beta_2\hbar\omega_2}{2}\right)\,.
\end{equation}

(4)  \textit{Cold isochore} $D(\omega_1,\beta_2)\rightarrow A(\omega_1,\beta_1)$:  the system is weakly coupled to a reservoir at inverse temperature $\beta_1>\beta_2$  and quickly relaxes to the initial thermal state A during $\tau_4$. The frequency is again kept constant and an amount of heat $Q_4=\la H\ra_A-\la H\ra_D$ is transferred from the working medium.

During the first and third strokes (compression and expansion), the quantum oscillator is isolated,, and the corresponding work values are
\begin{equation}
\begin{split}
W_1 &= \frac{\hbar \omega_2}{2}\left(Q^\ast_\mrm{comp} - \frac{\omega_1}{\omega_2}\right)\coth\left(\frac{\beta_1\hbar\omega_1}{2}\right),\\
W_3&=  \frac{\hbar\omega_1}{2}\left(Q^\ast_\mrm{exp} - \frac{\omega_2}{\omega_1}\right)\coth\left(\frac{\beta_2\hbar\omega_2}{2}\right)\,.
\end{split}
\end{equation}
During the thermalization strokes (isochoric processes),  heat  is exchanged with the reservoirs, and we have 
\begin{equation}
\begin{split}
Q_2&= \frac{\hbar \omega_2}{2} \left[\coth\left(\frac{\beta_2\hbar\omega_2}{2}\right) - Q^\ast_\mathrm{comp}\coth\left(\frac{\beta_1\hbar\omega_1}{2}\right)\right],\\
Q_4&= \frac{\hbar \omega_1}{2} \left[\coth\left(\frac{\beta_1\hbar\omega_1}{2}\right) - Q^\ast_\mathrm{exp}\coth\left(\frac{\beta_2\hbar\omega_2}{2}\right)\right].
\end{split}
\label{eq3}
\end{equation}
The efficiency of this quantum engine, defined as the ratio of the total work per cycle and the heat received from the hot reservoir, then follows as \cite{Abah2012}
\begin{equation}
\label{eff_gen}
\begin{split}
 \eta &= -\frac{W_1+W_3}{Q_2}\\
&=1-\frac{\omega_1}{\omega_2}\,\frac{\ct{\beta_1\,\hbar\omega_1/2} - Q^\ast_\mathrm{exp}\,\ct{\beta_2\,\hbar\omega_2/2}}{Q^\ast_\mathrm{comp}\,\ct{\beta_1\,\hbar\omega_1/2}-\ct{\beta_2\,\hbar\omega_2/2}},
\end{split}
\end{equation}
and the  power output per cycle, $P=-(W_1 + W_3)/\tau_\mrm{cyc}$ becomes
\begin{equation}
\begin{split}
P&=\big[ \langle H\rangle_A \left(1 - Q_\mathrm{comp}^\ast \omega_2/\omega_1\right)\\
&\qquad + \left(1 - Q_\mathrm{exp}^\ast\omega_1/\omega_2\right) \langle H\rangle_C \big]/\tau_\mrm{cyc}\,.
\end{split}
\label{power_gen}
\end{equation}
It is easy to see that for slow driving (adiabatic limit), during the isentropic processes $Q_i^\ast=1$, the thermal machine efficiency is $\eta_O^\text{AD}=1-\omega_1/\omega_2$, whereas the power vanishes.  As we will see shortly,  this single particle engine was realized in ion traps based on a theoretical proposal~\cite{Abah2012}.

\subsubsection{Many particles in harmonic trap}

\begin{figure*}
	\subfigure[]{
		\includegraphics[width=.29\textwidth]{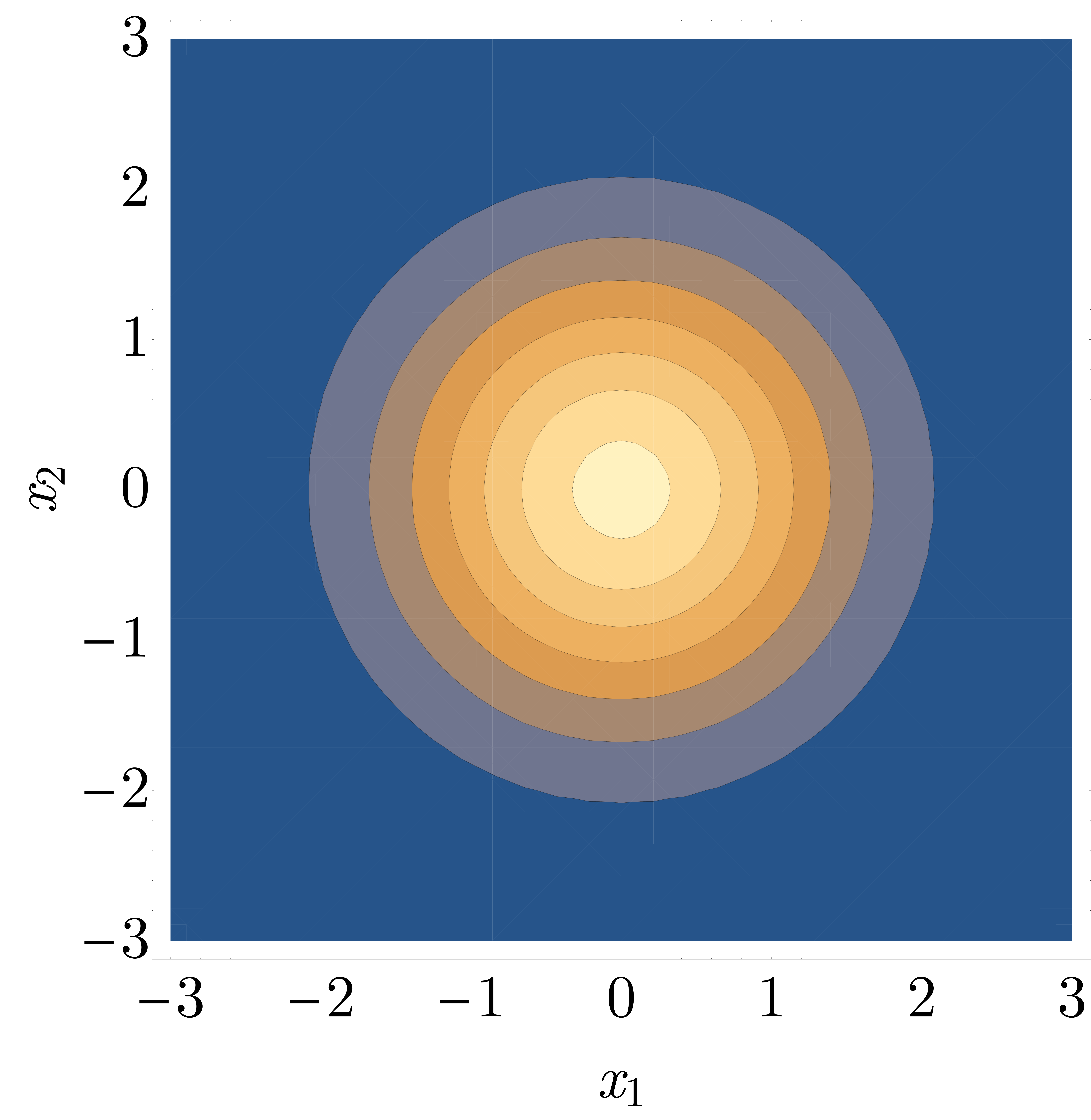}
	}
	\subfigure[]{
		\includegraphics[width=.29\textwidth]{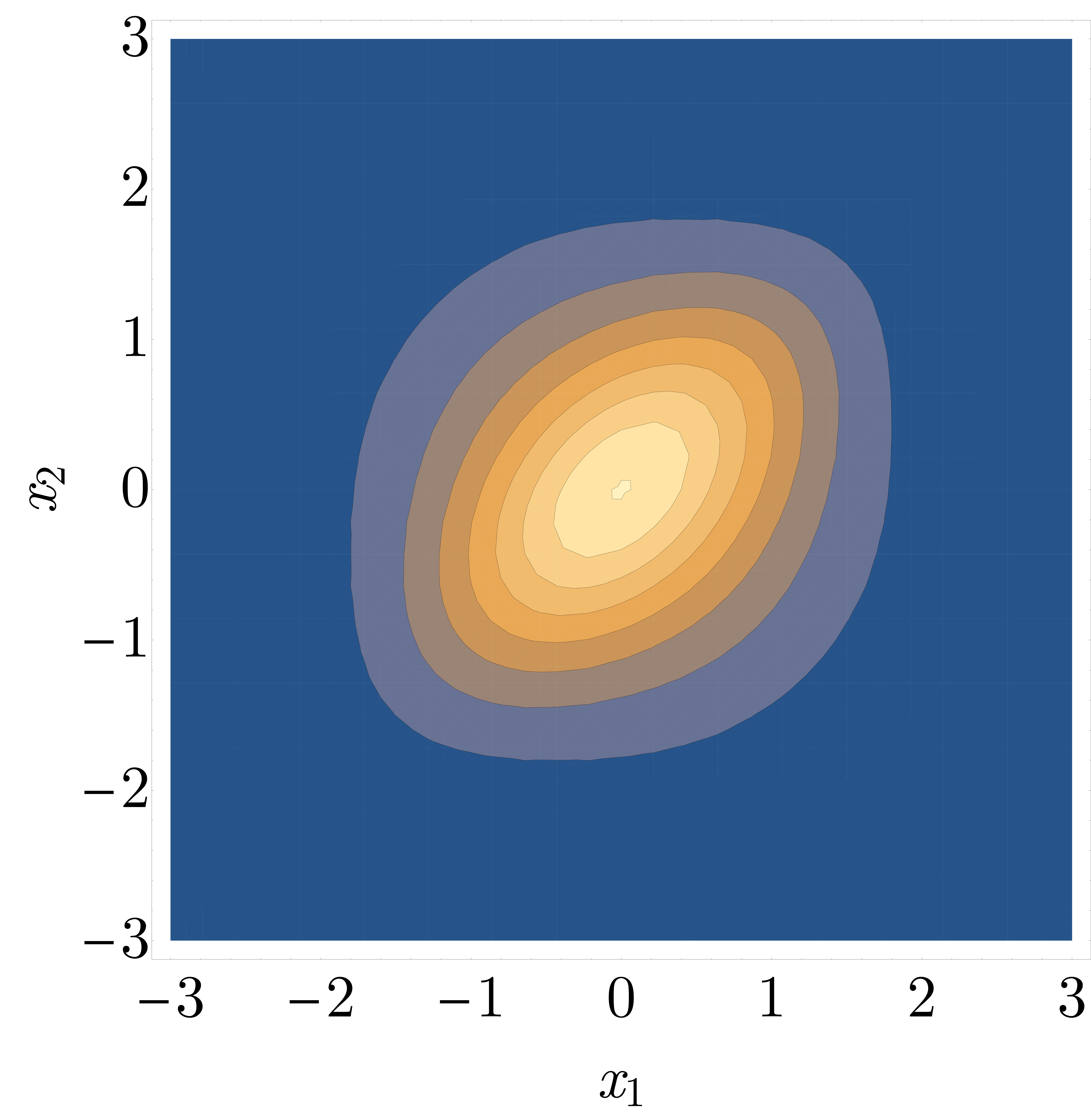}
	}
	\subfigure[]{
		\includegraphics[width=.29\textwidth]{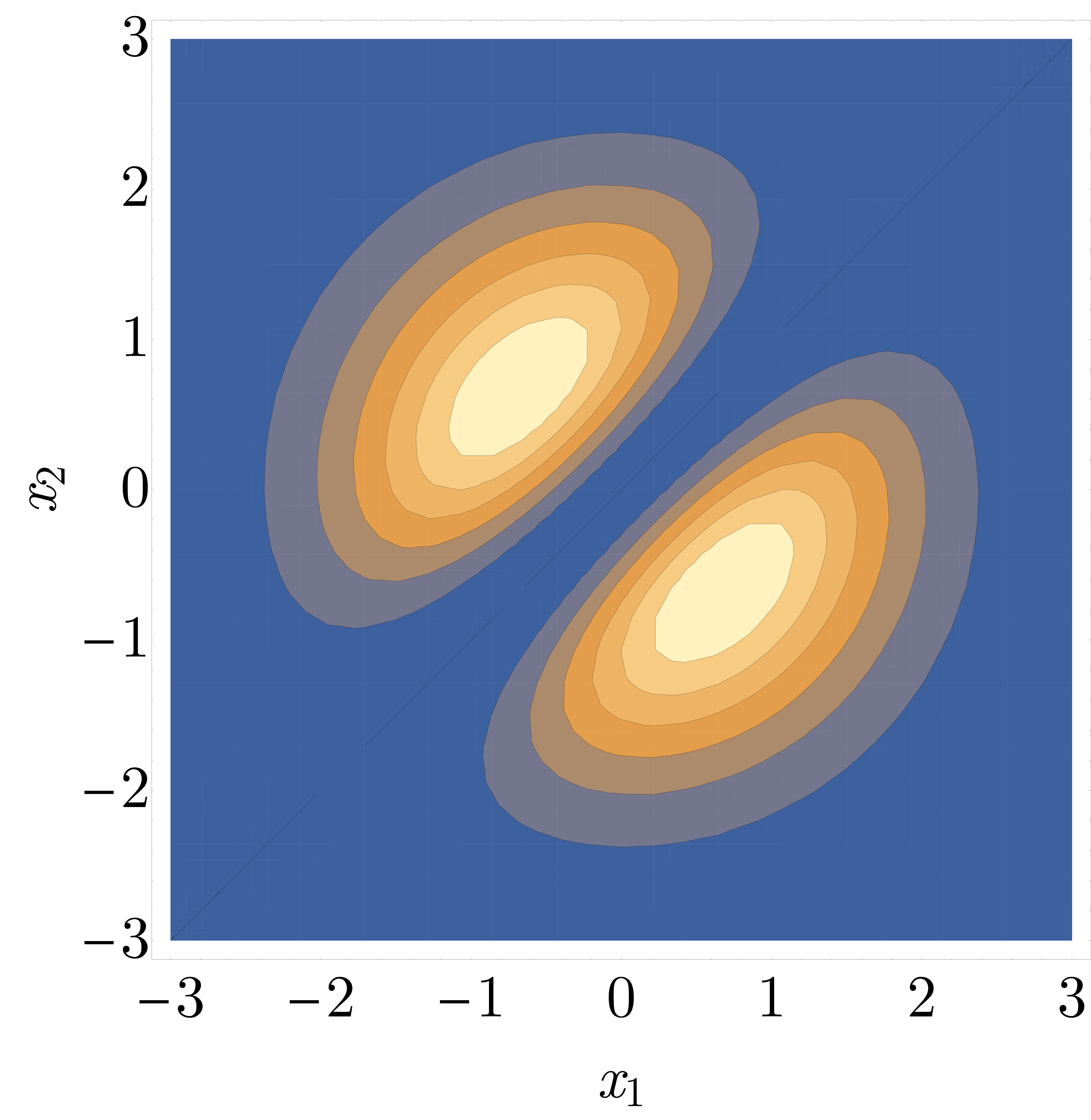}
	}
	\caption{\label{fig:thermalPosDist} Thermal state position distributions of the thermal state of (a) two distinguishable particles, (b) two bosons, and (c) two fermions in a harmonic potential. \textit{Figure adapted from Ref. \cite{Myers2020PRE}}.}
\end{figure*}

The natural question then is how the engine performance changes if not one, but two quantum particles are trapped in the harmonic oscillator. To this end, Ref.~\cite{Myers2020PRE} examined how ``exchange forces",  a collective phenomenon that arises from the fundamentally indistinguishable nature of quantum particles, affects the performance of this finite-time Otto cycle with a working medium of two particles, either bosons or fermions. Due to the underlying symmetry of the collective wave function, bosons are more likely to be found bunched together while fermions, with an underlying antisymmetric wave function, are more likely to be found spread apart. This can be clearly seen in the thermal state position distributions, as illustrated in Fig. \ref{fig:thermalPosDist}. Physically, these exchange forces manifest as an effective attraction between bosons and repulsion between fermions.

As the differences between bosons and fermions manifests entirely in terms of the symmetry of the total wave function, the Hamiltonian is identical for both working mediums,
\begin{equation}
	H(t) = \frac{p_1^2+p_2^2}{2m}+\frac{1}{2} m \omega^2 (t)\, \left(x_1^2+x_2^2\right).
\end{equation} 

By solving the full time-dependent dynamics for the unitary strokes, and assuming the heating and cooling strokes are of sufficient duration for the working medium to fully thermalize, the internal energy at each corner of the cycle is determined,
\begin{equation}
	\begin{split}
		\la H \ra_A &= \frac{\hbar \omega_1}{2}\,\left( 3 \mathrm{coth}(\beta_1 \hbar \omega_1) + \mathrm{csch}(\beta_1 \hbar \omega_1) \mp 1 \right), \\
		\la H \ra_B &= \frac{\hbar \omega_2}{2}\, Q^*_{\mathrm{comp}}\, \left( 3 \mathrm{coth}(\beta_1 \hbar \omega_1) + \mathrm{csch}(\beta_1 \hbar \omega_1) \mp 1 \right), \\
		\la H \ra_C &= \frac{\hbar \omega_2}{2}\,\left( 3 \mathrm{coth}(\beta_2 \hbar \omega_2) + \mathrm{csch}(\beta_2 \hbar \omega_2) \mp 1 \right), \\
		\la H \ra_D &= \frac{\hbar \omega_1}{2}\, Q^*_{\mathrm{exp}}\, \left( 3 \mathrm{coth}(\beta_2 \hbar \omega_2) + \mathrm{csch}(\beta_2 \hbar \omega_2) \mp 1 \right),
	\end{split}
\end{equation}
where the top sign corresponds to bosons and the bottom to fermions.

As in the single particle case, these internal energies are sufficient to characterize the finite-time performance of the cycle. Examining multiple figures of merit, including efficiency, power output, and efficiency at maximum power Ref. \cite{Myers2020PRE} showed that a bosonic working medium displays enhanced performance, in comparison to a working medium of distinguishable particles, while a fermionic working medium displays reduced performance. This ``bosonic enhancement'' even extends to the inherent trade-off between efficiency and power quantified by the efficient power, given by the product of efficiency and power. Furthermore, Ref. \cite{Myers2020PRE} demonstrated that the bosonic working medium functions as an engine or refrigerator over a wider range of the total parameter space in comparison to the fermionic medium. Notably, the differences in bosonic and fermionic performance are fundamentally an effect of the non-equilibrium nature of the cycle and vanish in the limit of long stroke times.                 

The impacts of wave function symmetry on heat engine performance were further explored in Ref. \cite{Myers2021PRXQ} which extended the results of Ref. \cite{Myers2020PRE} to anyonic statistics. Numerous other many-body phenomena have been explored in the context of heat engines, including quantum phase transitions \cite{Chand2018PRE,Campisi2016NC}, many-body localization \cite{Yunger2019PRB}, superradiance and collective enhancement of energy exchange \cite{Hardal2015SR, Niedenzu2018NJP, Kloc2019PRE}, inter-particle interactions \cite{Jaramillo2016, Myers2021Symm}, many-body quantum interference \cite{Hardal2018PRE}, and the non-Markovian anti-Zeno effect \cite{Mukherjee2020CP}.

%% file: sections/theory_STA.tex
\subsection{Optimized engine cycles: shortcuts to adiabaticity}
\label{theory_STA}

Equation~\eqref{power_gen} highlights that quantum engines also generally fail to produce finite work in infinitely slow processes.  A possible way to achieve finite power in fast processes has been sought in so-called shortcuts to adiabaticity (STA) \cite{Chen2010,MugaJPhysB2010,Torrontegui2013,OdelinRMP,Deffner2020EPL}. In particular, it has been found that STA methods enhance the performance of thermal devices by reducing irreversible losses that suppress efficiency and power \cite{Deng2013,DelCampo2014,Jaramillo2016,Abah2016EPL,Abah2017}. Note, however,  that this is typically only true if the energy balance of the external controller can be disregarded \cite{Tobalina2019}.

A particularly powerful technique for STA is called counterdiabatic driving \cite{Demirplak2003,Demirplak2005,Berry2009,Deffner2013PRX}. Within this paradigm all transitions away from the adiabatic manifold of a time-dependent Hamiltonian, $H(t)$, are suppressed by applying the \emph{counterdiabatic} field.  Thus, a quantum system evolves under a new Hamiltonian $H_\mrm{eff}(t)=H(t)+H_\mrm{CD}(t)$, which reads explicitly
\begin{equation}
H_\mrm{eff}(t) = H(t) + i\hbar \sum_n(\ket{\partial_t n}\bra{n} - \braket{n}{\partial_t n} \ket{n}\bra{n})
\end{equation}
where $\ket{n} \equiv \ket{n(t)}$ denotes the $n$th eigenstate of the original Hamiltonian $H(t)$.

For instance,  for a   time-dependent harmonic oscillator, the counterdiabatic term becomes \cite{Deffner2013PRX}
\begin{equation}
H_\mrm{CD}(t) = -\frac{\dot{\omega_t}}{4\, \omega_t}({x}{p}+ {p} {x}).
\end{equation}
Note that for implementation in heat engine cycles boundary conditions  are chosen to ensure $\la H_\mrm{CD}(0)\ra=\la H_\mrm{CD}(\tau)\ra$ for all strokes.

There has been some debate in the literature over how to appropriately assess the thermodynamic cost of STA, see for instance \cite{Zheng2015,Abah2017,KosloffEntropy,Campbell2017PRL,Li2018,Tobalina2019}.  Reference~\cite{Abah2017} took the arguably simplest approach by considering a modified efficiency of the corresponding engine,  which can be expressed as
\begin{equation}
\label{eq:eff_CD}
\eta = -\frac{\sum_j W_{2j+1}}{Q_2 + \sum_j \la H^{2j+1}\ra_\tau}\,,
\end{equation} 
where $W_i= \la H_\mrm{CD}(\tau)\ra - \la H(0)\ra$ is corresponding mean work of the STA protocol and  $\la H^i\ra_\tau=(1/\tau) \int_0^\tau dt\, \la H_\mrm{CD}^i(t)\ra$ is the time-average of the mean STA driving term.  Note that Eq.~\eqref{eq:eff_CD} is nothing but the ratio of output and input average energies.  Accordingly,  the power output becomes \cite{Li2018}
\begin{equation}
P_\mrm{CD} = -\frac{1}{\tau_\text{cyc}}\sum_j \left(W_{2j+1}  - \langle H^{2j+1}_\mrm{CD}\rangle_\tau\right)\,.
\end{equation}
It has been noted that the such defined efficiency and power represent the ``true'' performance of the Otto cycle~\cite{Abah2019}. 

Similar studies looked at thermodynamic cycles with STA in multiferroic material \cite{Chotorlishvili2016}, harmonic oscillators~\cite{Abah2018,Abah2020PRR},  anyonic working mediums \cite{Myers2021PRXQ}, BECs with nonlinear interactions\cite{Li2018}, spin-1/2 systems~\cite{Cakmak2019PRE,Insinga2020,Erdman2019NJP}, superconducting qubits \cite{Funo2019PRB}, and multi-spin systems \cite{Hartmann2020PRR,Hartmann2020Q} as working mediums.  More recently, techniques from machine learning have been employed to find optimal STA \cite{Erdman2021arXiv,Khait2021arXiv}.

\subsection{Fluctuations in finite-time quantum engines}
 
We conclude this section by briefly commenting on fluctuations in the performance of quantum thermodynamic devices.  For a comprehensive discussion of notions such as quantum work \cite{Campisi2011,Talkner2016,Deffner2016_work} and their corresponding fluctuation theorems we refer to the literature \cite{Talkner2007,Deffner2019book}. For our present purposes it is interesting to note that similar consideration were published in the context of engine cycles \cite{Campisi2014JPA,Verley2014,Martinez2016,Manikandan2019,Jiao2021,Saryal2021}.

Within the two-time energy measurement paradigm for quantum work,  the stochastic performance for the finite-time Otto cycle quantum heat engine, is evaluated with the joint probability distribution $\mc{P}(W_1,Q_2,W_3)$ \cite{Campisi2014JPA}.  To this end, projective measurements are taken at the end points of all strokes, and for instance the probability distribution of the compression work $W_1$ reads~\cite{Talkner2007},
\begin{equation}
    \mc{P}(W_1)=\sum_{j,k} \delta \left[W_1 - (E_k^\tau - E_j^0) \right] \wp_{jk}^\text{com} \wp_j^0\,.
\end{equation}
Here, $E_j^0$ and $E_k^\tau$ are the respective initial and final energy eigenvalues, $\wp_j^0\!=\!\exp\left(-\beta_1 E_j^0\right)/Z^0$ is the initial thermal occupation probability with partition function $Z^0$ and $\wp_{jk}^\text{com}=|\bra{j} U_\text{com}(\tau)\ket{k}|^2$ denotes the transition probability between the instantaneous eigenstates $\ket{j}$ and $\ket{k}$ in time $\tau$ with the corresponding unitary $U_\text{com}$. The probability distribution of the heat $Q_2$ after the second stroke, given the compression work $W_1$, is equal to the conditional distribution,
\begin{equation}
    \mc{P}(Q|W_1)=\sum_{i,l} \delta\left[Q - \left(E_l^\tau - E_i^\tau\right)\right] \wp_{il}^\text{heat} \wp_i^\tau,
\end{equation}
where the occupation probability $\wp_i^\tau=\delta_{ki}$ at time $\tau$ after the second projective energy measurement. The quantum work distribution for the expansion stroke, given the compression work $W_1$ and the heat $Q$, is\cite{Jarzynski2004}
\begin{equation}
    \mc{P}(W_3|Q,W_1) = \sum_{r,m}\delta\left[W_3 - \left(E_m^0 - E_r^\tau\right)\right] \wp_{rm}^\text{exp} \wp_r^\tau,
\end{equation}
where the probability  for finding the system in eigenstate $\ket{l}$ after the third projective measurement simply reads $\wp_r^\tau\!=\!\delta_{rl}$, and the transition probability $\wp_{rm}^\text{com}$ is determined by the unitary time evolution for expansion $U_\text{exp}$.

Based on the chain rule for conditional probabilities, the joint probability of a cycle of the quantum engine reads \cite{Denzler2020},
\begin{equation}
    \mc{P}(W_1,Q,W_3)=\mc{P}(W_3|Q,W_1) \mc{P}(Q|W_1) \mc{P}(W_1).
\end{equation}
By defining the stochastic total extracted work $W=-(W_1 + W_3)$, the joint distribution for work and heat is given by
\begin{equation}
\begin{split}
   & \mc{P}(W,Q)\\
   &\quad=\int dW_1 \int dW_3\, \delta[W + (W_1 + W_3)]\, \mc{P}(W_1,Q,W_3).
  \end{split}
\end{equation} 
Then,  defining the stochastic efficiency $\eta=W/Q$, its probability distribution is obtained by integrating over $W$ and $Q$ as~\cite{Denzler2020}
\begin{equation}
    \mc{P}(\eta) = \int dQ \int dW\, \delta\left(\eta - W/Q\right)\, \mc{P}(W,Q).
\end{equation} 
This probability distribution can be exploited to analyze the effects of thermal and quantum fluctuations on the efficiency of a quantum Otto heat engine.

The study of fluctuations for finite-time quantum heat engines has been attracting some attention mainly focused on two-level systems or harmonic oscillators~\cite{Denzler2020,Jiao2021,Jiao2021NJP,Denzler2021,Saryal2021}.  However, its stands to reason that interesting physics is encoded in the fluctuations of any working medium.

%% file: sections/experiment_ion.tex
\section{\label{sec:exp}Physical platforms}

The remainder of this review is dedicated to highlighting a selection particularly important studies in a variety of physical platforms.  We collecting the various publication and scenarios,  special emphasis  was put on giving a broad overview of the many different systems at the expense of being rather brief with regards to technical, physical, and mathematical details. 

\subsection{Ion traps}

We begin with recent implementations of quantum thermodynamic devices in ion traps.  It is interesting to note that while laser-cooled ions in linear Paul traps were originally invented for the experimental study of quantum computation and information processing applications,  they have become a prominent testbed for quantum thermodynamic notions \cite{Huber2008,Abah2012,Rossnagel2014,An2015,Rossnagel2016,Lindenfels2019PRL,Maslennikov2019}.

\subsubsection{Single-atom heat engine}

\begin{figure*}
\includegraphics[width=.88\textwidth]{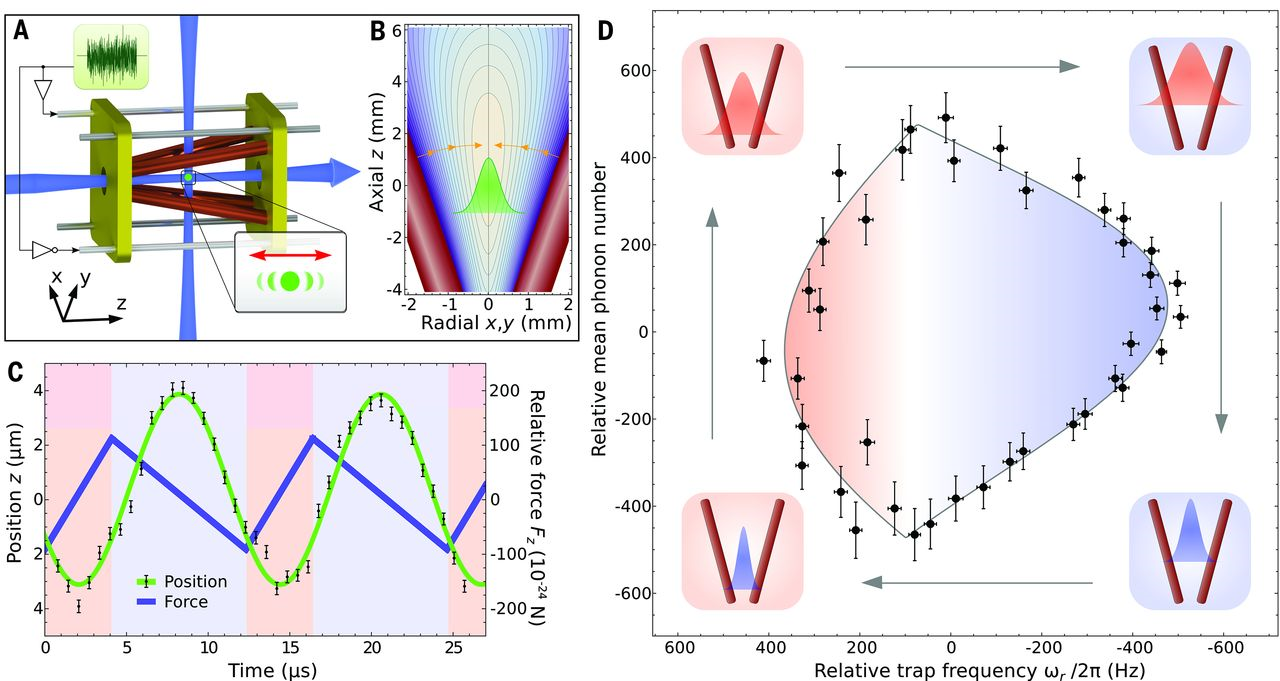}
\caption{\label{fig:expt_iontrap} (a) Schematic representation of the single-atom heat engine; (b) Position of the ion at each time step; (c) Thermodynamic cycle of the engine for one radial direction. The  area enclosed by the cycle corresponds to the work performed by the engine. \textit{Figure adopted from Ref.~\cite{Rossnagel2016}}.}
\end{figure*}

A particularly important breakthrough for the field was achieved with the realization of a heat engine that operates with a single ion as a working medium  \cite{Rossnagel2016}.  To achieve this, a single $^{40}\text{Ca}^+$ ion was trapped in a linear Paul trap with a funnel-shaped electrode geometry, as illustrated in Fig.~\ref{fig:expt_iontrap}. The authors of Ref.~\cite{Rossnagel2016} engineered cold and hot reservoirs by using laser cooling and electric field noise respectively, and the temperature of the ion was determined by fast thermometry methods \cite{Rossnagel2015}. The realized thermodynamic cycle resembles a Stirling engine and its thermodynamic properties were analyzed in this context.

The tapered trap realized a  harmonic pseudopotential of the form 
\begin{equation}
V(x,y)=\frac{1}{2} m \left(\omega_x^2 x^2+\omega_y^2 y^2\right)\,,
\end{equation}
where $m$ is the atomic mass and ${x,y}$ denote the trap axes. The electrodes were driven symmetrically at a radio frequency voltage of 830 $V_{pp}$ at 21 MHz.  Applying constant voltages on the two end-cap electrodes resulted in the axial confinement with a trap frequency of $\omega_z/2\pi\!=\!81$ kHz. The resulting radial trap frequencies $\omega_{x,y}$ reads
\begin{equation}
    \omega_{x,y} = \frac{\omega_{0x,0y}}{(1+ z \tan \theta/r_0)^2},
\end{equation}
where $\omega_{0x,0y}$ are the eigenfrequencies in the radial directions at the trap minimum $z\!=\!0$ and satisfy the cylindrical symmetry with $r^2\!=\!x^2+y^2$ with a mean radial trap frequency $\omega_r$. The electrode angle is denoted by $\theta$ while the radial distance of the electrode is $r_0$ at axial position $z=0$. 

To compensate for stray fields, a second set of outer electrodes was used. Then the trapped ion was cooled by a laser beam and the resulting fluorescence was recorded by a rapidly-gated intensified charge-coupled device  camera. An equilibrium  cold bath was realized by exposing the ion to a laser cooling beam at temperature $T_C=3.4$ mK \cite{Rossnagel2016} while a hot reservoir with  finite temperature $T_H$ was mimicked by exposing the ion to additional white noise from the electric field. 

In this setup, heating and cooling act on the radial degrees of freedom. Based on the geometry of the tapered potential, the ion experiences a temperature-dependent average force in axial direction is given by,
\begin{equation}
    F_z=-\int_0^\infty d\phi dr\, \xi_r(r,\phi,T) \frac{dU}{dz} ,
\end{equation} 
where the time-averaged spatial distribution of the ion thermal state is described by
\begin{equation}
    \xi_r(r,T)=\frac{1}{2\pi \sigma_r^2(T)}\exp \left[\frac{-(r-r_0)^2}{2 \sigma_r^2} \right].
\end{equation}
The time-averaged width $\sigma(T)$ of this two-dimensional Gaussian probability distribution depends on the temperature $T$ as $\sigma_r=\sqrt{k_BT/m\omega_r^2}$, where $k_B$ is the Boltzmann constant. The heat engine is driven by alternately heating and cooling the ion in radial direction.

The dynamics of the ion when driven in a four stroke thermodynamic cycle is depicted in Fig.~\ref{fig:expt_iontrap}. In the first part of the cycle the ion is heated, which results in the width $\sigma_r(T)$ increasing. During the second step, the ion moves along the $z$-axis to a weaker radial confinement which leads to the increase of total potential energy of the ion and thereby produces work. During the third step the ion is cooled to the initial temperature as the radial width of its probability distribution is reduced. For the final step, the ion moves back to its initial position due to the restoring force of the axial potential. However, the resulting cycle deviates from an ideal Stirling cycle due to the fact that full thermalization with the reservoirs is not reached. For each radial cycle, work produced is transferred to the axial degree of freedom and stored in the amplitude of oscillation.  

Rossnagel \textit{et al.} \cite{Rossnagel2016} computed the power  output,  $P=W/t_\mrm{cyc}$, during steady-state operation in three independent ways. First, the power is determined from the area enclosed by the engine cycle where $W=\int d\omega_r\,2\hbar \bar{n}_r  $ is the work computed from the area of the cycles from the two radial directions and $t_\mrm{cyc}=2\pi/\omega_z$ is the cycle time. 

In the second approach, the power was directly deduced from the measurement of the axial oscillation amplitudes $A_z$. The dissipated power of a driven damped harmonic oscillator at steady state is \cite{Taylor2005}
\begin{equation}
    P_\mrm{osc}=\gamma m \omega_z^2 A_z^2,
\end{equation}
with the damping parameter $\gamma$ determined from the oscillation decay. 

The third approach involved  the analytical  calculation of the engine power output using the expression for work performed over one cycle period $t_\mrm{cyc}$,
\begin{equation}
    W_\mrm{ana}=-\int_\mrm{cyc} dt\, F_z(t)  \dot{z}(t).
\end{equation}
The analytical power output $P_\mrm{ana}=W_\mrm{ana}/t_\mrm{cyc}$ depends only on the temperature variation and the trap geometry.  The analysis of Rossnagel \textit{et al.} \cite{Rossnagel2016} showed a good agreement between the measured and the analytical power output as a function of temperature difference, see Fig.~\ref{fig:expt_iontrap}. 

Furthermore, the engine efficiency from the measured data was evaluated as $\eta=W/Q_H$, by deducing the heat absorbed from the hot reservoir, $Q_H=\int_H\,dS\, T$. The entropy $S$ of the harmonic oscillator engine is calculated using $S=k_B\left[1+\ln(k_BT/\hbar\omega_r) \right]$. This basically involves transforming the heat engine cycle from the \{$n,\omega$\}-diagram to a temperature-entropy representation. Likewise, an analytical expression of the engine efficiency was derived which depends solely on temperature variation and trap geometry by calculating the heat absorbed from the reservoir as $Q_H=\int_H dt\, T(t)\,\dot{S}(t)$.The analysis of Ref. ~\cite{Rossnagel2016} showed that the engine operates at efficiency of 0.28\% and could be increased by varying either the angle of the taper or the absolute radial trap frequencies.

This experimental realization has inspired  investigations of quantum thermodynamic devices such as, e.g., quantum absorption refrigerators \cite{Maslennikov2019}, spin heat engines with quantum flywheels \cite{Lindenfels2019PRL,VanHorne2020},  and potential applications as sensitive thermal probes \cite{Levy2020}.

\subsubsection{Quantum flywheels}

\begin{figure}
\includegraphics[width=.48\textwidth]{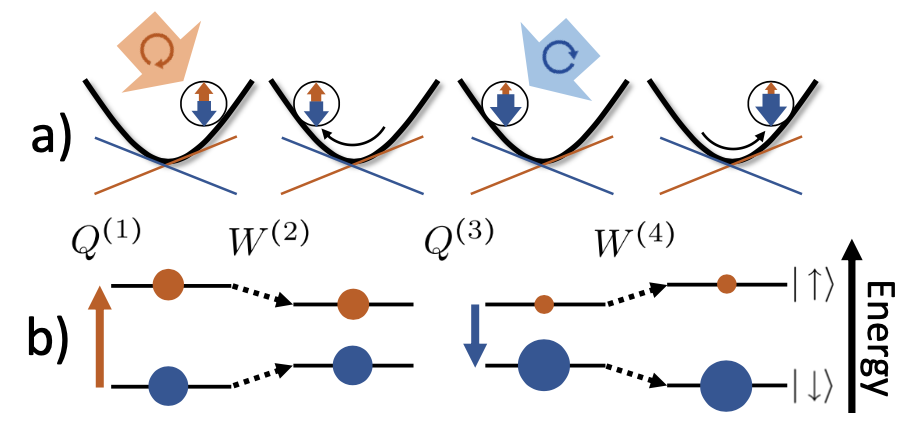}
\caption{\label{fig:expt_iontrap2} (a) Schematic representation of the quantum four-stroke engine with a single electron spin coupled to an oscillator; (b) energy level representation during the four-stroke Otto engine cycle. \textit{Figure adopted from Ref.~\cite{Lindenfels2019PRL}}.}
\end{figure}

An attempt to unravel the role of thermodynamic fluctuations  and quantum effects in the performance of quantum thermal devices led to the experimental study of a spin heat engine coupled to a harmonic-oscillator flywheel (load) \cite{Lindenfels2019PRL}.  Lindenfels \textit{et al.} \cite{Lindenfels2019PRL} experimentally realized a heat engine using the valence electron spin of a trapped $^{40}\mathrm{Ca}^+$ ion as a working medium,  and the heat reservoirs were mimicked by controlling the spin polarization via optical pumping. The  role of the flywheel coupled to the working medium was to absorb the output energy of the engine. The four-stroke Otto cycle operation is illustrated in Fig.~\ref{fig:expt_iontrap2}.

The authors \cite{Lindenfels2019PRL} trapped an ion in a Paul trap at a secular trap frequency $\omega_t\sim 2\pi \times 1.4$ MHz along the $x$-axis. The ion was placed in an optical standing wave (SW) which mediates the coupling between the engine and flywheel via a spin-dependent optical dipole force. The spin-oscillator system is described by
\begin{equation}
    H=H_\mathrm{HO}+\hbar\,\left[\omega_z + \Delta_S\sin(k_\mathrm{SW} x)\right]\,\sigma_z/2,
\end{equation}
where $\omega_z$ represents the Zeeman splitting of the spin, $\Delta_S$ is the amplitude of the SW depending on the ac-Stark shift variation, $k_\mathrm{SW}$ is the effective wavenumber and $\sigma_z$ is the Pauli $z$-operator. The bare flywheel Hamiltonian is  simply $H_\mathrm{HO}=\hbar \omega(t)\,(n+1/2)$, where $\omega(t)$ is the time-dependent trapping frequency and $n$ is the number operator. Optically pumping at the trap frequency $\omega(t)$ plays the role of reservoirs, with the spin population corresponding to the temperature.  The cold and hot reservoirs' temperatures are deduced from the population of the Zeeman sublevel that can be expressed as
\begin{equation}
    \langle \sigma_z\rangle = - \tanh (\hbar \omega_z^\prime/2k_B T),
\end{equation}
where $\omega_z^\prime=\omega_z+\Delta_S k_\mathrm{SW}\langle x\rangle$, $\langle \sigma_z\rangle \gtrsim -1$, and $\langle \sigma_z\rangle \lesssim 0$ corresponds to the cold (hot) reservoir temperature $T_C$ ($T_H$). The engine's cyclic operation results in an increasing amplitude of the harmonic oscillation,  stored as energy in the flywheel.

Lindenfels \textit{et al.}\cite{Lindenfels2019PRL}  characterized the state and energetics of the flywheel both theoretically and  experimentally. For this purpose, they evaluated the ergotropy of the flywheel, i.e, the maximum work that can be extracted via a cyclic unitary  transformation \cite{Pusz1978CMP,Allahverdyan2004,Touil2021JPA,Sone2021entropy}. This can be described as
\begin{equation}
    \mathcal{E}=\tr{H_\mathrm{HO}\rho} -\tr{H_\mathrm{HO}\, \rho_p},
\end{equation} 
where $\rho$ is the state of the flywheel and $\rho_p$ is the passive state unitarily related to $\rho$ \cite{Pusz1978CMP}. They found that the flywheel's ergotropy $\mathcal{E}$ is always less than the mean energy $E=\la H_\mathrm{HO}\ra$ due intrinsic fluctuations in machines operating with single atomic degrees of freedom.

The experiment of a spin-1/2 heat engine coupled to a harmonic oscillator flywheel agrees well with the theoretical analysis \cite{Lindenfels2019PRL} and opens the door for more investigations of quantum thermodynamic devices with a load attached \cite{VanHorne2020}. In addition,  a recent experimental study of Van Horne \textit{et al.}\cite{VanHorne2020}  considered a quantum refrigerator coupled to a load within the same framework of trapped ions technology.

\subsubsection{Continuous system-bath interaction} 

Another experimentally feasible single-ion quantum heat engine was proposed by Chand and Biswas \cite{Chand2017EPL}. Their design aims at a quantum Otto engine that mimics continuous heat engine cycles \cite{Mitchison2019}. To this end,  they considered an Otto model of a single-ion quantum heat engine with continuous interaction between the working medium and a thermal environment. The proposal relies on controlling the magnetic field adiabatically and performing a projective measurement of the electronic states.

Chand and Biswas \cite{Chand2017EPL} imagined the internal state of a single-ion as a the engine's working medium. In the Lamb-Dicke limit, the Hamiltonian describing the interaction between the internal and motional states of the ion is written as
\begin{equation}
    H= g\sigma_x + B\, \sigma_z + \omega a^\dagger a + k(a^\dagger \sigma_{-} + \sigma_{+} a),
\end{equation}
where $H_S=g\sigma_x + B\, \sigma_z$ describes the internal states of the ion with $2g$ being the Rabi frequency and $B$ is the strength of the magnetic field. $H_\text{ph}= \omega a^\dagger a$ represents the vibrational degree of freedom and $H_\text{int}=k(a^\dagger \sigma_{-} + \sigma_{+} a)$ is the interaction between the internal and the vibrational degrees of freedom of the ion. Thus, the eigenvalues of the working medium Hamiltonian are given by $E_{1,2}=\pm \sqrt{g^2 + B^2}$.

For this heat engine design, the quantum Otto engine cycle is the implemented as follows: 

(1) \textit{Ignition stroke}: During this isochoric process, the working medium thermalizes  with the hot bath at temperature $T_H$. In this stroke, the magnetic field is fixed at $B=B_H$ and  the amount of heat transferred can be written as $ Q_H=\sum_{n=1}^2 E_n^H \left(\wp_n^H - \wp_n^L\right)$
where, as before, $E_n^H$ is the $n$th energy eigenvalue and $\wp_n^L$ is the initial probability of occupying the $n$th energy eigenstates and $\wp_n^H$ describes the final probability after the thermalization.

(2) \textit{Expansion stroke}: Here, the magnetic field is adiabatically changed from $B_H$ to $B_L$. No heat exchange occurs with the heat bath or the phonon modes, whereas the work performed by the system is $W_1 = \sum_{n=1}^2 \wp_n^H \left(E_n^L - E_n^H\right)$,  where $E_{1,2}^H=\pm \sqrt{g^2 + B_H^2}$ are the eigenvalues of $H_S$ at the initial stage of the stroke.

(3) \textit{Exhaust stroke}: This isochoric process results in transferring the amount of heat $Q_L$ out of the system as the local driving fields are kept fixed. Chand and Biswas\cite{Chand2017EPL} proposed to employ a projective measurement of the state of the system $S$ to enable the release of heat. This purification process is equivalent to cooling down the system. Assuming that the system is prepared in the ground state $\ket{g}$, the heat removed from the system reads $Q_L=\sum_{n=1}^2 E_n^L \left(\wp_n^L - \wp_n^H\right)$.

(4) \textit{Compression stroke}: In this final stroke, the magnetic field strength is adiabatically changed from $B_L$ to $B_H$. The occupation probabilities remain unchanged, i.e, the system remains in the ground state, and the work done during this stroke is $W_2 = \sum_{n=1}^2 \wp_n^L \left(E_n^H - E_n^L\right)$.

Accounting for the measurement back action on the working medium, the effective efficiency of the engine is written as
\begin{equation}
    \eta_M=\frac{Q_H - |Q_L|}{Q_H + M},
\end{equation}
where the energy cost associated with the projective measurement of the qubit is $M\le k_\text{B}T_H \ln 2$. The equality sign holds for a maximally mixed state and corresponds to maximal change in entropy. Chand and Biswas~\cite{Chand2017EPL} further outlined how to prevent heating of the system when averaged over many cycles using feedback control.

This theoretical proposal  motivated measurement-based engines \cite{Das2018,Ding2018PRE} that offer the possibility of enhanced engine performance, which we will elaborate on next.

%% file: sections/experiment_measurement.tex
\subsection{\label{sec:measure} Engines driven by quantum measurements} 

From the point of view of quantum information theory thermal environments do nothing but perform measurements on a system of interest \cite{Zurek1991,ZurekRMP,Nielsen2010,BreuerBook,ZurekPhysReport2018}. Thus,  it becomes easy to recognize that quantum heat engine cycles can be designed in which ``thermodynamic strokes'' are replaced by quantum measurements. Beyond the work of Chand and Biswas~\cite{Chand2017EPL}  the following studies are particularly noteworthy.

\subsubsection{Harmonic oscillator and measurements}

Ding {\textit{et al.}}\cite{Ding2018PRE} proposed to construct a thermal device that operates between a measurement apparatus and a single heat bath.  The working medium is again the parametric harmonic oscillator \eqref{eq:harm_osc}. The engine operates in a four-stroke cycle:

(1) \emph{Adiabatic compression}: The engine is initialized in thermal equilibrium at the inverse temperature $\beta$,  before the working substance is compressed by changing the frequency from initial $\omega_1$ to final $\omega_2$. During this stroke no heat is added to the working medium and the work simply reads,  $W_I = E_2 - E_1$, where $E_1$ and $E_2$ are the initial and final eigenvalues of the Hamiltonian respectively. The energies at the initial and final state are deduced using projective measurements. 

(2) \emph{Isochoric heating}: In the second stroke, a measurement of the oscillator position is performed at a constant frequency $\omega_2$. This stroke provides an input energy into the system by ensuring that the  measured observable does not commute with the working substance Hamiltonian. Accordingly, the energy change $E_\mathcal{M}$ induced by the measurement is $ E_\mathcal{M}=E_3 - E_2$, where $E_3$ is the energy eigenvalue at the end of the stroke. 

(3) \emph{Adiabatic expansion}: The third stroke involves expanding the working substance back to $H(\omega_1)$. The amount of work performed on the working substance during this stroke is $ W_{II}=E_4 - E_3$,  where $E_4$ is the corresponding eigenvalue at the stroke completion. 

(4) \emph{Isochoric cooling}: In the final stroke, the working substance is brought back to it initial state by weakly coupling with the thermal reservoir at the initial inverse temperature $\beta$. The heat exchange is the thermal reservoir is $Q=E_1^\prime - E_4$, where $\ket{1^\prime}$ is the initial state of the next cycle.

Using the positive sign convention, the averages of the total work $\la W\ra = \la W_I\ra +\la W_{II}\ra$ and the supplied energy $\langle E_\mathcal{M}\rangle$ are calculated for the large class of minimally disturbing generalized measurements \cite{YiPRE2017}. As usual, the efficiency is $ \eta=-\langle W\rangle/\langle E_\mathcal{M}\rangle$. Interestingly, the efficiency takes the universal form $\eta=1-1/\kappa$, where $\kappa$ is again the compression ratio, for a uniform adiabatic compression and expansion of the working substance. 

Ding \textit{et al.}\cite{Ding2018PRE} analyzed the averages and fluctuations of the total work, supplied energy and performance of this  measurement-driven engine.  They showed that the power of a measurement engine is considerably larger than that of the Otto engine running with the same average work output because of the shorter measurement engine cycle time. This theoretical proposal has spurred interest to potential devices using qubits as working substance \cite{Das2018,Buffoni2019,Bresque2021PRL}. 

\subsubsection{Two-stroke two-qubit cooler}

A more versatile device was proposed by Buffoni \textit{et al.}\cite{Buffoni2019} They considered a prototype two-qubit device that exploits quantum measurements as a fuel. The model consist of two qubits governed by the Hamiltonian $H=H_1+H_2$, and each prepared by thermalization with a thermal bath at positive inverse temperatures $\beta_1$ and $\beta_2$ respectively. Thus, the initial state reads,
\begin{equation}
    \rho=\frac{\e{-\beta_1 H_1}}{Z_1}\otimes \frac{\e{-\beta_2 H_2}}{Z_2},
\end{equation}
where $H_i=\hbar \omega_i\sigma_z^i/2$ denotes the Hamiltonian of $i$th qubit in terms of its Pauli matrix $\sigma_z^i$ and its resonance frequency $\omega_i$.  Finally, $Z_i=\tr{\e{-\beta_iH_i}}$ is the canonical partition function.

\begin{figure}
\includegraphics[width=.48\textwidth]{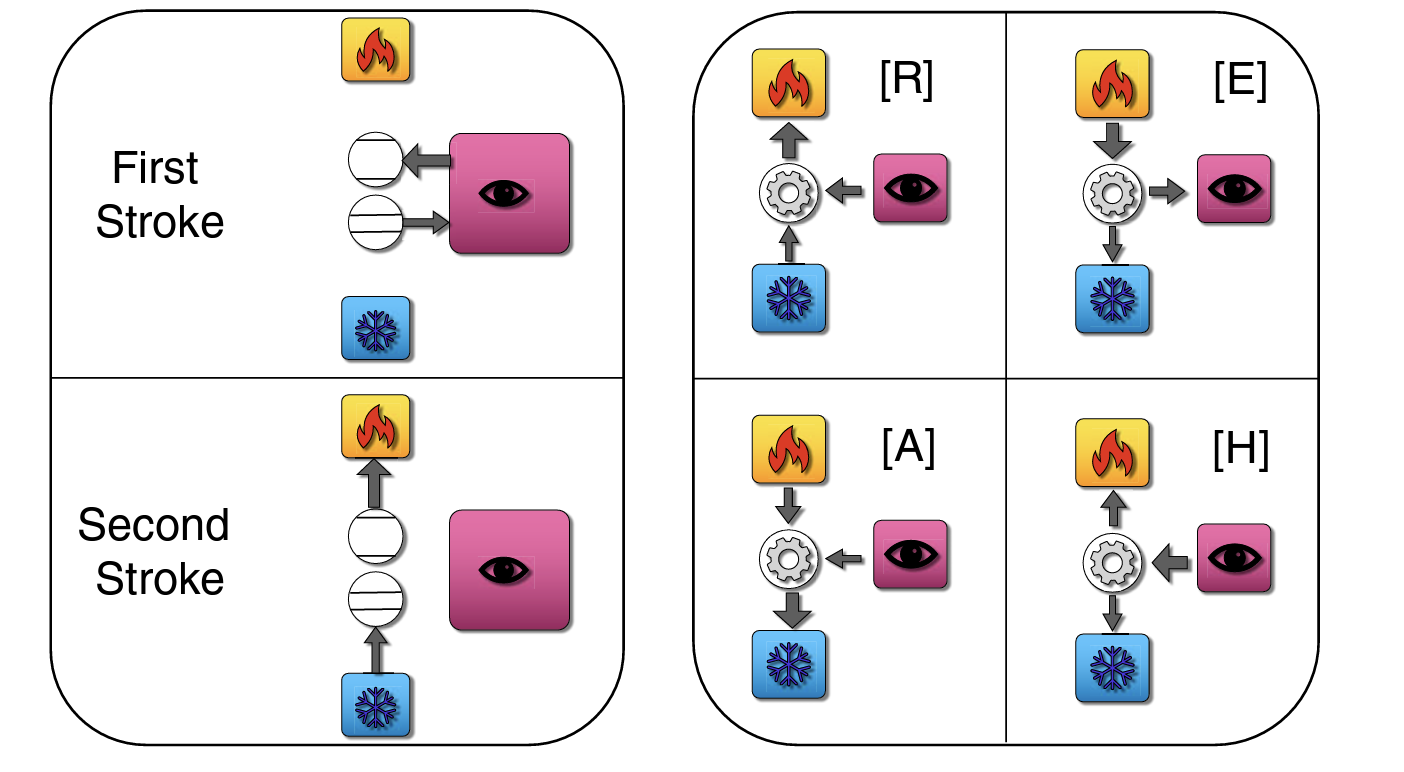}
\caption{\label{fig:expt_QMC} (a) Schematic representation of the two-stroke two-qubit quantum  device; (b) the four possible operations based on the thermodynamic laws.  \textit{Figure adopted from Ref.~\cite{Buffoni2019}}.}
\end{figure}

The two-stroke cycle is schematically depicted in Fig.~\ref{fig:expt_QMC}. In the first stroke, the two-qubit system interacts with a measurement apparatus. This stroke erases all coherences of the two qubit compound state in the measurement basis and the postmeasurement state reads $\rho^\prime=\Phi[\rho]=\sum_k \pi_k\rho\pi_k$, where $\pi_k$ is a projector \cite{Nielsen2010}. The change in the expectation energy of the $i$th qubit is $\la \Delta E_i\ra=\tr{H_i\,(\Phi[\rho]-\rho)}$. Based on the unital property of $\Phi$, the second law of thermodynamics can then be expressed as
\begin{equation}
    \beta_1\langle \Delta E_1\rangle + \beta_2 \langle \Delta E_2\rangle \ge 0.
    \label{Buffoni_eq1}
\end{equation}
In the second stroke, each qubit is restored to its initial Gibbs state by putting it back in contact with its thermal bath. During this stroke, each qubit releases an average energy $\langle \Delta E_i\rangle$, gained during the first stroke. Combining the energy conservation $\langle \Delta E\rangle=\langle \Delta E_1\rangle +\langle \Delta E_2\rangle$ and the second law relation (Eq.~\ref{Buffoni_eq1}) as well as assuming the condition $0<\beta_1<\beta_2$, Buffoni \textit{et al.}\cite{Buffoni2019} identified the operation regime where the device can be function as; a refrigerator [R], heat engine [E], thermal accelerator [A] or heater [H].

Buffoni \textit{et al.}\cite{Buffoni2019} further elaborated on how to experimentally realize the two-stroke quantum cooler using solid-state superconducting circuitry by a combination of circuit QED and circuit quantum thermodynamics. Specifically, they suggested a device that comprises two superconducting qubits coupled to an on-chip microwave line resonator\cite{Filipp2009}. The first stroke can be implemented by the combination of two-state manipulation and standard measurement while the second stroke can be implemented by inductively coupling each qubit to an on-chip resistor kept at inverse temperature $\beta_i$. This study has attracted a follow-up proposal on the design of quantum magnetometry using a two-stroke engine \cite{Bhattacharjee2020}. Similar conceptual tools also proved useful in characterizing the D-Wave machine as a thermodynamic device \cite{Buffoni2020QST}.

\subsubsection{Measurement powered entangled qubits}

Even more recently,  Bresque \textit{et al.}\cite{Bresque2021PRL} proposed a thermal device exploiting quantum entanglement to deepen the understanding of measurement as fuel. The operation of the quantum engine demonstrates that both local measurements and entanglement are crucial for work extraction. The working medium involves two qubits $A$ and $B$, whose Hamiltonian, $H_\mathrm{2qb}=H_\mathrm{loc}+ V$, reads explicitly
\begin{equation}
    H_\mathrm{2qb}=\sum_{i=A,B}\hbar \omega_i \sigma_i^\dagger \sigma_i + \frac{1}{2}\,\hbar g(t)(\sigma_A^\dagger\sigma_B + \sigma_B^\dagger \sigma_A)  ,
\end{equation}
where $\omega_i$ is the qubit frequency, $\sigma_i=\ket{0_i}\bra{1_i}$ is the lowering operator for the qubit $i \in \{A,B\}$ and $g(t)$ is the time-dependent coupling strength. 

\begin{figure}
\includegraphics[width=.48\textwidth]{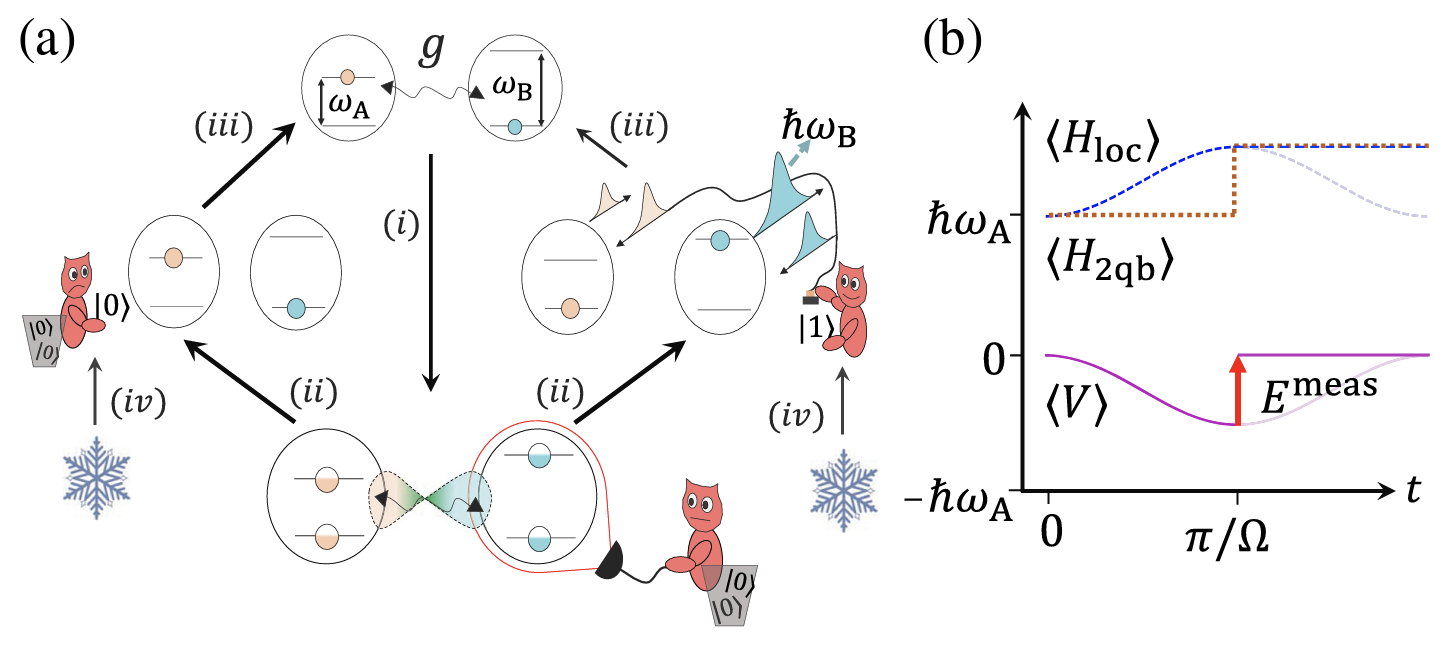}
\caption{\label{fig:expt_entangled_engine} (a) Schematic representation of the two-qubit  engine; (b) average energies during entangling evolution. \textit{Figure adopted from Ref.~\cite{Bresque2021PRL}}.}
\end{figure}

The four steps of the engine cycle are schematically sketched in Fig.~\ref{fig:expt_entangled_engine}.  The measurement powered engine cycle has the following four strokes:

(1) \textit{Entanglement creation}: Starting from the qubits prepared in the state $\ket{\psi_0}=\ket{10}$ of mean energy $\langle H_\mathrm{2qb}\rangle=\hbar\omega_A$, the qubits' state evolves into an entangled state $\ket{\psi(t)}$ by switching on the strength $g$. The resulting periodic exchange between the two qubits is
\begin{equation}
\begin{split}
    \ket{\psi(t)}&=\left(c^2_\theta\, \e{i\Omega t/2} + s^2_\theta\, \e{-i\Omega t/2}\right) \ket{10} \\
     &\quad- c_\theta s_\theta\left(\e{i \Omega t/2}- \e{-i\Omega t/2}\right) \ket{01},
\end{split}
\end{equation}
where $c_\theta=\cos(\theta/2)$, $s_\theta=\sin(\theta/2)$, $\theta=\arctan(g/\delta)$, $\Omega=\sqrt{g^2+\delta^2}$ is the generalized Rabi frequency,  and the parameter $\delta=\omega_B-\omega_A$ is the positive detuning. The sum of the average energies of free Hamiltonian of the qubits $\langle H_\mathrm{loc}(t)\rangle$ and the interaction Hamiltonian $\langle V(t)\rangle$ remains constant during this stroke.

(2) \textit{Measurement}:  At time $t_0=\pi/\Omega$, corresponding to the maximum value of $\langle H_\mathrm{loc}\rangle$ and $|\langle V\rangle(t)|$, a local projective energy measurement is performed on qubit $B$, and its outcome is recorded in a classical memory $M$. This step erases the quantum correlations between the qubits and results in a statistical mixture of the average qubits' state $\rho(\theta)=\cos^2(\theta)\ket{10}\bra{10} + \sin^2(\theta)\ket{01}\bra{01}$. The average energy input, $E^\mathrm{m} $,  and the von Neumann entropy, $S^\mathrm{m} $,  of the qubits become
\begin{equation}
\begin{split}
    E^\mathrm{m}&=-\langle V(t_0)\rangle= \Delta\langle H_\mathrm{2qb}\rangle = \hbar\delta\sin^2(\theta)\ge 0, \\
    S^\mathrm{m}&= -\cos^2(\theta)\log_2[\cos^2(\theta)]-\sin^2(\theta) \log_2[\sin^2(\theta)].
\end{split}
\end{equation}
Note that the energy and entropy increase simultaneously and depend on the coupling and detuning parameters. 

(3) \textit{Feedback}: At this step, the coupling term is switched off at time $t_0^+$, to allow the conversion of the acquired information into work. The amount of extractable work $W$ depends on the qubit in which the the excitation is measured and the qubits' entropy vanishes at this step. 

(4) \textit{Erasure}: In this final step, the memory is erased by a cold bath. The minimal work cost of this process is proportional to $S^m$.

The engine performance is defined as the work extraction ratio $\eta=W/E^\mathrm{m}$, which is less than unity for nonoptimal work extraction. Bresque \textit{et al.}\cite{Bresque2021PRL} further analyzed the source of measurement energy as well as the role of increasing the number of qubits on the device performance. 

%% file: sections/experiment_optomech.tex
\subsection{Light-matter interaction}

Originating in the field of quantum optics \cite{Scully1999,Schleich2011},  research in \emph{light-matter interaction} has grown into an important branch quantum physics. Thus, rather naturally,  quantum heat engines have also been proposed and realized that exploit the unique features of such systems, see for instance Refs.~\cite{Altintas2015,Barrios2017}.

\subsubsection{Quantum optomechanical Otto engine}

\begin{figure}
\begin{center}
\includegraphics[width=.48\textwidth]{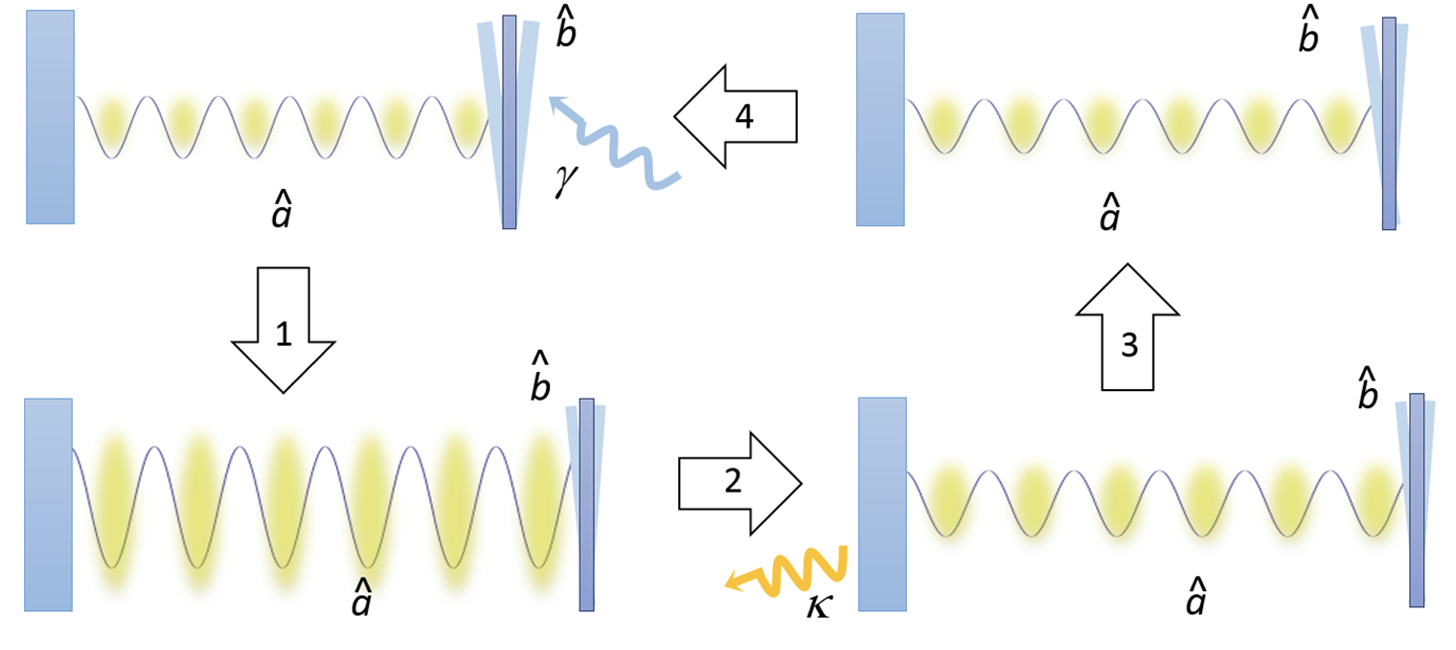}
\caption{\label{fig:expt_opto} Schematic representation of the quantum Otto heat engine cycle with a optomechanical set up.  \textit{Figure adopted from Ref.~\cite{Zhang2014PRL}}.}
\end{center}
\end{figure}

\emph{Quantum optomechanics} is a particularly instructive branch of light-matter interaction,  with many potential applications in quantum technologies \cite{Barzanjeh2021}. This is illustrated well by a proposal for a simple mechanical heat engine with potential to operate in the deep quantum regime by Zhang, Bariani and Meystre\cite{Zhang2014PRL,Zhang2014PRA}. They considered a cavity optomechanical setup at mode frequency $\omega_c$ coupled to a mechanical resonator at frequency $\omega_m$ with single photon coupling strength $g$. A typical example of such resonator is the harmonically bound end mirror of a Fabry-P\'erot resonator. An optical pump field with strength $\alpha_{in}$ and frequency $\omega_p$ is used to drive the resonator. Specifically, they numerically showed that a mechanical resonator of frequency $\omega_m=2\times 10^8$ Hz and quality factor $Q=10^5$ coupled to an optical cavity of linewidth $\kappa=10^6$ Hz and a steady-state occupation of $|\alpha|^2=10^{10}$ via an optomechanical coupling $g\!=\!10^2$ Hz can be used to realize an Otto cycle.  The set-up is schematically depicted in Fig.~\ref{fig:expt_opto}.

Assuming that the intracavity field is strong, the linearized Hamiltonian of the entire system  is given by \cite{Paternostro2006}
\begin{equation}
    H=-\hbar \Delta a^\dagger a + \hbar \omega_m b^\dagger b + \hbar G (b+b^\dagger)(a + a^\dagger),
\end{equation}
where $a$ and $b$ are the bosonic annihilation operators associated to the fluctuations of the photon and phonon mode annihilation operators around their mean amplitudes $\alpha$ and $\beta$, $G=\alpha g$ and the detuning $\Delta=\omega_p - \omega_c-2\beta g$. The quadratic Hamiltonian $H$ describes two linearly coupled harmonic oscillators, which can  result in sideband cooling when $\Delta<0$, i.e., in the red detuned regime.
 
To allow for the analysis of the energy conversion between photons and phonons, the system can be expressed  in its normal mode representation. The resulting Hamiltonian in the diagonal form is 
\begin{equation}
    H=\hbar\omega_A A^\dagger A + \hbar \omega_B B^\dagger B + \text{const.},
\end{equation}
where the operators $A$ and $B$ are the boson annihilation operators for the normal-mode excitations of the system with frequencies
\begin{equation}
    \omega_{A,B}=\sqrt{\frac{1}{2}\left(\Delta^2 + \omega^2_m \pm \sqrt{(\Delta^2 - \omega^2_m)^2 - 16 G^2 \Delta \omega_m}\right)}.
\end{equation}
The bosonic annihilation operator $B$ describes phononlike excitations in the low-energy polariton branch ($\Delta\!\ll\!-\omega_m$) or photon-like excitations on the other side of the avoided crossing ($-\omega_m\ll \Delta <0$). 

In addition to the coherent dynamics, the system is subject to thermalization due to damping and decoherence of the excitations. The optical and mechanical dissipation is characterzied by  the cavity decay rate $\kappa$ and mechanical damping rate $\gamma$, which allows the construction of a heat engine with two thermal reservoirs. The hot thermal bath is responsible for the relaxation of the phonon mode, whereas the cold thermal bath arises due to the damping of the optical mode.  Both the normal-mode excitations and their reservoir temperatures are controlled by the detuning $\Delta$. Thus, the proposed quantum Otto engine cycle operates by varying $\Delta$ to alternate between phononlike and photonlike nature of the polariton such that work is extracted from the system after a complete cycle. 

Consider a situation in which the optomechanical system is initially prepared in thermal equilibrium at large red detuning, $\Delta\ll -\omega_m$, and the phononlike lower polariton branch $B$ is in thermal equilibrium with a reservoir at effective temperature $T_{Bi}$.  Similarly, at optical frequencies, the photonlike upper polariton branch $A$ is in thermal equilibrium with a reservoir at temperature $T_{Ai}\simeq 0$ K. Then, $T_{Bi}\gg T_{Ai}$ and for the initial polariton population we have $N_{Bi} \gg N_{Ai}$. 

The four-stroke Otto engine cycle is realized as follows: 

(1) \emph{Isentropic expansion}: First, the detuning $\Delta$ is adiabatically varied from its initial value $\delta_i=\Delta_i\ll -\omega_m$ to a final value $\delta_f\equiv-\omega_m\ll\Delta_f<0$ for a time interval $\tau_1$. This step is fast enough that the interaction of the system with the thermal reservoirs can be neglected and at the same time be slow enough to avoid transitions between the two polariton branches. 

(2) \emph{Isochoric heating}: In the second stroke, the photonlike polariton $B$ is coupled to the photon reservoir at temperature $T_{Bf}$ and allowed to thermalize over a time $\tau_2$. During this step, $1/\tau_2<\kappa$ and the thermal occupation adjusts from $N_i$ to $N_f$. 

(3) \emph{Isentropic compression}: The third stroke of the cycle involves changing the detuning back to its initial large negative value at fixed $N_f$. The duration $\tau_3$ must satisfy the same conditions as $\tau_1$. 

(4) \emph{Isochoric cooling}: The final stroke is the rethermalization with the phonon reservoir at a fixed frequency $\delta_i$. The step duration $1/\tau_4<\gamma$ and its thermal population becomes $\hat{N}_i$. 

Thus, the total work per cycle becomes
\begin{equation}
    W_\mrm{cyc}=\hbar (\omega_i - \omega_f)(N_i - N_f),
\end{equation}
where $\omega_{i,f}$ are frequencies of the polariton modes at the initial and final detunings. To obtain positive work $W_\mrm{cyc}$,  we need $\omega_i>\omega_f$ and $N_i>N_f$.  Accordingly, the heat received by the working medium is
\begin{equation}
Q=\hbar\omega_i (N_i - N_f)\,,
\end{equation}
and hence the engine operates at the Otto efficiency $  \eta = 1-\omega_f/\omega_i$.

Zhang, Bariani and Meystre\cite{Zhang2014PRL} further derived the analytical solutions for $W_\mrm{cyc}$ and $\eta$ in the limit $(G/\omega_m,-\Delta_f/\omega_m)\ll 1$ by means of perturbation theory.  In this limit, the total work can be expressed as
\begin{equation}
    \frac{W}{\hbar\omega_m}=\left(\frac{\Delta_f}{\omega_m}+\frac{2G^2}{\omega^2_m}+1\right)\left[\left(1-\frac{2G^2}{\omega^2_m}\right)n_b - \frac{G^2}{\omega^2_m}\right]\,,
\end{equation}
where $n_b$ is the occupation of the mechanical mode.  The corresponding efficiency reads
\begin{equation}
    \eta_W < 1-\sqrt{\frac{-\hbar \Delta_f}{2 k_B T_b}},
\end{equation}
where the lower classical thermal energy $k_BT_a$ has been replaced by the ground state energy of a quantum oscillator of frequency $-\Delta_f$.

The quantum optomechanical heat engine proposed by Zhang, Bariani and Meystre has inspired several other designs \cite{Mari2015JPB,Abari2019,Bennett2020NJP}, work extraction optimization\cite{Bathaee2016} and possible applications in phonon cooling \cite{Dong2015}.  Other applications include optomechanical engines based on a cascade setup\cite{Mari2015JPB} and the implementation of an engine based on feedback control\cite{Abari2019,Serafini2020}.

\subsubsection{Superradiance in heat engines}

Exploring an even more unique quantum feature, Hardal and M{\"u}stecaplio{\u g}lu\cite{Hardal2015SR} proposed to exploit superradiance phenomena, i.e.,  cooperative emission of light from an ensemble of excited two level atoms in a small volume relative to emission wavelength, to enhance the performance of a quantum heat engine. To this end, they considered a four-stroke photonic quantum Otto cycle which comprises the photons inside the cavity. That is, a single mode optical cavity is let to interact with a cluster of $N$ two-level atoms. The interaction is described by the Tavis-Cummings Hamiltonian,
\begin{equation}
\label{eq:tavis}
    H=\hbar\omega_f a^\dagger a +\hbar\omega_a S_z + g(a S^\dagger + a^\dagger S^-)\,,
\end{equation}
where $\omega_f$ is the cavity photon frequency, $\omega_a$ is the transition frequency of the atoms, and $g$ is the uniform interaction strength. The bosonic photon annihilation operators are denoted by $a$, while the atomic cluster is represented by collective spin operators $\left(S^\pm,S_z\right)=\left(\sum_i\sigma_i^\pm,\sum_i\sigma_i^z\right)$, where $\sigma_i^\pm$ and $\sigma_i^z$ are the Pauli spin matrices. The initial state of the cavity field is a thermal state  at temperature $T_c$,  $\rho_f(T_c)$.  The initial state of the atomic cluster $\rho_a(0)$ is prepared in a thermal coherent spin state,  and the collective atomic coherent state is related to the Dicke states and is superradiant.  After the cavity and atomic ensemble have interacted over a period of time, assuming $\omega_f=\omega_a$, the  cavity is characterized by an effective temperature $T_\mathrm{eff}$ obeying the relation $\la n\ra=1/[\exp(1/T_\mathrm{eff})-1]$, which is monotonically increasing as a function of number of atoms $N$.

(1) \emph{Ignition stroke}: During the first step of the engine cycle,  the photon gas is initially prepared in the thermal state $\rho_f(T_c)$ and then heated to evolve into a coherent thermal state $\rho_f(T_c,\alpha)$. The coherence parameter is denoted by $\alpha$ and the final state after thermalization can be described with an effective temperature. 

(2) \emph{Expansion stroke}: The second step involves adiabatically changing the photon gas frequency at constant occupation probabilities. The density matrix of the field at the end of this stroke is $\rho_f(T^\prime,\alpha^\prime)$ and only work is done by the gas. 

(3) \emph{Exhaust stroke}: In the third step the photon gas is transformed into a thermal state $\rho(T_L)$ by transferring some coherence to the environment. 

(4) \emph{Compression stroke}: The final step the brings the photon frequency back to its initial value. 
 
Exploiting the the second law of thermodynamics, Hardal and M{\"u}stecaplio{\u g}lu\cite{Hardal2015SR} then showed that the work output is maximal at maximum efficiency and obeys a power law.  Concretely, the work output of the photonic Otto engine scales quadratically with the number of atoms in the cluster,  $W_\mathrm{max}\sim N^2$.

\subsubsection{Polaritonic heat engine}

Fully exploiting the tool-kit of light-matter interaction, Song \textit{et al.}\cite{Song2016PRA} noticed that quantum systems can be described by polaritons. These quasipatricle excitations are quantum superpositions of the system's constituents with relative weights that depend on some coupling parameter.  Coupling these constituents to reservoirs at different temperatures, quantum thermodynamic cycles can be realized.

Song \textit{et al.}~\cite{Song2016PRA} proposed a working medium that consists of a single qubit trapped inside a high-Q single-mode resonator in a circuit QED geometry. The system is again described by the Jaynes-Cummings Hamiltonian \eqref{eq:jaynes} written here as,
\begin{equation}
    H=\hbar \omega (\sigma_z + 1)/2 + \hbar \omega_L a^\dagger a + \hbar g (a\sigma_+ + \sigma_{-} a^\dagger),
\end{equation}
where $\omega$ and $\omega_L$  are the frequencies of the two-level  and bosons, respectively. Moreover,  $g$ is the interaction strength.  In complete analogy to above, the two-level system is characterized by the ladder operators $\sigma^+=\sigma^{-\dagger}=\ket{e}\bra{g}$, $\sigma_z=\ket{e}\bra{e}-\ket{g}\bra{g}$,  where $\ket{g}$ and $\ket{e}$ are again the ground and excited states of the two-level system, while the bosonic mode is described by the annihilation operators $a$.

The dressed states associated with one of the eigenstates are photon-like for large positive detunings and qubit-like for large negative detunings, and the opposite is true for the second dressed states. The qubit and optical mode  are separately coupled to thermal reservoirs at temperatures $T_a$ and $T_f$, respectively. The qubit-field system density operator is described well by the quantum master equation of the form Eq.~\eqref{eq:open}. Song \textit{et al.}~\cite{Song2016PRA} then proposed a single atom-single photon heat engine  utilizing the difference in temperatures of thermal reservoirs for the qubit and the photon field as well as controlling the detuning parameter $\Delta=\omega - \omega_L$.  

The engine operations rely on a four-stroke  quantum Otto cycle. The system is initially prepared in the ground state $\ket{g,0}$ with transition frequency $\omega=\omega_1 <\omega_L$ and detuning $\Delta_1=\omega_1-\omega_L<0$, in thermal equilibrium at $T_a\simeq0$. 

(1) \emph{Isentropic compression}: The first  stroke involves changing the frequency $\omega$ from $\omega_1$ to a new value $\omega_2>\omega_L$ and detuning $\Delta_2>0$.  

(2) \emph{Isochoric heating}: During  the second stroke, the system  thermalizes with the two thermal reservoirs. At the end of the stroke, the hybrid system is left in a mixed state. 

(3) \emph{Isentropic expansion}: In the third stroke the frequency $\omega$ is returned to its initial value $\omega_1$, and the corresponding dressed state goes from its approximate photon-like nature to qubit-like form. This step needs to be carried out slowly to avoid the nonadiabatic transitions. 

(4) \emph{Isochoric cooling}: The final stroke is facilitated by the spontaneous decay of the qubit-like state to the ground state.  Hence, the average work associated with a complete Otto cycle is $ W = \wp_1 [E_{2,0}(\omega_1) - E_{2,0}(\omega_2)]$, where $\wp_1$ is the transition probability, and  $E_{2,0}(\omega_{1,2})$ are the energy eigenvalues at initial and final state of the third stroke.

Furthermore, Song \textit{et al.}\cite{Song2016PRA} studied the multiphoton case as well as the two-qubit case. Their analysis showed that the amount of extracted work is determined by the photon field. They further establish the hierarchies required for the operation of the heat engine, aka the positive work condition, based on the system parameters and the time spent on each $i$th stroke as
\begin{equation}
  \omega, \omega_L \gg |\Delta_{1,2}| \gg g
\end{equation}
and
\begin{equation}
   \tau_2 \gg \kappa^{-1}\gg \tau_4 \gg \gamma^{-1}\gg \tau_3 \gg g^{-1}.
\end{equation}
Based on the above constraints, devices based on circuit QED~\cite{Schmidt2013} are identified as a promising choice to realize the proposed quantum heat engine. Finally, the influence of quantum measurements on the operation of the heat engine can be directly addressed. The theoretical proposal of Song \textit{et al.}\cite{Song2016PRA} has spurred further quantum heat engine designs based on light-matter quantum systems~\cite{Barrios2017,Dodonov2018}.

\subsubsection{Nonstationary cavity QED}

As a last example of employing light-matter interaction we turn to cavity quantum electrodynamics (QED).  Dodonov, Valente and Werlang\cite{Dodonov2018} theoretically showed how to boost the power of a heat engine by means of quantum coherence in the nonstationary regime of cavity QED. This regime involves an external control of one or several system parameters in a time-dependent manner. Dodonov, Valente and Werlang\cite{Dodonov2018}  considered a two-level atom weakly interacting with a single-mode cavity which is described well by the  nonstationary Jaynes-Cummings model
 \begin{equation}
 \label{eq:jaynes}
     H(t)/\hbar = \hbar\omega a^\dagger a + \hbar\Omega_t \omega_z/2 + g_t (a\sigma_+ + a^\dagger \sigma_-),
 \end{equation}
where $a$ is again the cavity annihilation operator,  and as usual $\sigma_z$ is the Pauli spin matrix, $\sigma_+=\ket{e}\bra{g}$ and $\sigma_{-}=\sigma_{+}^\dagger$ are the atomic ladder operators, with the ground $\ket{g} $ and excited $\ket{e}$ state.  The cavity frequency is $\omega$, the time-dependent atomic transition frequency is $\Omega_t$, and atom-cavity coupling strength is $g_t\ll \omega$. Assuming an external  sinusoidal form of frequency modulation and the switching on/off of the atom-cavity coupling  monotonically,  Dodonov, Valente and Werlang\cite{Dodonov2018}  devised an Otto cycle. The system dynamics of the interaction between the atom (cavity) and its thermal reservoir was modeled by the microscopic Markovian master equation \cite{BreuerBook},
\begin{equation}
\label{eq:open}
    \dot{\rho} = -i/\hbar\, [H(t),\rho] + \mathcal{L}_a[\rho] + \mathcal{L}_f[\rho] \,,
\end{equation}
where $\mathcal{L}_{a/f}$ is the Liouvillian superoperator characterized by decay rate $\Gamma$ and $\kappa$. The cold and hot thermal reservoirs are coupled to the cavity (atom) and the engine function by cyclically operating between the two reservoirs.


The  working medium is initially prepared in the ground state $\ket{g,0}$. The Otto engine cycle is implemented as follows: first, the atom is coupled to the hot bath, $\Gamma\ne 0$, while the cavity is isolated, $\kappa=0$. The system's internal energy, $U(t)=\tr{\rho(t) H(t)}$, and its entropy, $S(t)=-\tr{\rho(t)\ln \rho(t)}$, are increased due to the heat supplied by the hot bath, $Q_{in}$. Then, in the second stroke, the system is isolated from the thermal reservoirs and  the atom-cavity coupling is switched on while modulating the atomic transition frequency, $\ket{e,0}\rightarrow \ket{g,1}$ . In the third stroke, the atom-cavity system is coupled to the cold bath, $\kappa \ne 0$ and $\Gamma=0$. The atom-field system is thermalized to the ground state $\ket{g,0}$ of the time-independent Jaynes-Cummings Hamiltonian. During this stage, heat is transferred from the working medium to the cold bath. During the last stroke, an isentropic reset,  the system is again decoupled from the two reservoirs and the coupling is monotonically switched off at a constant atomic transition frequency. No work is performed by the modulation of $g_t$, since the system remains in its ground state.

Dodonov \textit{et. al.}\cite{Dodonov2018} then showed that the average quantum power always exceed the average classical power due to quantum coherences. Their numerical analysis demonstrated that such a quantum boost is achieved by controlling the frequency modulations to specific frequencies that induce transition between particular pairs of the system's dressed states. They further demonstrated that the quantum power boost can be realized in both the Jaynes-Cummings and antidynamical Casimir effect regimes of the nonstationary Rabi model.

\subsubsection{Hybrid quantum-classical engine}

\begin{figure}
\includegraphics[width=.48\textwidth]{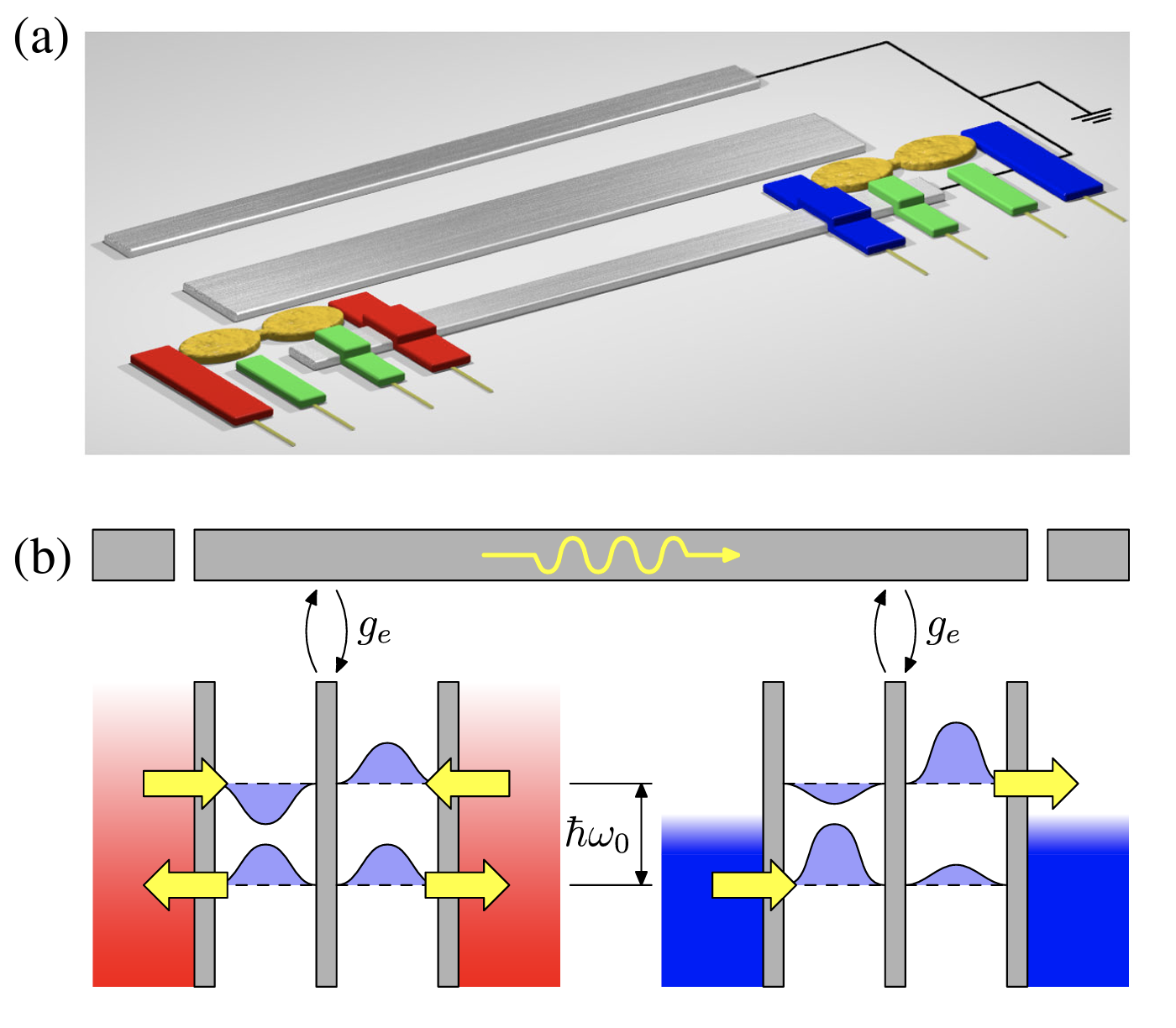}
\caption{\label{fig:hybrid} (a) Schematic representation of the hybrid quantum engine.  Double quantum dots (yellow) are coupled to the ends of a microwave cavity (gray). External gates (green) are used the tune the quantum dots,  and hot (red) and cold (blue) electrodes act as heat reservoirs; (b) effective energy diagram of the quantum dots, which are in resonance with the cavity at frequency $\omega_0$.  \textit{Figure adopted from Ref.~\cite{Bergenfeldt2014PRL}}.}
\end{figure}

Light-matter interaction is also perfectly suited to explore ``hybrid'' quantum-classical systems, which are physical systems that exhibit both quantum and classical characteristics \cite{Clerck2020NP}.  Such situations are highly sought after, as they may provide a bridge technology in the quest to realize genuine quantum computers.  Rather naturally, such hybrid systems have also be studied as a platform for hybrid quantum-classical heat engines.

Based on a related design \cite{Sothmann2012EPL}, a theoretical proposal for such an engine was put forward by Bergenfeldt \etal \cite{Bergenfeldt2014PRL}. This system consists of two macroscopically separated quantum-dot conductors, which are capacitively coupled to the fundamental model of a microwave cavity.  This is schematically depicted in Fig.~\ref{fig:hybrid}(a).  Controllable coupling of quantum dots to microwave cavities has been demonstrated in several experiments \cite{Meschke2006Nature,Savin2006JAP,Delbecq2011PRL,Frey2012PRL,Petersson2012Nature,Toida2013PRL,Deng2015NL}, with potential applications in, for instance, quantum information processes \cite{Trif2008PRB}, micromasers \cite{Childress2004PRA}, and  quantum-dot lasers \cite{Jin2011PRB}. Hence,  a comprehensive understanding of the thermodynamics of such hybrid systems appears instrumental, in particular since real systems may exhibit unwanted heat leaks \cite{Dresselhaus2007AM}.

Similarly to the superradiant engine \cite{Hardal2015SR} discussed above, the double quantum dot-cavity system is described by a generalized Tavis-Cummings Hamiltonian \cite{Bergenfeldt2013PRB}, and the open system dynamics is given by a Lindblad master equation, cf. Eqs.~\eqref{eq:tavis} and \eqref{eq:open}.   Bergenfeldt \etal \cite{Bergenfeldt2014PRL} solved the dynamics numerically exactly, and fully characterized the thermoelectric properties of the system including power output and efficiency. As a main result, they found that the maximal efficiency roughly coincides with the same bias voltage that yields the maximum power output.  Carnot efficiency is achieved at the stopping voltage, which however leads to vanishing power. In conclusion, Bergenfeldt \etal \cite{Bergenfeldt2014PRL}  demonstrated that a hybrid quantum engine could realistically work, and that it would operate at a significant fraction of the Carnot efficiency.

The theoretical proposal by Bergenfeldt \etal \cite{Bergenfeldt2014PRL}  has had significant impact.  In particular, the hybrid engine found attention in work on thermoelectrics with quantum dots \cite{Sothmann2013NJP,Jiang2014JAP,Thierschmann2015NN,Karwacki2015PRB,Thierschmann2016CRP,Karwacki2018PRB,Lu2019PRB,Jaliel2019PRL,Zhang2020PhysE,Jaliel2021} and other quantum heat engine designs \cite{Sothmann2014EPL,Schiro2014PRB,Sanchez2015PRL,Sanchez2015NJP,Wong2017PRA}.

%% file: sections/experiment_mems.tex
\subsection{Micro- and nano-electromechanical systems}

Micro and Nanoelectromechanical Systems (MEMS/NEMS) are miniaturized systems ranging between millimeter and nanometer \cite{Blick2005NJP}. From a technological standpoint, MEMS/NEMS are very attractive since such systems may find various applications,  including,  for instance,  high-frequency filters and switches in signal processing circuits, and ultra-sensitive sensors.  More fundamentally,  NEMS permit the investigation of electron–phonon coupling, the study of single electrons being shuttled via mechanical motion, and the manipulation of single molecules with nano-mechanical tweezers. Thus,  quantum thermodynamic devices have also been explored in MEMS/NEMS.

\subsubsection{Thermal reservoirs with squeezing}

\begin{figure}
\includegraphics[width=.48\textwidth]{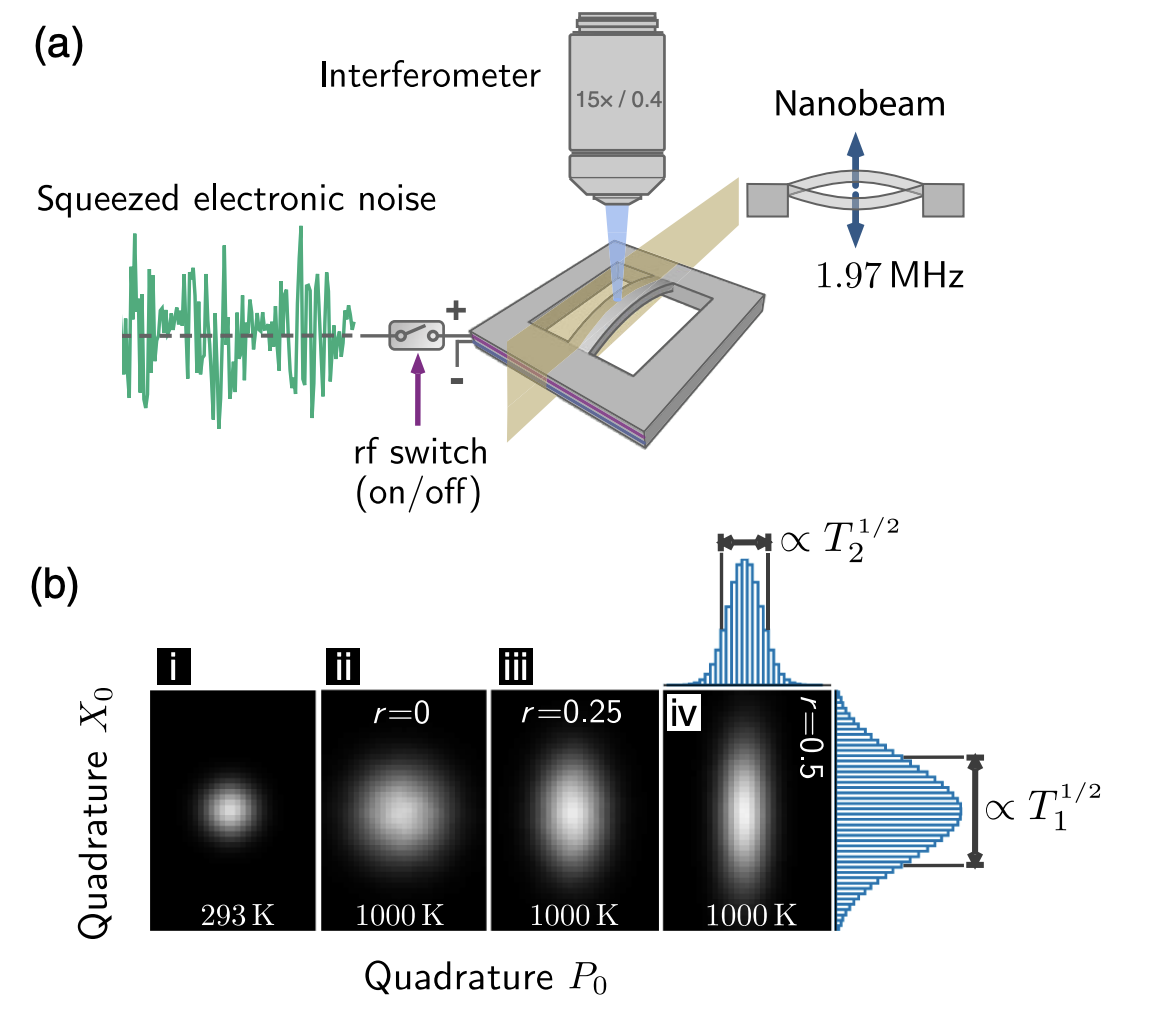}
\caption{\label{fig:expt_nems} (a) Schematic representation of the nanobeam heat engine; (b) phase-space density of the nanobeam motion when no noise is applied (i) and when squeezed noise is applied (ii) - (iv).  \textit{Figure adopted from Ref.~\cite{Klaers2017}}.}
\end{figure}

The first quantum thermodynamic experiment with NEMS sought to verify the theoretical prediction that squeezed thermal reservoirs allow to outperform the Carnot efficiency \cite{Rossnagel2014}. To this end, Klaers \textit{et. al.}~\cite{Klaers2017} realized an Otto heat engine, whose working medium consists of a vibrating nanobeam with reservoirs engineered by driven squeezed electronic noise. The working medium is described well by a harmoinic oscillator \eqref{eq:harm_osc}, which was fabricated via conventional nanostructuring techniques as a doubly clamped (GaAs) nanobeam structure with eigenfrequency $\nu=\omega/2\pi=1.97$ MHz, and a quality factor $Q\simeq 10^3$ at room temperature. The beam contained two doped layers facilitating the application of electric fields across the beam, see Fig.~\ref{fig:expt_nems}. 

To mimic an engineered thermal reservoir,   the beam was driven by a random force generated through a noisy waveform. By combination of two independent white noise signals $\xi_{1,2}(t)$, the waveform generates a stochastic force of the form 
\begin{equation}
f(t)=a_0\left[\exp({+r})\xi_1(t)\cos(\omega t) + \exp({-r})\xi_2(t)\sin(\omega t)\right]\,.
\end{equation}
The positive squeezing parameter $r$ corresponds to an amplified cosine and attenuated sine component at frequency $\nu$ in the thermal bath, while the overall strength of the noise can be determined by the amplitude $a_0$.  The motional state of the nanobeam was measured by Mach-Zehnder interferometry and employed to deduce the corresponding phase-space probability distribution. Increasing the noise amplitude ($a_0>0$) at $r=0$ results in the preparation of the nanobeam in a thermal state at higher temperature.  Moreover, the application of squeezed noise leads to an elliptical phase-space probability distribution.

\begin{figure*}
\includegraphics[width=.88\textwidth]{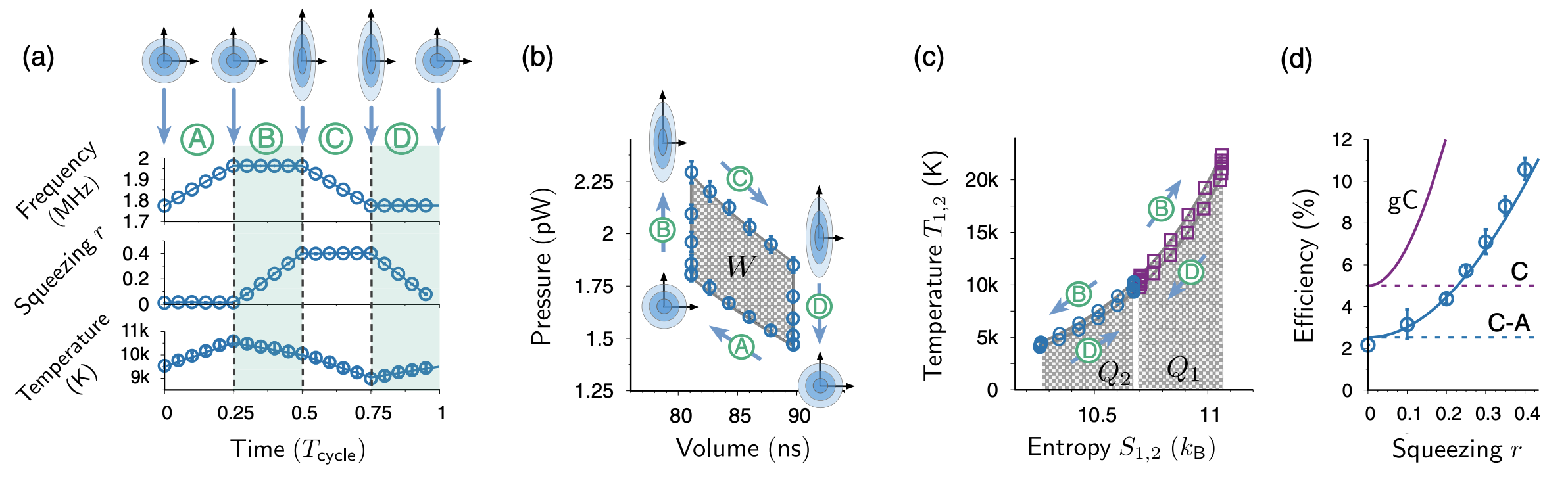}
\caption{\label{fig:expt_nems2}  Schematic representation of the Otto heat engine with squeezed reservoir. (a) Frequency, squeezing, and temperature of the nanobeam throughout one cycle.  The four strokes: Isentropic compression (A), hot isochoric (B), isentropic expansion (C), and cold isochoric (D); (b) pressure-volume diagram of the Otto cycle; (c) temperature-entropy diagram; (d) efficiency as a function of the squeezing parameter.  \textit{Figure adopted from Ref.~\cite{Klaers2017}}.}
\end{figure*}

In position-momentum representation, the observed probability densities closely follow the theoretical Gaussian distribution,
\begin{equation}
    \rho(x_0,p_0) \propto \exp\left(-\frac{\hbar \omega x^2_0}{2 k_B T_1}-\frac{\hbar \omega p^2_0}{2 k_B T_2}\right),
\end{equation}
where $x_0,$ and $p_0$ are dimensionless position and momentum variables, and $T_1$ and $T_2$ are two temperatures describing the level of fluctuations in the quadratures. The squeezed thermal state is characterized by an effective temperature $T$ and squeezing parameter $r$ which obey the relation $T_{1,2}=T\exp(\pm 2 r)$ \cite{Tucci1991}.  The mean energy of the nanobeam motion as a function of the squeezing parameter and fixed  temperature reads 
\begin{equation}
    U=k_B T\,(1 + 2 \sinh^2r).
\end{equation}
In Fig.~\ref{fig:expt_nems2}, the phase-space density of the nanobeam motion is illustrated.  Klaers \textit{et. al.}~\cite{Klaers2017} verified that the energy of the nanobeam increases as a function of the squeezing.

To implement an Otto engine, Klaers \textit{et. al.}\cite{Klaers2017} modulated the eigenfrequency of the fundamental flexural mode of the nanobeam over a few 100 kHz by applying a dc electrical potential. The change in trapping frequency $\omega$ results in expansion and compression of the working medium and corresponds work output.  Accordingly,  the first law of thermodynamics can be expressed as
\begin{equation}
    dU=T_1dS_1+ T_2dS_2 -P dV,
\end{equation}
where $S_1=-k_B\int dx_0 \rho(x_0)\ln[\rho(x_0)]$ is the entropy of the antisqueezed quadrature, $S_2$ is obtained by replacing $x_0$ with $p_0$ and $P=-(\partial U/\partial V)_{S_1,S_2}$ is the pressure \cite{Callen1985}.  The engine's work output produced per cycle was found to be $W\simeq 26$ meV and the corresponding efficiency $\eta=W/Q_h \simeq 10.6\pm 0.5\%$, which is roughly twice the standard Carnot cycle efficiency.

Finally, the nanomechanical engine implemented by Klaers \textit{et al.}\cite{Klaers2017} provides an important perspective on the role of nonequilibrium reservoir engineering.  This experimental demonstration has  already inspired a quantum heat engine proposal based  on two superconducting transmission line resonators interacting via an optomechanical-like coupling \cite{Hardal2017PRE}.

\subsubsection{Casimir interaction}

\begin{figure}
\includegraphics[width=.48\textwidth]{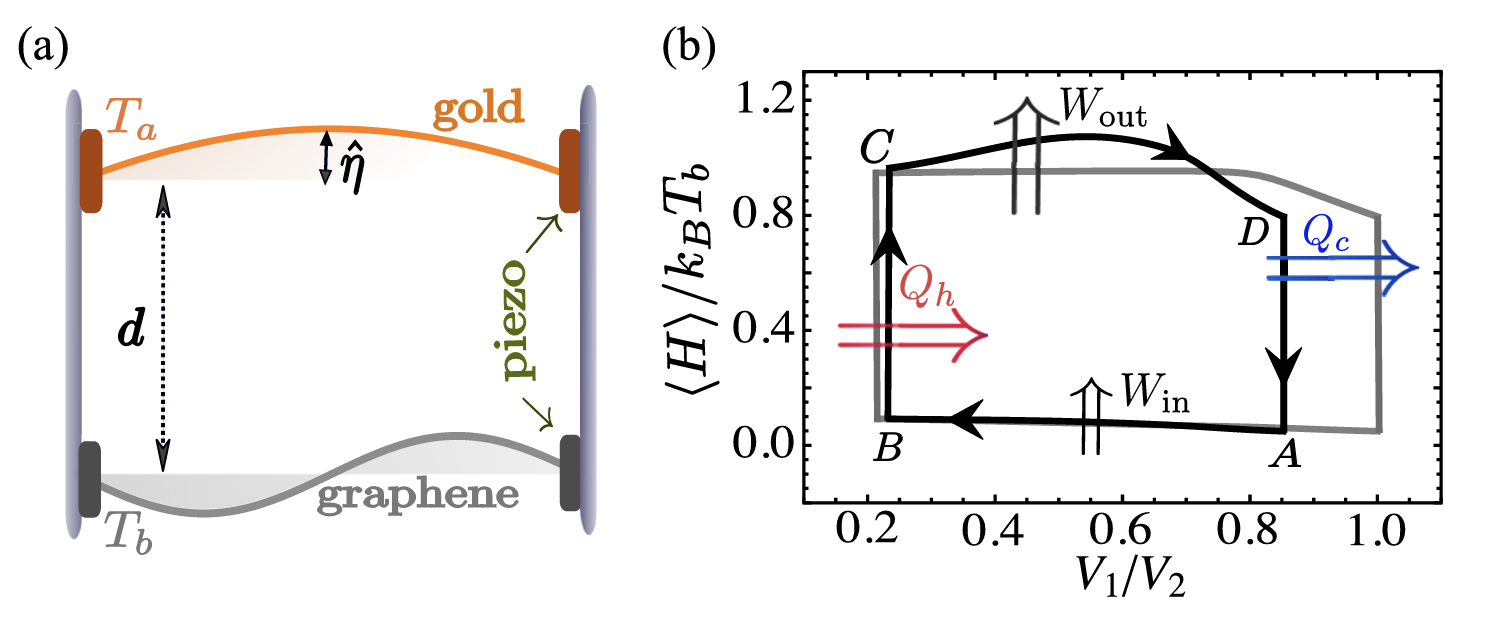}
\caption{\label{fig:NEM} (a) Schematic representation of the quantum thermal machine  with two nanobeams clamped in cryogenic environment; (b) graphic illustration of the quantum Otto cycle engine. \textit{Figure adopted from Ref.~\cite{Tercas2017PRE}}.}
\end{figure}

Another MEMS/NEMS inspired thermal device proposal is a quantum Otto engine cycle fueled by Casimir interaction between two nanomechanical resonators separated by a few $\mu$m from each other, put forward by Ter\c{c}as \textit{et al.} \cite{Tercas2017PRE} The energy associated with the out-of-plane (flexural) phonons can be exploited to produce work. The proposed setup, schematically illustrated in Fig.~\ref{fig:NEM}, consists of gold and graphene membranes placed at different temperatures in a cryogenic environment. A piezoelectric cell, acting as a piston, is proposed to select and control the flexural modes in a cyclic fashion.  By combining the direction of the cycle and changing the membrane temperatures, the setup could function either as a quantum heat engine or as a quantum refrigerator.

\begin{figure}
\includegraphics[width=.48\textwidth]{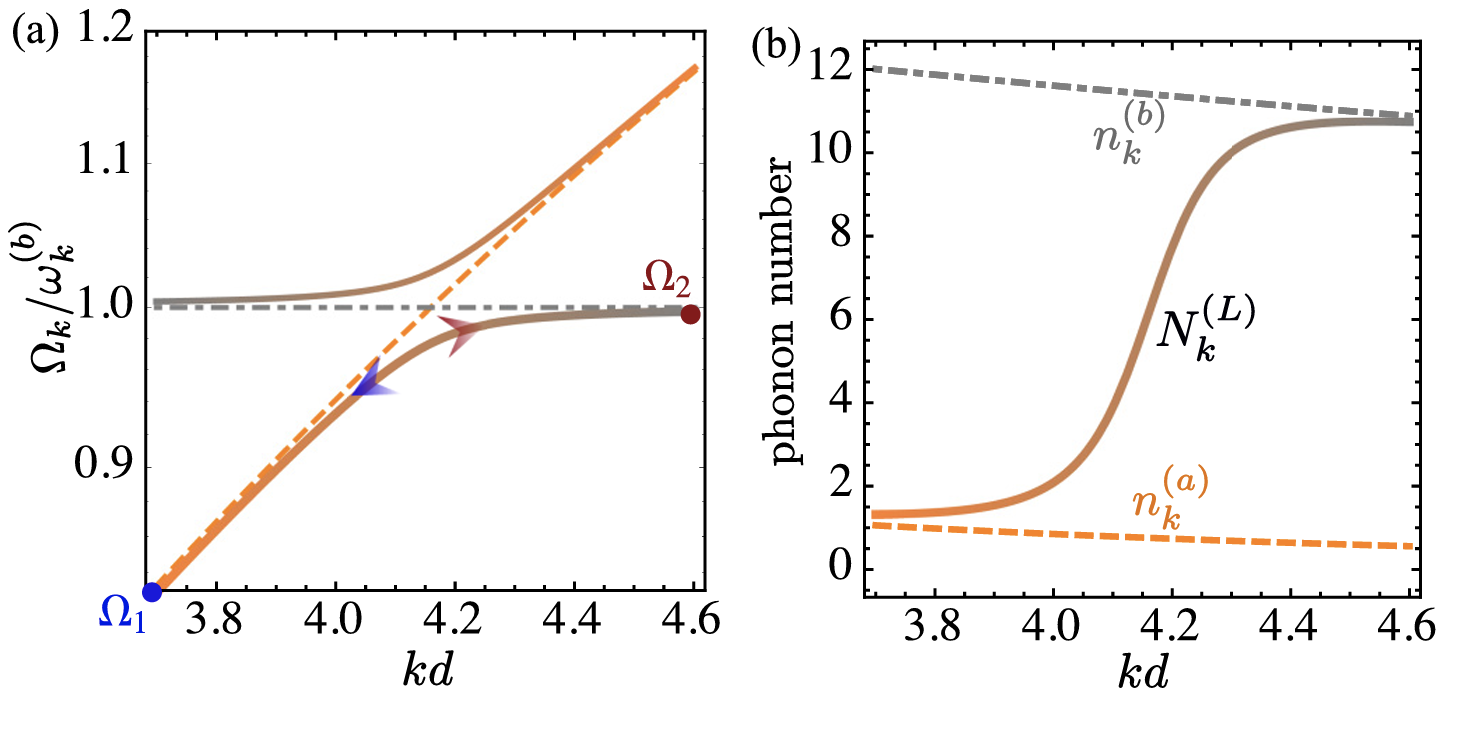}
\caption{\label{fig:NEM2} (a) Graphical illustration of the Hamiltonian spectrum near the avoided-crossing region; (b) bare and lower-polariton [$N_k^L$] phonon number for $T_b>T_a$.  \textit{Figure adopted from Ref.~\cite{Tercas2017PRE}}.}
\end{figure}

Following the Kirchhoff-Love plate theory describing the mechanical vibrations of the system, the flexuron Hamiltonian can be written as
\begin{equation}
    H_0=\hbar \sum_{k,\sigma} \left(\omega_k^{(a)} a_{k,\sigma}^\dagger a_{k,\sigma} + \omega_k^{(b)} b_{k,\sigma}^\dagger b_{k,\sigma} \right),
\end{equation}
where $\omega_k^{a,b}$ denotes the bare-mode frequency.
The interaction between the graphene and gold flexural modes due to the fluctuation of the Casimir potential leads to their hybridization. The interaction Hamiltonian reads
\begin{equation}
    H_\text{int} = \sum_k g_k a_k^\dagger b_k + h.c.,
\end{equation}
where $g_k$ is the coupling strength. Assuming that $g_k\ll \omega_k^{a,b}$, the total Hamiltonian $H=H_0 + H_\text{int}$ becomes in the rotating-wave approximation
\begin{equation}
    H=\hbar \sum_k\left[ \omega_k^{(a)} a_k^\dagger a_k + \omega_k^{(b)} b_k^\dagger b_k\right] + \sum_k g_k a_k^\dagger b_k,
\end{equation}
where there is an induced Stark shift to the bare frequencies as $\omega_k^{(j)} \rightarrow \omega_k^{(j)} + g_k$. The Hamiltonian, after diagonalization, in the polariton operator basis can be written as
\begin{equation}
    H=\hbar \sum_k\left[\Omega_k^{(L)} A_k^\dagger A_k + \Omega_k^{(U)} B_k^\dagger B_k\right],
\end{equation}
where the lower (L) and upper (U) polariton frequencies are given by
\begin{equation}
    \Omega_k^{(U,L)} = \frac{1}{2} \left[\omega_k^{(a)} + \omega_k^{(b)} \pm \sqrt{\left(\omega_k^{(a)} - \omega_k^{(b)}\right)^2 + 4\, |g_k|^2}\, \right].
\end{equation}
The resulting spectrum of the Hamiltonian is depicted in Fig.~\ref{fig:NEM2}. To achieve strong coupling, the Rabi frequency $\Lambda$ has to be chosen much larger than the decoherence rate $\Gamma$. The polariton decay rate is attributed to polariton-flexural phonon scattering. For the graphene-gold interface setup, Ter\c{c}as \textit{et al.} \cite{Tercas2017PRE} obtained $\Lambda\sim 5$ MHz and the mode frequency at the avoided-crossing $\omega_c\equiv \omega_{k_c}^{(a)}=\omega_{k_c}^{(b)} \sim 100$ MHz, which satisfies the strong coupling condition $\Lambda\gg \Gamma$.

Tre\c{c}as \textit{et al.} proposed a quantum Otto cycle using the lower polaritonic mode as the working medium~\cite{Tercas2017PRE}. The four stroke thermodynamic cycle, see Fig.~\ref{fig:NEM}, is as follows:

(1) \emph{Isentropic compression}: The frequency of the working medium initially prepared at the temperature $T_1\sim T_a$ and frequency $\Omega_1$ (mode $k_1$) is changed to $\Omega_2$ (mode $k_2$) using the piezoelectric cells. During this stroke, only work is performed by the system. The polariton number $N_1\equiv N_k^L$ is fixed by satisfying the constraint $\Lambda\gg 1/\tau_1\gg \Gamma$, where $\tau_1$ is the stroke duration.

(2) \emph{Isochoric heating}: The system is let to thermalize with the hot source at $T_2\sim T_b$ while the frequency is fixed. In this step, the polariton number increases from $N_1$ to its final value $N_2\equiv N_{k_2}^L \sim n_b$. The system receives heat $Q_h$ from the hot source during $\tau_2\sim 1/\Gamma\gg \tau_1$.

(3) \emph{Isentropic expansion}: During this third stroke, the frequency is changed from $\Omega_2$ to its initial value $\Omega_1$. No heat is exchanged during this process and the duration is $\tau_3\sim \tau_1$.

(4) \emph{Isochoric cooling}: Finally, the system transfers heat to the cold reservoir at constant frequency $\Omega_1$.

Based on the standard thermodynamic analysis, the engine total work output $W$ and efficiency $\eta$ become~\cite{Tercas2017PRE}
\begin{equation}
W = \hbar \Omega_2(N_2 - Q^\ast N_1) + \hbar \Omega_1 (N_1 -Q^\ast N_2)\,,
\end{equation}
and
\begin{equation}
\eta = 1- \frac{\Omega_1}{\Omega_2}\frac{N_2 Q_2^\ast - N_1}{N_2 - N_1 Q_1^\ast},
\end{equation}
where $Q_{1,2}^\ast$ is  again the adiabaticity parameter of the parametric harmonic oscillator, see Sec.~\ref{sec:many}.

Ter\c{c}as \textit{et al.}\cite{Tercas2017PRE} further discussed the performance, when the system is functioning as a quantum refrigerator. In addition, realistic parameters to implement the thermal device in an experiment are provided.  This theoretical proposal extends the application of controlled vacuum forces to quantum thermal devices, and the proposal by Ter\c{c}as \textit{et al.} has inspired the design of optomechanical straight-twin engine~\cite{Zhang2017}.

%% file: sections/experiment_NMR.tex
\subsection{Nuclear magnetic resonance}

Due to its relatively low cost and excellent controllability nuclear magnetic resonance (NMR) has found important applications in the verification of quantum thermodynamic relations \cite{Batalhao2014,Micadei2019NC}.  To highlight its utility we continue with discussing the implementation of two unique thermodynamic devices.

\subsubsection{Spin quantum heat engine}\label{expt_NMR1}

Peterson \textit{et al.}\cite{Peterson2019} implemented a proof-of-concept experiment of a quantum heat engine operating at finite time using a $^{13}$C-labeled CHCl$_3$ liquid sample diluted in Acetone-D6 and a 500 MHz Varian NMR spectrometer.  The heat engine cycle is described well by a quantum Otto cycle with the spin-1/2 of the $^{13}$C nucleus as the working medium. The hot reservoir is engineered by a high rf mode close to the $^1$H Larmor frequency while the low reservoir is realized by a low rf mode near the $^{13}$C resonance frequency.  Using an interferometric method, previously employed to verify quantum fluctuation theorems \cite{Batalhao2014}, the authors characterized the work and heat statistics as well as the engine performance with associated irriversibility. 

The four stroke finite-time Otto engine cycle is implemented as follows:  

(1) \emph{Isochoric cooling}: First,  the $^{13}$C nuclear spin is prepared in a cold pseudo-thermal state, $\rho_0^{eq,1}=\e{-\beta_1 H_1^C}/Z_1$ , where $\beta_1=(k_B T_1)^{-1}$ is the cold, inverse spin temperature, and the Hamiltonian 
\begin{equation}
H^C_1=-\hbar\omega_1\sigma_y^C/2\,,
\end{equation} 
at a nuclear spin energy gap $2\pi\omega_1=2.0$kHz.

(2) \emph{Isentropic expansion}: During the second stroke, the working medium is driven by a time-modulated rf field for time duration $t=\tau$. The resulting nuclear spin Hamiltonian is of the form  
\begin{equation}
H_\mrm{exp}^C(t)=-\frac{1}{2}\hbar\omega(t)\left[\cos\left(\frac{\pi t}{2\tau}\right)\sigma_x^C +\sin\left(\frac{\pi t}{2\tau}\right)\sigma_y^C\right]\,,
\end{equation}
which has its energy gap  linearly expanded to $2\pi\omega_2=3.6$ kHZ. The driving time duration ($\sim 10^{-4}$ s) is much shorter than the typical decoherence times ($\sim $seconds),  and thus the process can be described as a unitary evolution $U_\tau$ which drives the $^{13}$C nuclear spin to an out-of-equilibrium state $\rho_\tau^C\equiv\rho_2$ with the final Hamiltonian 
\begin{equation}
H_2^C=H^C_\mrm{exp}(\tau)=-\hbar\omega_2\sigma_x^C/2\,.
\end{equation}
The characteristic function associated with this energy gap expansion stroke reads \cite{Campisi2011}
\begin{equation}
    \chi_\mrm{exp}(u) = \sum_{n,m=0}^1 \wp^0_n \wp^\tau_{m|n} \e{i u (\epsilon^\tau_m - \epsilon^0_n)},
\end{equation} 
where $\wp^0_n$ is again the occupation probability of the $n$th energy level in the initial thermal state $\rho_0^{eq,1}$,  and $\wp^\tau_{m|n}$ is the transition probability between the Hamiltonian eigenstates. Further, $\epsilon_m^\tau$ and $\epsilon_n^0$ are eigenenergies of the initial and final Hamiltonian respectively. The work probability distribution $\mc{P}_\mrm{exp}(W)$ for this  engine stroke is the inverse Fourier transform of the measurable $\chi_\mrm{exp}(u)$.

(3) \emph{Isochoric heating}: In the third stroke, the $^{13}$C nucleus is put in contact with a hot reservoir, namely the $^{1}$H nuclear spin prepared at the inverse spin temperature $\beta_2=(k_BT_2)^{-1}$.  Full thermalization is achieved by repeatedly implementing a sequence of free evolutions between both nuclei and rf pulses which  takes the $^{13}$C nuclei state to the hot equilibrium state $\rho_0^{eq,2}\equiv \rho_3=\e{-\beta_2 H_2^C}/Z_2$. In this stroke, no work is done and the stochastic heat absorbed from the hot reservoir is assessed by a two-time energy measurement scheme.  The mean heat from the hot reservoir can be written as
\begin{equation}
    \langle Q_h\rangle= \tr{H_\mrm{exp}^C(\tau)\, \left(\rho_0^{eq,2} - U_\tau \rho_0^{eq,1} U_\tau^\dagger\right)}\,.
\end{equation}

(4) \emph{Isentropic compression}: Then, the energy gap is compressed by performing the time-reversed protocol of the expansion such that the Hamiltonian $H_\mrm{comp}^C(t)=-H^C_\mrm{exp}(\tau - t)$. In full analogy to the expansion stroke, the mean work value is deduced from the characteristic function of the work probability distribution $\chi_\mrm{comp}(u)$. 

Peterson \textit{et al.}\cite{Peterson2019} measured the inverse Fourier transform of the characteristic functions, which was employed to deduce the engine performance, such as $\langle W\rangle$, $\langle Q_{hot}\rangle$, power and efficiency $\eta$  as a function of the driving protocol time duration.  Specifically, the average work is obtained from
\begin{equation}
    \la W\ra =\int dW\, \mc{P}(W)\, W, 
\end{equation}
where $\mc{P}(W)= \int du \chi (u) \e{i u W}$,  and the product of characteristic functions of expansion and compression protocols,  $\chi(u)=\chi_\mrm{comp}(u) \chi_\mrm{exp}(u)$.  The nuclear spin system operates as a heat engine when $\langle W\rangle > 0$
 and the efficiency, $\eta=W/Q_h$ can be computed in terms of the energy level transition probability as
\begin{equation}
    \eta = 1 - \frac{\omega_1 (1 - 2\xi \mc{F})}{\omega_2 (1 - 2\xi \mathcal{G})},
\end{equation}
where $\mathcal{F}=\tanh(\beta_2 \hbar\omega_2)/(\tanh(\beta_2 \hbar\omega_2)-\tanh(\beta_1 \hbar \omega_1))$, $\mathcal{G}=\mathcal{F}\tanh(\beta_1 \hbar\omega_1)/\tanh(\beta_2 \hbar \omega_2)$, and $\xi$ are the transition probabilities between the instantaneous
eigenstates of the Hamiltonian during the expansion and compression strokes.  In the finite-time regime the efficiency decreases as $\xi$ increases and the Otto efficiency limit, $\eta=1-\omega_1/\omega_2$, is recovered in the adiabatic regime, i.e.  $\xi= 0$. 

Finally,  Peterson \textit{et al.}\cite{Peterson2019}  analyzed the engine's irreversibility and efficiency lag from the statistics of energy fluctuations. The engine achieved efficiency at maximum power $\eta=42\pm 6\%$, which is very close to the expected ideal Otto engine efficiency $\eta=44\%$ based on the compression ratio implemented in the experiment.

Interestingly, this NMR engine provided a template for an algorithmic cooling cycle \cite{Kose2019PRE}, which was eventually realized with nitrogen vacancy centers (see Sec.~\ref{sec:algo_cool}).

\subsubsection{Effective negative temperatures}

\begin{figure}
\includegraphics[width=.48\textwidth]{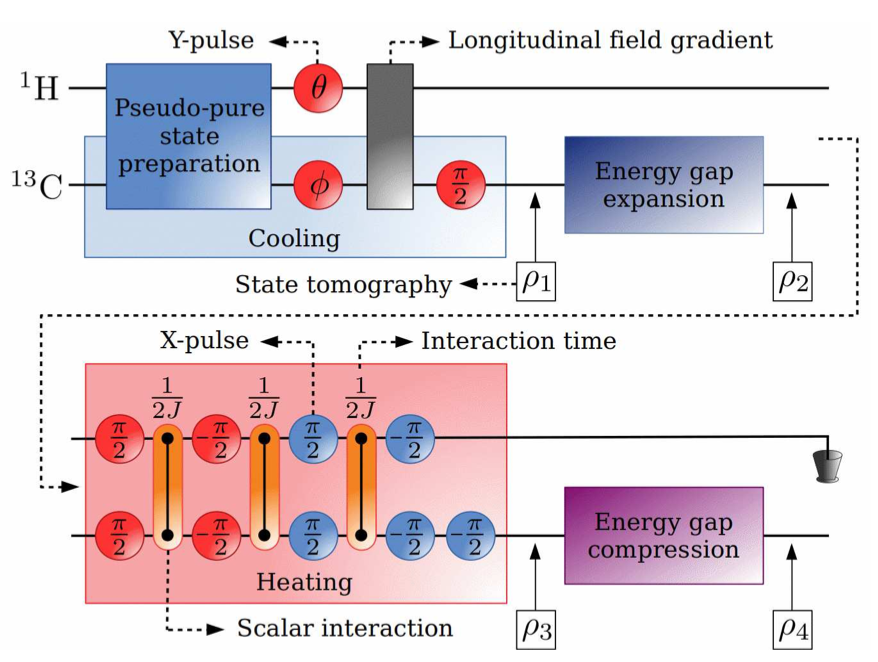}
\caption{\label{fig:expt_nmr2} Schematic representation of the quantum Otto spin heat engine with nuclear magnetic resonance. \textit{Figure adopted from Ref.~\cite{deAssis2019}}.}
\end{figure}

Returning to the origins of quantum thermodynamic devices, NMR was also exploited to realize a device with population inversion. Using the coherent control offered by the very sensitive  radio frequency and gradient field in the NMR setup, de Assis \textit{et al.}\cite{deAssis2019} implemented a quantum Otto heat engine that operates between a thermal reservoir at a positive spin temperature and a reservoir at an effective \emph{negative} spin temperature. Following the previous NMR spin engine experiment summarized above,  the inverse temperature $\beta_{1/2}$ of the working medium ($^{13}$C nuclear spin) can be determined by adjusting the population of its excited state $\wp^+_{i}$ ($i=1,2$) according to the relation
\begin{equation}
    \beta_i=\frac{1}{\hbar\omega_i}\ln \left(\frac{1-\wp^+_i}{\wp^+_i}\right),
\end{equation}
where $\wp^+_i\!=\!\bra{\Psi_{\pm}^i}\rho_0^{eq,i}\ket{\Psi_{\pm}^i}$,  and $\Psi_{\pm}^i$ is the eigenstate of the Hamiltonian $H_i^C$ with a positive eigenvalue.  Positive $\beta_i$ corresponds to $0<\wp^+_i<0.5$ value while negative $\beta_i$ correspond to $0.5<\wp^+_i<1$. In the experiment, de Assis \textit{et al.}\cite{deAssis2019} chose $\wp^+_1=0.261\pm 0.004$ while $\wp^+_2$ was varied from 0.5 to 1.0. The population was assessed by tomogrphy of states $\rho_0^{eq,i}$. The corresponding engine cycle is depicted in Fig.~\ref{fig:expt_nmr2}.

In the experiment by de Assis \textit{et al.}\cite{deAssis2019}, the authors analyzed  work, heat and efficiency of the heat engine when one of the thermal reservoirs is prepared at negative temperature.  Interestingly, it was found that $\beta_1\omega_1<|\beta_2|\omega_2$ implies $\eta<1-\omega_1/\omega_2$, whereas if $\beta_1\omega_1 \ge |\beta_2|\omega_2$  then $\eta \ge 1-\omega_1/\omega_2$.

The proof of concept experiments of a quantum Otto heat engine realized with NMR demonstrated their versatility of investigating the thermodynamic properties as well as the irreversibility of thermal devices. The NMR set up has recently been used to shed more light on understanding the performance of quantum engines with engineered reservoirs~\cite{Mendonca2020,Assis2020PRE} as well as experimentally demonstrating the influence of fluctuations in quantum heat engines~\cite{Denzler2021arXiv}.

%% file: sections/experiment_NV.tex
\subsection{Nitrogen vacancy centers}

A fundamentally different implementation of spin-1/2 systems are so-called nitrogen vacancy (NV) centers \cite{Herandez2021front}. In such systems,  diamond is doped with nitrogen atoms, which creates a ``vacancy'' in the valence electrons.  The effective two-state system (electron and hole pairs) offer great versatility and controllability.  For instance,  NV centers have been also bee used to verify fluctuation theorems \cite{Hernandaz2020PRR}.

\subsubsection{Genuine quantum effects in heat engines}

In a highly noticed experiment, Klatzow \etal \cite{Klatzow2019PRL} realized two types of heat engines in NV centers. The experiment was based on the theoretical proposal by Uzdin \etal \cite{Uzdin2015PRX}.  In particular, Uzdin \etal \cite{Uzdin2015PRX} had shown that internal, coherent superposition states can have a measurable effect on thermodynamic quantities, cf.  also the quantum afterburner \cite{Scully2002} in Sec.~\ref{sec:afterburner}.  

\begin{figure}
\includegraphics[width=.48\textwidth]{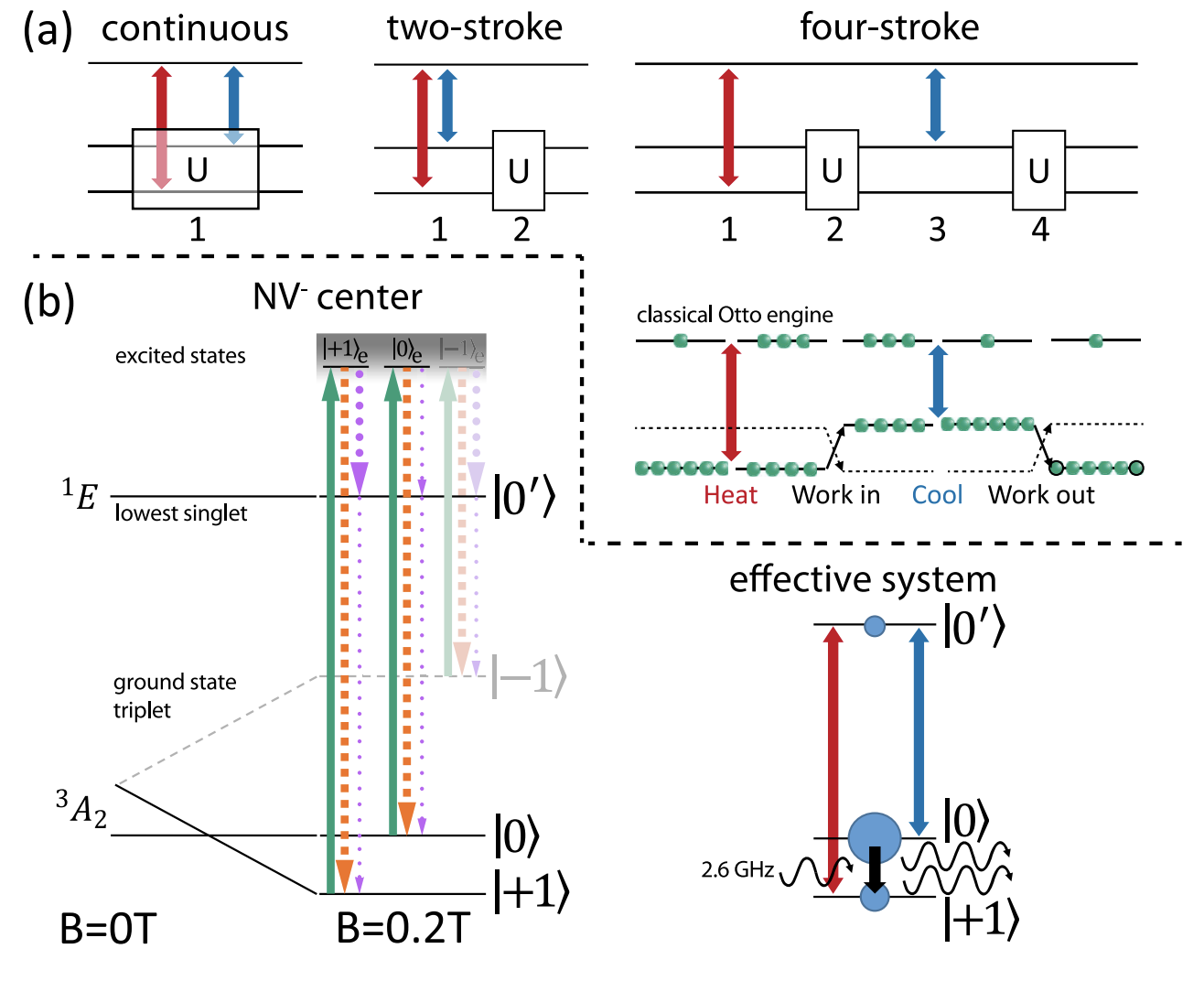}
\caption{\label{fig:NV_1} (a) Schematics of three basic heat engine types for a three-level system driven by a unitary $U$; (b) left: relevant energy levels in the NV center and optically induced transitions; right: schematics of the effective three-level NV$^-$ heat engine.  \textit{Figure adopted from Ref.~\cite{Klatzow2019PRL}}.}
\end{figure}

The experiment by Klatzow \etal \cite{Klatzow2019PRL} was the first to implement thermal machines that demonstrate this genuine quantum feature.  More specifically, they realized a two-stroke engine and a continuous engine, which are schematically depicted in Fig.~\ref{fig:NV_1}. 

Reference~\cite{Uzdin2015PRX} predicted two coherent superposition-related effects: (i) increased output power of a quantum engine, which is considered a quantum thermodynamic signature (QTS); and (ii) a convergence in output power of different types of quantum heat engines, i.e., a quantum heat machine equivalence (QHME).

It is important to note that coherence is created in unitarily driving a quantum system, whereas the interaction with a thermal bath is accompanied by decoherence.  Hence, coherence related effects are best studied in the quantum Otto cycle. However, since no realistic quantum system is ever fully isolated from its surroundings, different engine cycles can be implemented by varying the stroke times. As noted several times above, for instance, stroke times much shorter than the decoherence time effectively realize unitary dynamics.  For a three level system,  Klatzow \etal \cite{Klatzow2019PRL} considered three engine types, a continuous, two stroke, and four stroke engines, see Fig.~\ref{fig:NV_1}(a).

The engines were realized in an NV$^-$ center in diamond \cite{Doherty2013PR}.  Its ground state manifold is spanned by three spin states, $\ket{-1}$, $\ket{0}$, and $\ket{+1}$, which can maintain coherence even at room temperature. The work reservoir was realized through the coherent interaction with a microwave field. In operation, the system is optically excited and then decays back into the ground-state manifold through either spin-preserving radiative decay, or spin-non-preserving non-radiative channels through a meta-stable spin-singlet state. Consequently, the stationary state exhibits population differences in the three ground state spin components.  From a theoretical point of view, the dynamics can simply be described by a Lindblad master equation, which makes the comparison of theory and experiment rather manageable.

Finally, the spin-dependence of the non-radiative decay channels determines the flourescence intensity, which can be exploited to directly measure the ground-state populations. Actually, this optically detected magnetic resonance \cite{Doherty2013PR} provides better sensitivity than more standard methods. A sketch of the effective energy spectrum can be found in Fig.~\ref{fig:NV_1}(b).

Klatzow \etal \cite{Klatzow2019PRL} then used an ensemble of NV$^-$ centers in diamond to implement two types of heat engines, the 2-stroke engine and the continuous engine.  With very high accuracy, they were able to verify the theoretical predictions of Uzdin \etal \cite{Uzdin2015PRX}, and observed both QTS as well as QHME.

As a first experiment to demonstrate genuine quantum effects in heat engine performance, the work by Klatzow \etal \cite{Klatzow2019PRL} has already received significant attention.  Particular noteworthy are recent developments in optimal driving and shortcuts to adiabaticity \cite{Cakmak2019PRE,Funo2019PRB,Pancotti2020PRX,Hartmann2020PRR},  spin-based quantum devices \cite{Lindenfels2019PRL,Hong2020PRE},  signatures of quantum coherence in quantum engines \cite{Niedenzu2018NJP,Thomas2019PRE,Camati2019PRA,Gonzalez2019PRE,Dann2020NJP,Das2020PRR,Wiedmann2020NJP,Hammam2021NJP}, and nonequilibrium operation \cite{Manzano2019NJP,Kloc2019PRE,Newman2020PRE,Carollo2020PRL,Chiaracane2020PRR,Chiara2020PRR,Zhang2020PhysA}.

\subsubsection{\label{sec:algo_cool} Minimal algorithmic cooling refrigerator}

\begin{figure}
\includegraphics[width=.48\textwidth]{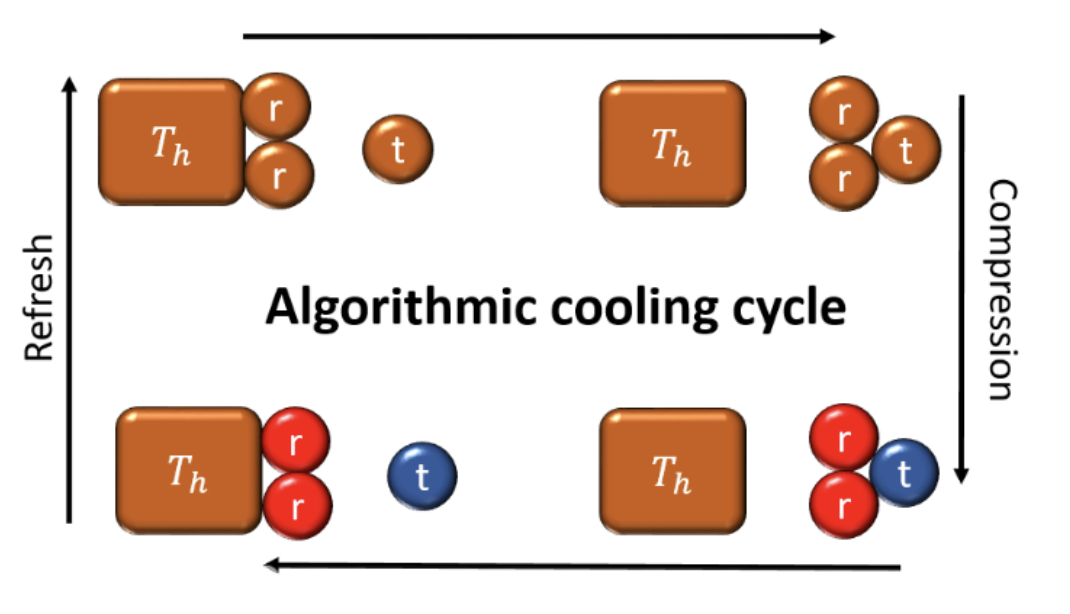}
\caption{\label{fig:NV_2} Schematic representation of an algorithmic cooling cycle.  Heat is extracted from the target spin during the compression stroke, which is exhausted into the heat bath during the refresh stroke. \textit{Figure adopted from Ref.~\cite{soldati2021arXiv}}.}
\end{figure}

An even more recent experiment with NV centers was reported by Soldati \etal\cite{soldati2021arXiv}.  Specifically, the experiment realized an algorithmic cooling refrigerator \cite{Park2016} made of three nuclear spins hyperfine coupled to the central electron spin of an NV center in diamond \cite{Doherty2013PR}. The device is schematically depicted in Fig.~\ref{fig:NV_2}.

Heat-bath algorithmic cooling reduces the entropy of a target spin by exploiting the fast relaxation of reset spins. The cooling cycle consists of (i) compression steps that cool the target spin and heat up the reset spins, and of (ii) refresh steps during which the reset spin rapidly relax back to the bath temperature.  In the experimental setting \cite{soldati2021arXiv}, the target spin and the two reset spins were chosen has the $^{14}$N and the two $^{13}$C nuclear spins, respectively. Thus, the central electron spin can act as (i) the heat bath as well as (ii) an ancillary spin that drives the interaction among the spins required to achieve the entropy compression on the target spin \cite{Zaiser2021NPJQI}.

Theoretically, the three spins are simply described by the Hamiltonian
\begin{equation}
H=\sum_i\omega_i \sigma_i^z\,,
\end{equation}
where $\omega_i$ is the Rabi frequency and $\sigma_i^z$ the Pauli-$z$-matrix. Initially the authors assumed the three spins to be prepared in a product state with
\begin{equation}
\rho(0)=\frac{1}{2}\,\otimes_i \left(\id-\epsilon_i(0)\, \sigma_i^z\right)\,,
\end{equation}
where $\epsilon_i(0)$ are the polarizations.

For the thermodynamic analysis, the authors then considered the heat extracted per $n$th cycle,
\begin{equation}
Q(n)=\omega_1\,\sigma_1^z\left(\rho_1(n)-\rho_1(n-1)\right)\,,
\end{equation}
for which the work
\begin{equation}
W(n)=\sum_i \tr{\omega_i\,\sigma_i^z\left(\rho_i(n)-\rho_i(n-1)\right)}
\end{equation}
has to be expended\cite{Rempp2007PRA}. The device is then full characterized by the coefficient of performance, $\zeta(n)$, and the cooling power $J(n)$. In formula, we have
\begin{equation}
\zeta(n)=-\frac{Q(n)}{W(n)}\quad\text{and}\quad J(n)=Q(n+1)-Q(n-1)\,.
\end{equation}

The authors then computed the coefficient of performance and cooling power from solving the dynamics analytically with the help of a standard description in terms of quantum channels. They found the maximal coefficient of performance as a function of the damping coefficient $\gamma$ to read \cite{soldati2021arXiv},
\begin{equation}
\zeta_\mrm{max}(n,\gamma= 0)=1\,,
\end{equation}
and the corresponding cooling power is
\begin{equation}
J_\mrm{max}(n,\gamma= 0)=\frac{\epsilon}{2}\,(1+\epsilon^2)\,\e{-(n-3/2) g(\pi/2,0)}\,.
\end{equation}
Here, $\epsilon_i(0)=\epsilon$ for all $i$, and $g(\theta,\gamma)=\lo{4/(1-\gamma) f(\theta)}$ with $f(\theta)=3+(1+\epsilon^2) \co{2\theta}-\epsilon^2$.

The theoretical predictions were verified to very high accuracy. More importantly, the experiment demonstrated that genuine quantum devices can be implemented in NV centers. Thus the algorithmic refrigerator of Soldati \etal \cite{soldati2021arXiv} complements the engine of Klatzow \etal \cite{Klatzow2019PRL} to give a fuller thermodynamic picture.

%% file: sections/experiment_BEC.tex
\subsection{Quantum gases and ultracold atoms}

So far we have discussed several physical platforms that allow the realization of genuine quantum working media.  Despite their physical differences, they have in common that the effective theoretical description is simple, in the sense that we were able to work with two-level systems or harmonic oscillators.  In the following, we will now discuss more complex quantum scenarios and we begin with (many) cold atoms.

\subsubsection{Thermoelectric engine with fermions}

\begin{figure}
\includegraphics[width=.48\textwidth]{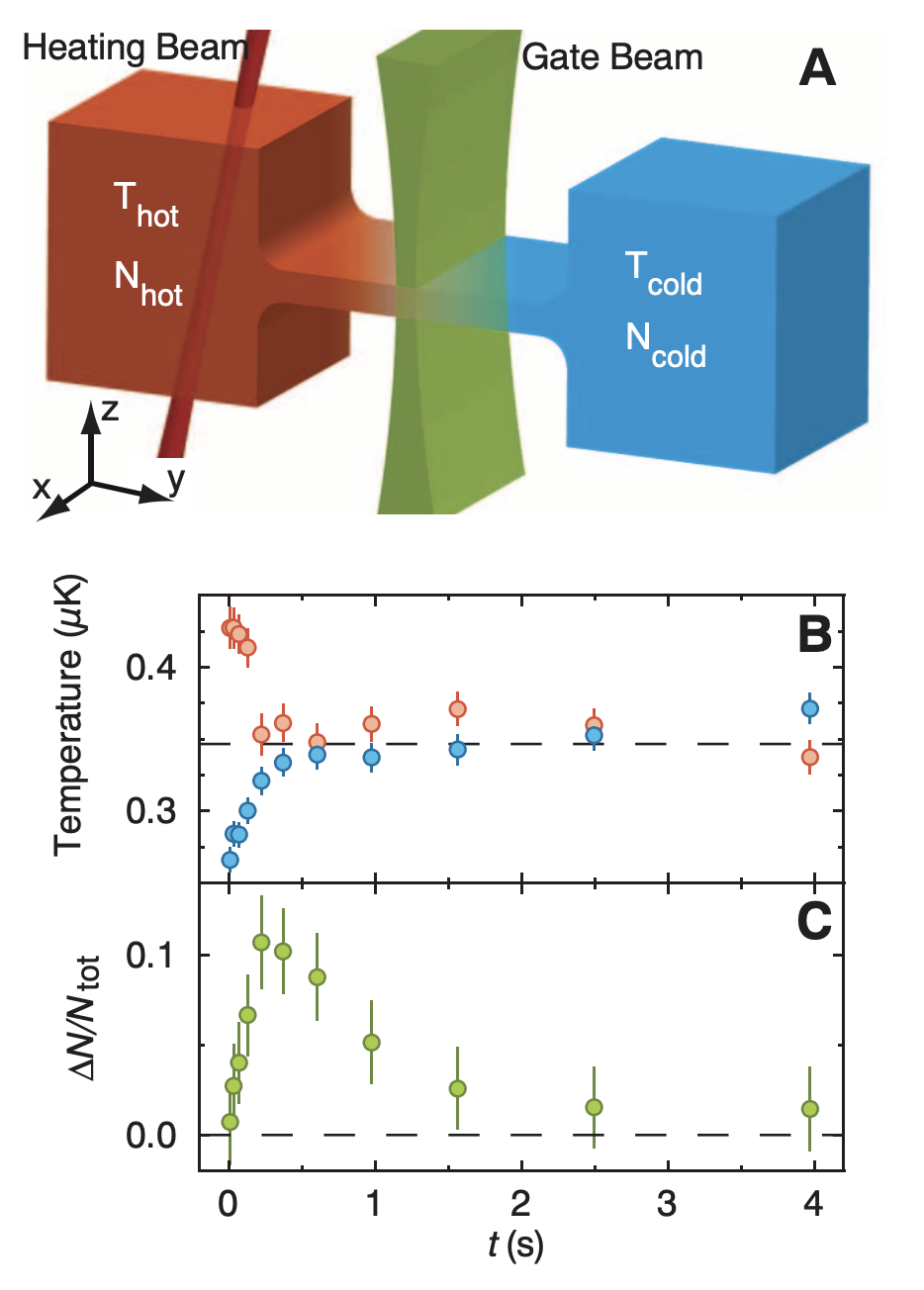}
\caption{\label{fig:thermo_electric} (A) Two atomic reservoirs connected by a quasi-2-dimensional channel. A gate beam acts a switch for particle and heat transfer, and a heating beam is shined on the left reservoir; (B) $T_h$ and $T_c$ as a function of time; (C) relative change in particle number $\Delta N/N_\mrm{tot}$ as a function of time.  \textit{Figure adopted from Ref.~\cite{Brantut2013Science}}.}
\end{figure}

Brantut \etal \cite{Brantut2013Science} realized a thermoelectric engine with ultracold $^6$Li atoms. To this end, they prepared $N_\mrm{tot}=3.1(4) \times 10^5$ weakly interacting atoms at $\sim$250(9)K in an elongated trap. Using a replusive laser they further separated the $^6$Li-atoms into two identical clouds that were connected by a quasi two-dimensional channel, see Fig.~\ref{fig:thermo_electric} for a schematic representation.  One of the clouds was then heated  by about 200nK.  Thus, they experimentally realized a hot and a cold reservoir, with temperatures $T_h$ and $T_c$, that are weakly coupled, and for which the thermodynamic properties can be fully analyzed.

Such scenarios can be accurately described by means of linear response theory \cite{Callen1985}. In particular, we are interested in particle and entropy  currents, 
\begin{equation}
I_N=\frac{d}{dt}\, \left(N_c-N_h\right)\quad\text{and}\quad I_S=\frac{d}{dt}\,\left(S_c-S_h\right)\,,
\end{equation}
respectively, for which we have\cite{Brantut2013Science}
\begin{equation}
\begin{pmatrix} I_N\\I_S\end{pmatrix}=-G\,\begin{pmatrix}1&\alpha_{ch}\\ \alpha_{ch}&L+\alpha_{ch}^2\end{pmatrix}\,\begin{pmatrix}\mu_c-\mu_h\\T_c-T_h\end{pmatrix}\,.
\end{equation}
Here, $\mu_c$ and $\mu_h$ are the chemical potentials of the cold and hot reservoirs. Further, $G$ is the conductance of the two-dimensional channel and $\alpha_{ch}$ is the thermal power. The Lorenz number $L$ can be determined from the thermal conductance $G_T$, 
\begin{equation}
L\equiv \frac{G_T}{G}\frac{1}{\bar{T}}\,,
\end{equation}
where $\bar{T}=(T_h+T_c)/2$ is the average temperature. Therefore,  the dynamics of the thermoelectric engine is fully described by the phenomenological equation\cite{Brantut2013Science}
\begin{equation}
\label{eq:flux_thermoelectric}
\tau_0\,\frac{d}{dt}\,\begin{pmatrix} \Delta N\\ \Delta S\end{pmatrix}=-\begin{pmatrix}1&\kappa\left(\alpha_{ch}-\alpha_{T}\right)\\ \frac{\alpha_{ch}-\alpha_T}{\ell \, \kappa}&\frac{L+\left(\alpha_{ch}-\alpha_T\right)^2}{\ell}\end{pmatrix}\,\begin{pmatrix} \Delta N\\ \Delta S\end{pmatrix}\,,
\end{equation}
where we introduced the usual thermodynamic quantities
\begin{equation}
\kappa=\frac{\pd N}{\pd \mu}\bigg|_T\,,\quad C_N=T\,\frac{\pd S}{\pd T}\bigg|_N\,,\quad\text{and}\quad \alpha_T=\frac{\pd S}{\pd N}\bigg|_T\,,
\end{equation}
namely compressibility, heat capacity, and dilation coefficient. Moreover,  $\ell\equiv C_N/\kappa \bar{T}$ is an analog of the Lorenz number for the reservoirs measuring the relative magnitude of the thermal fluctuations of entropy and atom number. Finally, $\tau_0\equiv\kappa/G$ is a timescale analogous to a capacitor's discharge time. 

For small enough coupling between the reservoirs, i.e., small enough two-dimensional channels, the atomic clouds remain approximately in equilibrium. Therefore,  the phenomenological coefficients $G$, $G_T$, and $\alpha_{ch}$ can be computed with the Landauer-B\"uttiker formalism \cite{Ashcroft1976}. Using the resulting values Brantut \etal \cite{Brantut2013Science} then found that Eq.~\eqref{eq:flux_thermoelectric} describes the experimentally measured fluxes to very high accuracy.

For our present purpose it is even more interesting that Brantut \etal \cite{Brantut2013Science} realized that the controlled exchange of heat between a hot and a cold reservoir can be used to produce work.  Thus, the two coupled clouds are a realization of an atomic heat engine.  The work output is given by
\begin{equation}
W=\frac{1}{2}\,\int_0^\infty dt \,\left(\mu_c(t)-\mu_h(t)\right)\,I_N(t)\,,
\end{equation}
and the heat can be written as
\begin{equation}
Q=-\frac{1}{2}\,\int_0^\infty dt\,\left(T_c(t)-T_h(t)\right)\,I_S(t)\,.
\end{equation}
Accordingly, the efficiency is $\eta=W/Q\leq 1$, where equality is achieved for quasistatic transport.

Brantut \etal \cite{Brantut2013Science} continued their analysis by measuring $W$ and $Q$ for a variety of different coupling strengths, and also for disordered systems. To this end,  they projected a blue detuned laser speckle pattern on the channel, which creates a random potential. As a main result they found that the efficiency $\eta$ is the largest in the strongly disordered regime, where the thermodynamic response is the largest. In contrast the average power $W/\tau_0$ has a maximum at low disorder, where the particles can flow freely and the transport is unobstructed.

In conclusion, the thermoeletric engine realized by Brantut \etal \cite{Brantut2013Science} is an important milestone in the development of technologies with ultracold atoms.  The experiment has inspired significant work on,  for instance,  thermoelectrics \cite{Rancon2014NJP,Ouerdane2015PRB,Esposito2015PRB,Yamamoto2015PRE,Sekera2016PRA,Dorda2016PRB,Eich2016JPCM,Bohling2018PRB,Marchegiani2020PRB}, quantum transport \cite{Liu2014PRA,Gallego2014PRA,Bhaseen2015NP,Chien2015NP,Gaspard2015NJP,Ludovico2016PRB,Filippone2016PRA,Dare2016PRB,Yao2018PRA}, and cooling of atomic gases \cite{Grenier2014PRL,Grenier2016CRP}.

\subsubsection{Quantum refrigerator for bosons}

Applying quantum effects to cool more efficiently has also led to the theoretical proposal of a quantum Otto refrigerator for bosons \cite{Niedenzu2019BEC,Deffner2019quantum}. Niedenzu \etal \cite{Niedenzu2019BEC} proposed to build a thermal device that operates on a mixture of two atomic species. One of the species represents the working medium that is alternately coupled to atomic cloud at high and low temperatures of the other species. 

More specifically, Niedenzu \etal \cite{Niedenzu2019BEC} imagined the following setup: The working mediums is formed by atoms in a narrow optical trap, such that the gas becomes effectively a two-level system.  The corresponding Hamiltonian is written as
\begin{equation}
H=\varepsilon\, \sigma_{+} \sigma_{-}\,,
\end{equation}
where $\sigma_{\pm}$ are the Pauli raising and lowering operators. The level spacing $\varepsilon$ can be controlled externally by shining in a resonant laser. 

The Otto refrigeration cycle is then implemented by four strokes:

(1) \emph{Isochoric cooling}: The working medium with level spacing $\varepsilon_c$ thermalizes with the cold reservoir at temperature $T_c$.  The thermalization is facilitated through atomic collisions, and hence no radiative processes need to be considered. During this stroke the working medium receives the heat 
\begin{equation}
Q_c=\varepsilon_c\,\left(\bar{n}_c-\bar{n}_c\right)\,,
\end{equation}
where $\bar{n}_{c,h}=1/\left[\e{\beta_{c,h}\,\varepsilon_{c,h}}+1\right]$ are the cold and hot thermal occupations, respectively. 

(2) \emph{Isentropic expansion}: In the second stroke the working medium is detached from the cold bath, and $\varepsilon$ is adiabatically raised from $\varepsilon_c$ to $\varepsilon_h$. For this the work
\begin{equation}
W_\mrm{in}=\left(\varepsilon_h-\varepsilon_c\right)\,\bar{n}_h
\end{equation}
is required. This work is provided by the optical trap holding the working medium in place. 

(3) \emph{Isochoric heating}: During the third stroke the working medium equilibrates with the hot reservoir, and exhausts the heat
\begin{equation}
Q_h=\varepsilon_h\,\left(\bar{n}_h-\bar{n}_c\right)\,.
\end{equation}

(4) \emph{Isentropic compression}: Finally,  the fourth stroke completes the cycle, during which the level spacing is adiabatically varied back to $\varepsilon_c$. This produces the work
\begin{equation}
W_\mrm{out}=\left(\varepsilon_c-\varepsilon_h\right)\,\bar{n}_c\,,
\end{equation}
which is radiated into the optical field.

Using standard considerations \cite{Callen1985}, it is obvious\cite{Niedenzu2019BEC} that such a setup will function as an Otto refrigerator for all
\begin{equation}
\bar{n}_c > \bar{n}_h\quad\text{and hence}\quad \frac{\varepsilon_h}{T_h}>\frac{\varepsilon_c}{T_c}\,.
\end{equation}
Therefore, the theoretical value for the minimal achievable temperature can be written as
\begin{equation}
T_c^\mrm{min}=\frac{\varepsilon_c}{\varepsilon_h}\, T_h\,,
\end{equation}
which is entirely determined by the range over which the level spacing can be tuned.

Niedenzu \etal\cite{Niedenzu2019BEC} then continued to elaborate on a proposal for a real experiment. To this end, they suggested that the spatially separated hot and cold reservoirs can be implemented by separate harmonic traps. The working medium itself is formed by means of a species-selective optical lattice or tweezer. Controlling the level spacing can be done by a variety of readily available experimental techniques. That such a scenario is realistic was then further demonstrated by a numerical analysis and simulation of such a two-species Otto device.  Specifically,  the bosonic baths were imagined as Cs atoms, and the working medium was made of Rb atoms. Niendenzu \etal\cite{Niedenzu2019BEC} demonstrate that for experimentally realistic parameters the Rb atoms can be cooled below the quantum critical point of the BEC phase transition.

This theoretical proposal opens the door to a variety of potential devices that operate with a BEC as a working medium. For instance, Keller \etal\cite{Keller2020PRR} analyzed only recently how work can be extracted from the Feshbach resonances.

\subsubsection{Bose-Einstein condensation}

The BEC phase is characterized by a large fraction of particles simultaneously occupying the zero-momentum state. As particles in the zero-momentum state cannot exert pressure, it is not easy to see how the typical paradigm for work extraction from thermal machines, involving pressure exerted against an external piston or potential, translates to a BEC working medium. This has led to inventive proposals for BEC-based engines, including extracting work by manipulating the interparticle interaction strength using Feshbach resonances \cite{Li2018, Keller2020PRR} and using BECs as the basis for thermal machines that act on a working medium of quantum fields \cite{Gluza2021PRXQuantum}. 

In Ref. ~\cite{MyersBECArXiv} it was shown that, even within the typical paradigm of work extraction, the phenomenon of Bose-Einstein condensation can be used to enhance engine performance by taking advantage of the transition of particles between the BEC and excited states in the thermal cloud. The proposed setup consists of a gas of bosons in an isotropic harmonic potential with frequency $\omega$. The system is considered to be in the thermodynamic limit, such that the number of particles, $N$, is large enough that $N\omega$ can be considered constant.

Reference~\cite{MyersBECArXiv} considers an Otto cycle operating in both the equilibrium and finite-time endoreversible paradigms. During the isochoric strokes the trapping potential is held constant while the working medium exchanges heat with a hot or cold reservoir. The isentropic strokes are implemented by disconnecting the working medium from the thermal reservoirs while the trapping potential is varied between $\omega_{1}$ and $\omega_{2}$.    

The performance of the cycle is explored in three different regimes, one in which the working medium remains below the critical temperature for the BEC phase transition over the full duration of the cycle, one in which the medium remains above the critical temperature over the full duration of the cycle, and one in which the working medium is driven across the BEC phase transition during each heating and cooling stroke. Notably, the condition that the compression and expansion strokes must be isentropic has the natural consequence that the fraction of the working medium particles in the ground state must remain constant during these strokes. Thus the phase transition between Bose gas and BEC can only occur during the heating and cooling strokes.     

In all three regimes, for both endoreversible and quasistatic operation, the efficiency is found to be given by the typical Otto efficiency,
\begin{equation}
	\eta_{\mathrm{below}} = \eta_{\mathrm{above}} = \eta_{\mathrm{transition}} = 1 - \kappa, 
\end{equation}
where $\kappa \equiv \omega_1/\omega_2$ is the compression ratio.

\begin{figure*}
	\centering
	\subfigure[]{
		\includegraphics[width=.28\textwidth]{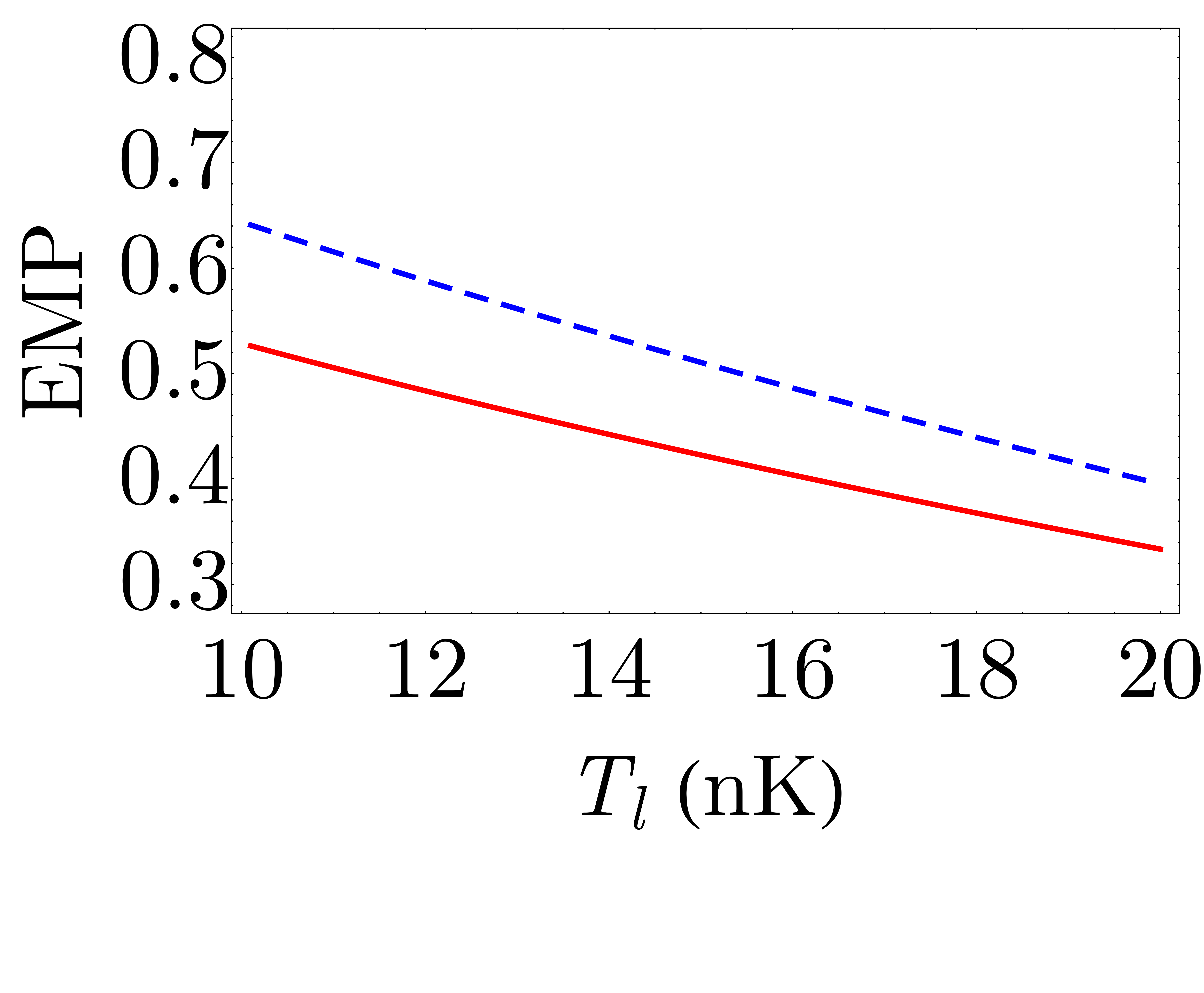}
	}
	\subfigure[]{
		\includegraphics[width=.28\textwidth]{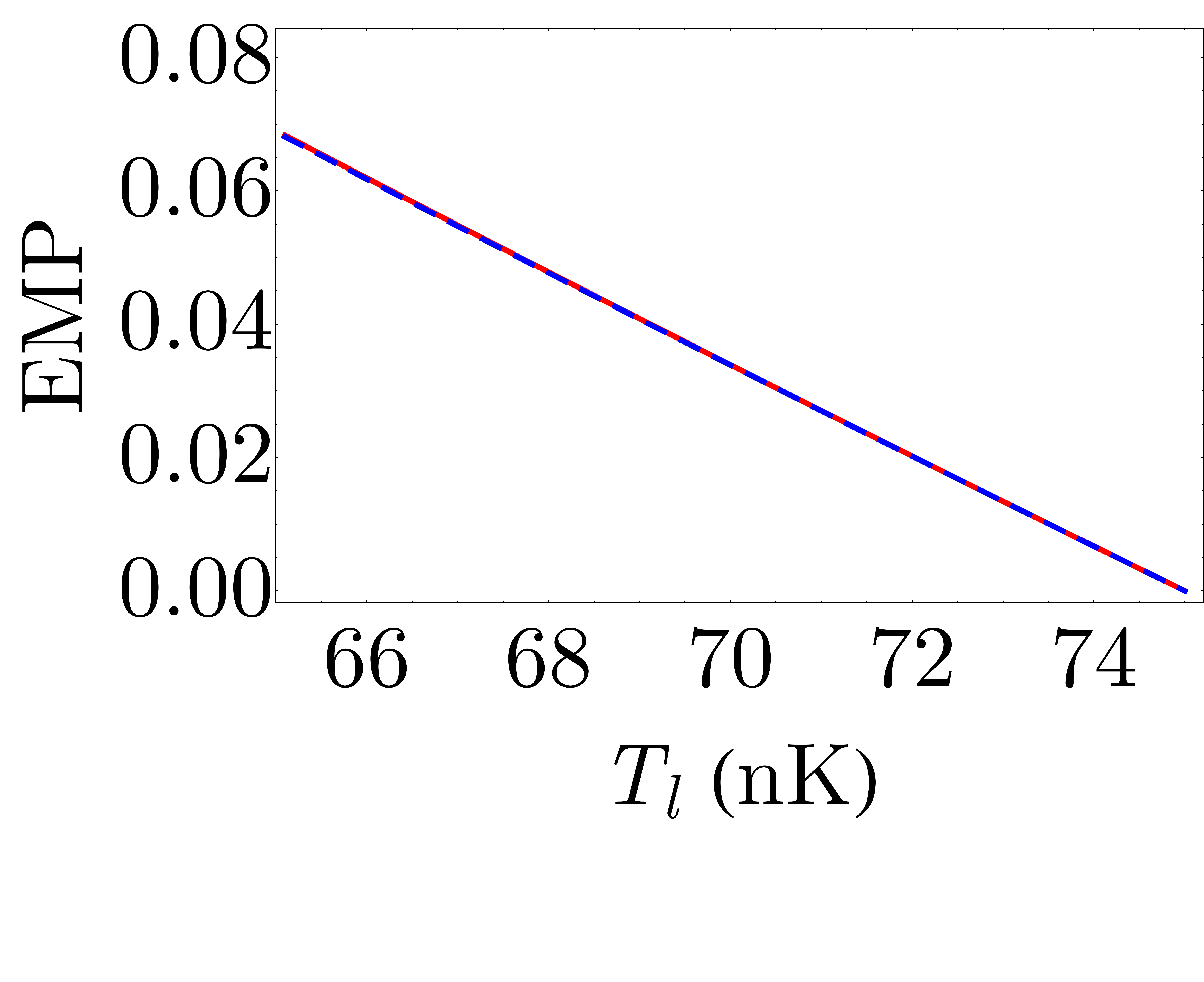}
	}
	\subfigure[]{
		\includegraphics[width=.28\textwidth]{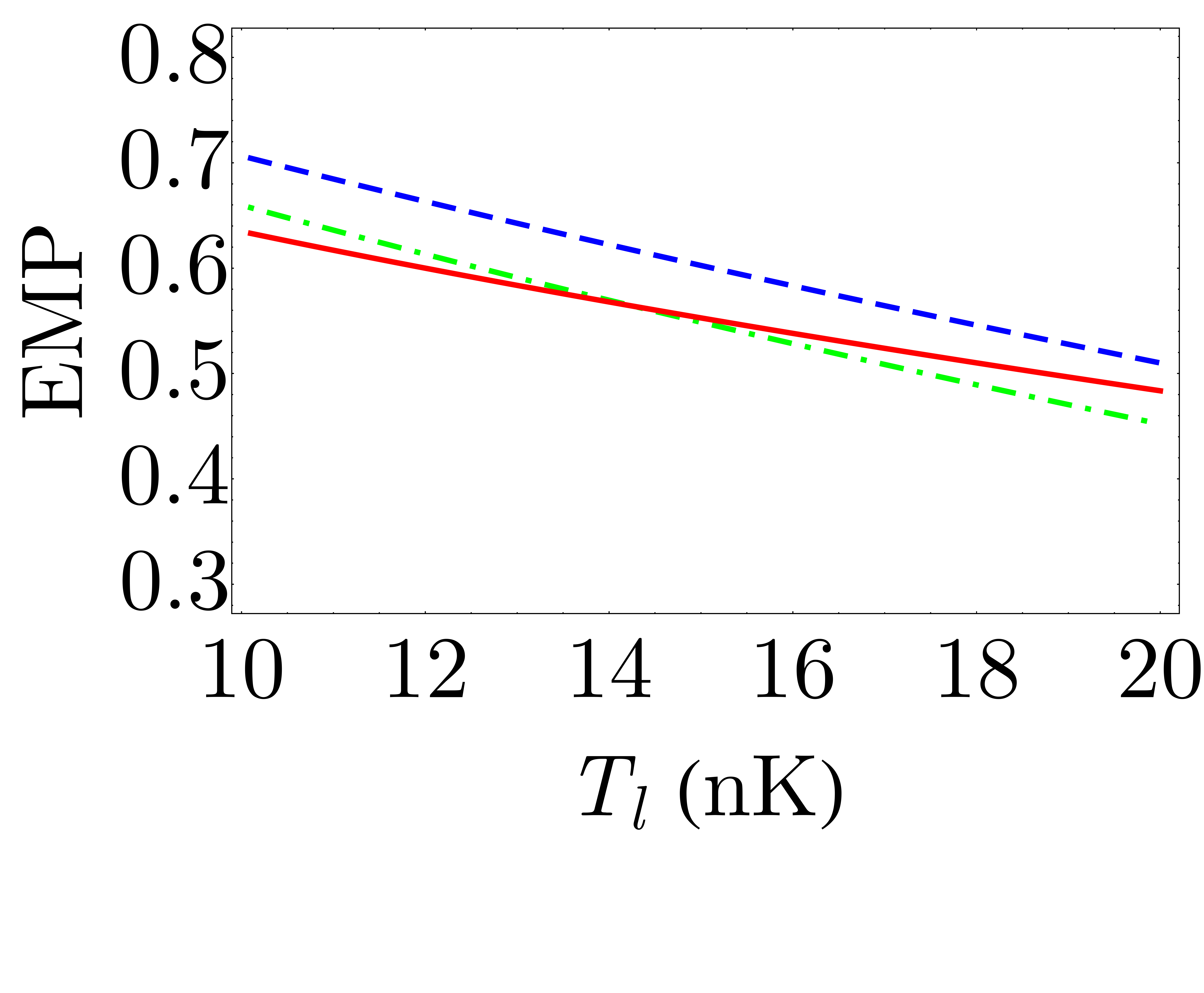}
	}
	\caption{\label{fig:EMPplotBEC} (a) EMP as a function of the cold bath temperature for a BEC working medium (blue, dashed). (b) EMP as a function of the cold bath temperature for a working medium of 60,000 bosons above the condensation threshold (blue, dashed). (c) EMP as a function of the cold bath temperature for a working medium of 60,000 bosons driven across the BEC phase transition with short stroke durations (blue, dashed) and long stroke durations (green, dot dashed). In each plot the Curzon-Ahlborn efficiency (red, solid) is given for comparison. \textit{Figure adopted from Ref. \cite{MyersBECArXiv}}.}
\end{figure*}

While the efficiency in each regime is identical, the endoreversible efficiency at maximum power (EMP), illustrated in Fig. ~\ref{fig:EMPplotBEC}, differs appreciably. For a working medium that remains in the BEC phase, the EMP is significantly higher than the Curzon-Ahlborn efficiency. As the Curzon-Ahlborn efficiency is the EMP for a classical harmonic Otto engine \cite{Deffner2018Entropy}, this demonstrates that the condensate working medium leads to a significant advantage in performance. Consistent with the classical limit, the EMP of a working medium above the critical temperature is found to be identical to the Curzon-Ahlborn efficiency. Finally, the EMP of an engine with the working medium driven across the BEC transition was found to exceed Curzon-Ahlborn when the temperature of the cold bath is low, but falls below Curzon-Ahlborn at higher cold bath temperatures. Notably, this behavior only emerges for short cycle durations. For longer cycle times, the EMP of the transition engine is found to exceed the Curzon-Ahlborn efficiency even at higher cold bath temperatures.    

For a working medium below the critical temperature the work, and thus power output, are shown to depend directly on the number of excited particles in the thermal cloud during the expansion and compression strokes,
\begin{equation}
	\label{eq:Npower}
	P = \frac{\pi^4 (1-\kappa)\hbar \omega_2}{30 \tau \zeta(3)^{4/3}}\left[\left(N^{\mathrm{exp}}_T\right)^{4/3} - \left(N^{\mathrm{comp}}_T\right)^{4/3} \right],
\end{equation}
where $\tau$ is the cycle duration and $\zeta$ is the Riemann-$\zeta$-function.  The number of excited particles is given by,          
\begin{equation}
	\label{eq:Nexcited}
	N_T = \left(\frac{k_B T}{\hbar \omega} \right)^3 g_3(z)\,,
\end{equation}
where $T$ is the temperature, $z$ is the fugacity, and $g_3(z)$ is the Bose function.

From Eq.~\eqref{eq:Npower} it is clear that the power is maximized when the number of particles in the thermal cloud during the expansion stroke is as large as possible, and the number of particles in the thermal cloud during the compression stroke is as small as possible. This occurs when the expansion stroke is as far below the critical temperature as possible, and the compression stroke occurs as close to the critical temperature as possible.

The physical interpretation of this behavior lies in the fact that, just as no work can be extracted from the bosons in the BEC during the expansion stroke, no work is needed to compress them during the compression stroke. However, after the isochoric heating stroke a fraction of the particles that were compressed ``for free" in the condensate will have been excited into the thermal cloud, allowing them to do work during the expansion process.  

This behavior of the work extraction is strikingly similar to the results found in Ref.~\cite{Fogarty2020QST} which examines the performance of a finite-time Otto engine with a working medium of a Tonks–Girardeau gas in a box driven between superfluid and insulating phases. Analogously to the BEC engine, the work extraction is determined by the number of particles excited across the energy gap between the ground and excited states.

\subsubsection{Engine driven by atomic collisions}

\begin{figure*}
\includegraphics[width=.88\textwidth]{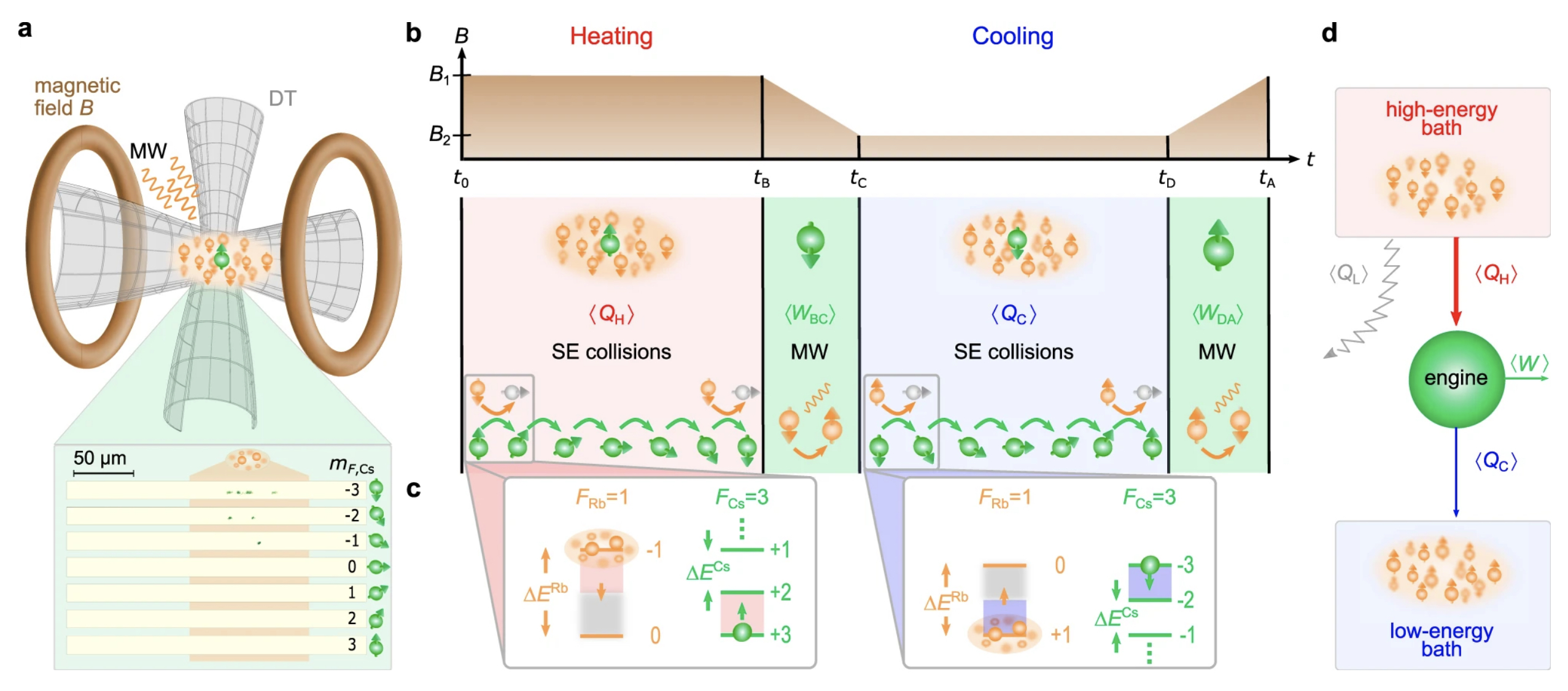}
\caption{\label{fig:atomic} (a) Schematic representation of the experimental set-up. Laser cooled Cs atoms (green) are immersed in an ultracould Rb gas (orange) and confined to an optical dipole trap.; (b) experimental Otto cycle implemented through a variation of the external magnetic field $B$. (c) relevant energy spectrum of the Rb (orange) and Cs (green) atoms during the spin-exchange collisions; (d) schematic representation of the Otto engine, depicting the heat leak $\la Q_L\ra$ rendering the engine endoreversible.   \textit{Figure adopted from Ref.~\cite{Bouton2021NC}}.}
\end{figure*}

Another experiment realized a fully quantum Otto engine with a cold quantum gas as a working medium.  Bouton \etal\cite{Bouton2021NC} considered an endoreversible Otto cycle in the large quasi-spin states of Cs impurities immersed in an ultracold Rb bath, see Fig.~\ref{fig:atomic}(a).

As discussed in Sec.~\ref{sec:endo}, the endoreversible Otto cycle is a prime example to investigate genuine quantum effects in the performance of engines. Bouton \etal\cite{Bouton2021NC} implemented the corresponding expansion and compression strokes by varying an external magnetic field, which changes the energy-level spacing of the working medium.  Heat exchange between the Cs atoms and the Rb atoms occurs via inelastic endoenergetic and exoenergetic spin-exchange collisions \cite{Schmidt2019PRL}. Moreover, the engine is fully controllable through the coherent spin-exchange process \cite{Sikorsky2018NC}, which determines the direction of the heat transfer between system and bath at the level of individual quanta of heat \cite{Schmidt2019PRL}.

More specifically,  Bouton \etal\cite{Bouton2021NC} immersed up to ten laser-cooled Cs atoms, into an ultracold Rb gas of up to $10^4$ atoms, see Fig.~\ref{fig:atomic}(a). The actual ``engine'' is run in the spin-state manifold of the seven Cs-hyperfine ground states. Heat is then exchanged at the microscopic level via inelastic spin-exchange collisions, see Fig.~\ref{fig:atomic}(c). Each collision changes the value of the Cs-atoms by $\Delta m_\mrm{Cs}=\pm \hbar$, which corresponds to a quantum of heat $\Delta E^\mrm{Cs}=\pm \lambda B$, where $B$ is the magnetic field and $\lambda=|g_F^\mrm{Cs}| \mu_B$. Here,  $g_F^\mrm{Cs}=-1/4$ is the Land\'{e} factor of the Cs-atoms, and $\mu_B$ is the Bohr magneton.

Thus, the spin populations are directly related to the energy exchanged between engine and reservoir at the level of single energy quanta. This makes it possible to determine and control the direction of the heat transfer through the spin polarization of the Rb atoms. The engine cycle was then implemented  as illustrated in Fig.~\ref{fig:atomic}(b). The engine's performance is determined from the heat exchanged with the hot and cold reservoirs, respectively.  We have
\begin{equation}
\la Q_h\ra=\sum_n n \left(\wp_n^B-\wp_n^A\right)\,\lambda\,B_1
\end{equation}
and
\begin{equation}
\la Q_c\ra=\sum_n n \left(\wp_n^D-\wp_n^C\right)\,\lambda\,B_2\,,
\end{equation}
where as always $A,B, C,$ and $D$ denote the four corners of the cycle, see also Fig.~\ref{fig:otto}.  Accounting for the heat leak (rendering the cycle endoreversible instead of fully reversible), it is then a simple exercise to show that the efficiency is given by
\begin{equation}
\eta=\frac{\gamma (B_1-B_2)}{B_1-B_2+\gamma B_2}\leq 1-\frac{B_2}{B_1}=\eta_\mrm{max}\,,
\end{equation}
where $\gamma=\lambda/\kappa$ and $\kappa=|g_F^\mrm{Rb}| \mu_B$ with the Rb Land\'{e} factor $g_F^\mrm{Rb}=-1/2$.

Moreover, the endoreversible power output over one cycle of duration $\tau_\mrm{cyc}$ becomes
\begin{equation}
\la P\ra\leq \frac{\la Q_h\ra}{\tau_\mrm{cyc}} \left(1-\frac{B_2}{B_1}\right)\,.
\end{equation}
The latter was measured directly using full counting statistics making it possible to track the time evolution of the cycle.

As main result, Bouton \etal\cite{Bouton2021NC} realized a fully quantum heat engine, that can be properly characterized as an endoreversible Otto cycle.  As such, the Cs-atom engine presents an important step towards more complex working mediums, as, for instance, with contact interactions \cite{Myers2021Symm}, at criticality \cite{Fogarty2020QST}, or in complex media \cite{Johal2021entropy}.

%% file: sections/experiment_JJ.tex
\subsection{On-chip and superconducting devices}

As a next physical platform, we will be focusing on the implementation of quantum thermodynamic devices ``on chip".  Such situations can eithr be found in circuit QED, or more generally built with superconducting materials~\cite{Leivo1996APL,Nguyen2016PRApp,Quan2006PRL,Solinas2016PRB,Marchegiani2018EPL,Hussein2019PRB,Marchegiani2020PRL}.

\subsubsection{Superconducting qubits}

\begin{figure}
\includegraphics[width=.48\textwidth]{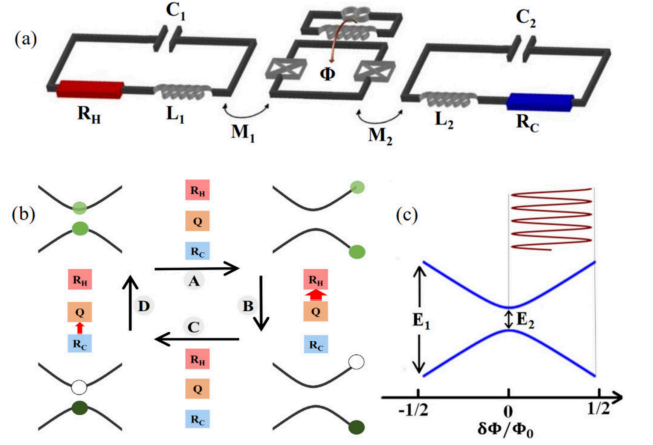}
\caption{\label{fig:superOtto} (a) Schematic representation of the quantum refrigerator with superconducting qubits; (b) effective Otto refrigeration cycle; (c) energy levels under external, sinusoidal driving. \textit{Figure adopted from Ref.~\cite{Karimi2016PRB}}.}
\end{figure}

Based on their extensive experimental experience with employing superconducting circuits in (quantum) thermodynamic studies  \cite{Niskanen2007PRB,Pekola2010PRL,Koski2015PRL,Campisi2015NJP,Pekola2016PRB}, Karimi and Pekola \cite{Karimi2016PRB,Karimi2017PRB} proposed an Otto refrigerator that can exhibit both, quantum as well as classical performance. The setup and the operation are schematically illustrated in Fig.~\ref{fig:superOtto}. The working medium is a superconducting qubit, which is comprised of a loop interrupted by two Josephson junctions.  It is inductively coupled to two heat reservoirs, which are realized as RLC circuits.  Using an intuitive notation, $M$ denotes the coupling inductance,  $R$ is the resistance, $L$ is the inductance, and $C$ is the capacitance. The magnetic flux through the qubit is $\Phi$, which can be tuned externally by $q\equiv \Delta\Phi/\Phi_0$ and $\Phi_0=h/2e$.

Hence, the two resistors $R_C$ and $R_H$ can be identified as cold and hot reservoirs, respectively.  More precisely,  the temperatures are set by the resonance frequencies, where a high frequency denotes hot, and a low frequency is cold.  Karimi and Pekola\cite{Karimi2016PRB} then outlined how such a setup may operate in an Otto refrigeration cycle, cf. Fig.~\ref{fig:superOtto}. In the isentropic ``expansion'' the magnetic flux is increased, $q=0\rightarrow q=1/2$, while the qubit is isolated from the two baths. The the qubit is let to equilibrate with the hot reservoir, before the magnetic flux is reduced again in an isentropic ``compression'', $q=1/2\rightarrow q=0$. The cycle is closed by letting the qubit equilibrate with the cold reservoir.

The proposed system has been well-characterized experimentally,  and its properties can be theoretically described by a quantum master equation of the Lindblad form \cite{BreuerBook}. Using these tools Karimi and Pekola \cite{Karimi2016PRB} showed that within experimentally accessible parameter regimes the device could indeed operate as refrigerator.  Moreover, at high driving frequencies the power output exhibits signatures of coherent oscillations, which may be interpreted as a distinct evidence of quantum performance.  Additional theoretical analysis of the nonadiabatic characteristics of the setup was reported in Ref.~\cite{Solfanelli2020PRB}.

\subsubsection{Self-oscillating Josephson engine}

\begin{figure}
\includegraphics[trim=1cm 3cm 1cm 3cm, clip, width=.48\textwidth]{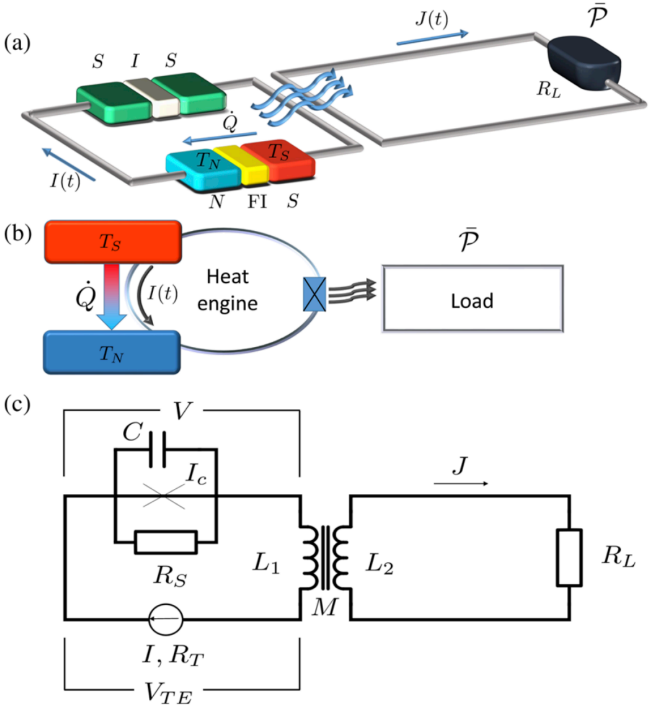}
\caption{\label{fig:self_oscill}  (a) Schematic representation of the quantum heat engine; (b) thermodynamic block diagram of the set-up; (c) circuit model of the proposed engine.  \textit{Figure adopted from Ref.~\cite{Marchegiani2016}}.}
\end{figure}

Taking things one step further,  while working with a similar setup,  Marchegiani \etal \cite{Marchegiani2016} proposed a device,  whose operation is fully sustained by quantum effects.  A quantum heat engine is realized as a closed circuit that comprises an $N$-$FI$-$S$ (normal metal - ferromagnetic insulator - superconductor) tunnel junction, which is connected to a Josephson weak link. The work reservoir is realized as a generic LC circuit,  that is coupled to the ``engine'' via electromagnetic induction.  A schematic representation of the device is depicted in Fig.~\ref{fig:self_oscill}.

The device is designed (i) to sustain persistent work generation from the self-oscillations, i.e.,  the Josephson effect, (ii) to be easily implementable with available experimental techniques,  and (iii) and to have a wide range of potential applications.  For instance, Marchegiani \etal \cite{Marchegiani2016} suggested that their device could be directly realized in caloritronic platforms \cite{Giazotto2012Nature,Martinez2014JLTP}.

Similarly to the above described Otto refrigerator \cite{Karimi2016PRB}, the self-oscillating Josephson engine can be fully described by simple theoretical means.  The key ingredient is the $FI$ layer, which  breaks the particle-hole symmetry of the superconductor's density of states.  For spin-up ($\uparrow$) and spin-down ($\downarrow$) bands  at energy $E$ (relative to the chemical potential) we have
\begin{equation}
N_{\uparrow,\downarrow}(E)=\frac{1}{2}\,\left|\mrm{Re}\left\{\frac{E+i\Gamma\pm h_\mrm{exc}}{\sqrt{\left(E+i\Gamma\pm h_\mrm{exc}\right)^2-\Delta^2}}\right\}\right|\,,
\end{equation}
where $h_\mrm{exc}$ is the exchange field induced by the $FI$ layer, $\Gamma$ is the phenomenological broadening parameter \cite{Dynes1984PRL}, and $\Delta$ is the pairing superconducting potential \cite{Giazotto2008PRB,Giazotto2015PRAppl}.

In the tunneling limit, the thermodynamic behavior of the device is fully determined by the dc current. It can be written as \cite{Ozaeta2014PRL}
\begin{equation}
I_{TE}(V_{TE})=\frac{1}{e R_T}\,\int dE\,\mc{N}(E)\left[f_S(T_S)-F_N(V_{TE},T_N)\right]\,,
\end{equation}
where $e$ is the elementary charge, $R_T$ denotes the normal-state resistance, and $V_{TE}$ is the bias voltage.  Further, $\mc{N}=N_+ +\Pi\, N_-$ with $N_\pm=N_\uparrow \pm N_\downarrow$, and with the polarization $\Pi=(G_\uparrow-G_\downarrow)/(G_\uparrow+G_\downarrow)$, where $G$ is the normal state conductance.  Moreover, Marchegiani \etal \cite{Marchegiani2016} assumed the device to be in thermal equilibrium, and hence we have for the normal conductor
\begin{equation}
f_N(V_{TE},T_N)=\frac{1}{1+\e{\beta_N \left(E+e V_{TE}\right)}}\,,
\end{equation}
and for the superconductor
\begin{equation}
f_S(T_S)=\frac{1}{1+\e{\beta_S E}}\,.
\end{equation}

Like all devices, also the self-oscillating Josephson engine has a positive work condition.  For the present set-up it translates to the condition that the critical current $I_c$ is less than $|I_{TE}(0)|$.  If the critical current becomes larger, the terminals of the $N$-$FI$-$S$ junction are effectively short-circuited and no work can be performed on the LC circuit.

The dynamics of the device can then be fully analyzed with the Josephson equations\cite{Marchegiani2016}. In particular,  Marchegiani \etal \cite{Marchegiani2016} solved the dynamics numerically to compute current $J(t)$ in the load circuit, i.e., the work reservoir.  This current determines the average power output,
\begin{equation}
\label{eq:power_selfosc}
\bar{P}=\frac{1}{\tau} \int_0^\tau dt\,R_L J(\tau)^2\,.
\end{equation}
In the limit of high frequencies,  $\bar{P}$ \eqref{eq:power_selfosc} can be simplified to read
\begin{equation}
\bar{P}=\left(\frac{M I_c R_S}{2 L_1}\right)^2\frac{2 R_L}{\left(\omega L_e\right)^2+R_L^2}\,,
\end{equation}
where $M$ is the mutual inductance, $L_{1,2}$ are the self-inductances with $L_e=(L_1 L_2-M^2)/L_1$,  and $\omega=2 \pi \left| I_{dc} \right| R_S/\Phi_0$ for some applied dc current $I_{dc}$,  see also Fig.~\ref{fig:self_oscill}. As before, $\Phi_0=h/2e$ is the flux quantum.

Remarkably, such a setup leads indeed to persistent quantum performance and the power output can achieve approximately 1 pW.  However, the corresponding efficiency is rather low, and Marchegiani \etal \cite{Marchegiani2016} estimated $\eta=\bar{P}/\dot{Q}\lesssim 10^{-6}$ for experimentally realistic parameters. The reason for the very low efficiency lies in the resistive heating at $R_S$. However,  Marchegiani \etal \cite{Marchegiani2016} noted that the low efficiency is compensated by the comparatively high power output, and the versatility of the device. At the very least, the latter claim proved to be true as following the proposal by Marchegiani \etal\cite{Marchegiani2016} several other designs were developed such as, e.g.,  alternative heat engines \cite{Hofer2016PRB,Haack2019PRB,Vischi2019entropy,Marchegiani2020PRB} or thermal switches \cite{Sothmann2017NJP,Karimi2017QST,Farsani2019PLA}.

\subsubsection{Topological Josephson heat engine}

One of the most recently proposed designs relies on exploiting even more striking quantum effects. Topological Josephson junctions have a characteristic  4$\pi$-periodic ground state fermion parity in the superconducting phase difference \cite{Deacon2017PRX,Kayyalha2019PRL,Laroche2019NatComm}.  Scharf \etal\cite{Scharf2020} demonstrated that such topological effects can be exploited in a Josephson-Stirling cycle, and that this thermodynamic cycle exhibits the 4$\pi$-periodicity.

\begin{figure}
\includegraphics[width=.48\textwidth]{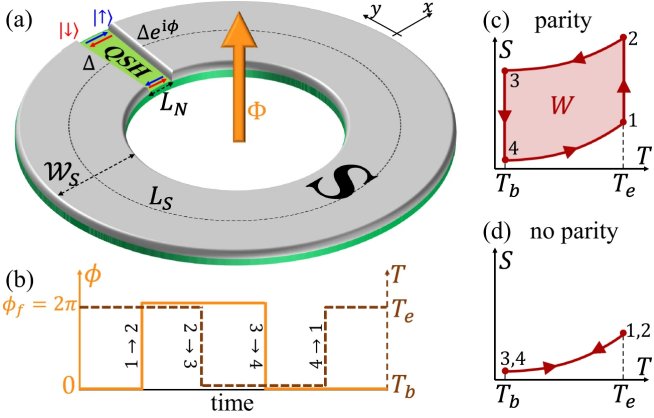}
\caption{\label{fig:topological_engine}  (a) Schematic representation of the QSH insulator, which is partially covered by an $s$-wave superconductor; (b) driving protocols of phase $\phi$ and temperature $T$ to implement the Josephson-Stirling cycle; (c) $TS$ diagram with enforced fermion parity for $\phi_f=2\pi$; (d) $TS$ diagram without enforced fermion parity for $\phi_f=2\pi$.  \textit{Figure adopted from Ref.~\cite{Scharf2020}}.}
\end{figure}

The setup is schematically depicted in Fig.~\ref{fig:topological_engine}.  A Josephson junction based on a quantum spin Hall (QSH) insulator is driven through a Josephson-Stirling cycle \cite{Vischi2019entropy,Vischi2019SR}.  This cycle consists of (i) an isothermal phase change, $\phi\rightarrow \phi_f$ at temperature $T=T_e$, (ii) isophasic cooling $T_e\rightarrow T_b$ at constant $\phi=\phi_f$, (iii) an isothermal phase change $\phi_f\rightarrow 0$ at temperature $T_b$, and (iv) isophasic heating, $T_b\rightarrow T_e$ at constant phase $\phi=0$, cf. Fig.~\ref{fig:topological_engine}. 

If the maximal phase $\phi_f$ is chosen as an integer multiple of 2$\pi$, then the work output strongly depends on the presence of constraints on the fermion-parity. This can be seen from the corresponding expression for the free energy. Under the absence of constraints,  Scharf \etal\cite{Scharf2020} used a scatting approach to determine  that the Andreev bound states we have
\begin{equation}
\label{eq:F_topo_no}
F_0(\phi, T)=-\frac{2}{\beta} \mrm{ln}\left[2\,\mrm{cosh}\left(\frac{2\Delta}{\beta}\,\cos{\left(\frac{\phi}{2}\right)}\right)\right]\,,
\end{equation}
where $\Delta$ is here the proximity-induced pairing amplitude. Note that the pre-factor 2 accounts for the energetically degenerate top and bottom edges.

If, however, the fermion-parity is enforced, the Helmholtz free energy becomes\cite{Scharf2020} 
\begin{equation}
\label{eq:F_topo}
\begin{split}
F_p(\phi, T)&=-\frac{2}{\beta} \mrm{ln}\left[2\,\mrm{cosh}\left(\frac{2\Delta}{\beta}\,\cos{\left(\frac{\phi}{2}\right)}\right)\right.\\
&\quad+\left.\mathfrak{p} \e{J_s(T)}\, \mrm{sinh}\left(\frac{2\Delta}{\beta}\,\cos{\left(\frac{\phi}{2}\right)}\right)\right]\,,
\end{split}
\end{equation}
where $\mathfrak{p}=\pm 1$ denotes  even and odd ground-state parity,. Moreover,
\begin{equation}
J_S(T)=-\frac{2\beta}{\pi E_s}\,\int_\Delta^\infty d\epsilon\,\frac{\sqrt{\epsilon^2-\Delta^2}}{\mrm{sinh}\left(\beta \epsilon\right)}\,,
\end{equation}
and $E_s=\hbar \nu_F/L_s$, where $L_s$ is the total length of the superconducting QSH edge.  The frequency $\nu_f$ determines the energy gap in $z$-direction.

From the expressions for the free energy \eqref{eq:F_topo_no} and \eqref{eq:F_topo} it is then only a simple thermodynamic exercise to compute the work output, efficiency, etc.  Scharf \etal\cite{Scharf2020}  found that, indeed, the 4$\pi$-periodicity is reflected in the free energy as well as entropy changes, and they identified the positive work conditions. In particular, they found that for $T_e>T_b$ the device operates as an engine, and for $T_e<T_b$ one can realize a refrigerator. The actual phase diagrams depend sensitively on whether or not the phase parity is enforced, and thus the device could actually be used as a diagnostics tool for the presence of 4$\pi$-periodicity in Josephson junctions.  Indeed,  the proposal by Scharf \etal\cite{Scharf2020} has already attracted follow up work in the design of engines \cite{Blasi2021PRB,Chatterjee2021PRE}, and in quantum metrology and sensing \cite{Blasi2020PRB,Gresta2021PRB,Marchegiani2021APL}.

\subsubsection{Silicon tunnel field-effect transistor}

Whereas the preceding examples are theoretical proposals ready to be implemented in experiments, Ono \etal \cite{Ono2020PRL} reported an actual experiment in silicon tunnel-effect transistors. Their design rests on the dynamics of modulated dissipative qubits, which has been shown to be the minimal working medium for quantum heat engines \cite{Gelbwasser2013PRE,Erdman2019NJP,Dann2020entropy}.  The working principle of the realized device is akin to so-called Sisyphus lasing and cooling cycles.  Such processes occur if the Rabi frequency becomes comparable to one over the relaxation time.

More specifically, Ono \etal \cite{Ono2020PRL} realized the following four stroke cycle: (i) resonant excitation into the upper state,  (ii) adiabatic evolution within this state,  (iii) relaxation to the ground state,  and (iv) adiabatic evolution in the ground state. Moreover, distinct quantum features are expected to be present if the driving period is shorter than the decoherence time.

Ono \etal \cite{Ono2020PRL} provide both,  the full theoretical description as well as the experimental realization.  The spin-qubit device is implemented in a short-channel silicon tunnel-effect transistor, whose deep impurities are intensively ion implanted. These impurities act as quantum dots \cite{Ono2019SR} at room temperature.  Such a device can be tuned into a regime,  in which the source-to-drain conduction is dominated by the tunneling current between two impurities \cite{Ono2020PRL}. A time-ensemble measure of the electron spin is then facilitated by the spin-blockade phenomena exhibited by such devices \cite{Ono2002Science,Liu2008PRB,Giavaras2013PRB,Ono2017PRL}.

Remarkably, the corresponding theoretical description becomes rather simple. In fact,  the corresponding, time-dependent Hamiltonian simply reads \cite{Ono2020PRL},
\begin{equation}
H(t)=B_z(t) \sigma_z/2+B_x(t) \sigma_x/2\,,
\end{equation}
which describes here nothing but a single qubit subject to a fast microwave driving and a slow rf modulation of both amplitude and frequency. 

Ono \etal \cite{Ono2020PRL} then demonstrated, theoretically as well as experimentally, that the spin-qubit device can operate as both, engine as well as refrigerator, in the classical and in the fully quantum coherent regime. Despite its rather recent publication, this study has already inspired follow-up work in the design of heat engines \cite{guthrie2021arXiv,Johal2021entropy} and in the observation of the spin blockade \cite{Kondo2021PRB}.

%% file: sections/experiment_photo.tex
\subsection{Photovoltaics}

We finally return to photo-Carnot engines. Whereas  so far we have described physical platforms that exhibit obvious quantum features, this was not immediately clear in \emph{photovoltaics}. However, once quantum effects had been demonstrated \cite{Scully2010PRL,Alharbi2015} it was only a small step to a fully quantum thermodynamic analysis of light harvesting systems.

\subsubsection{Solar cell as a heat engine}

Solar cells, also known as photovoltaic converters,  are simple engines capable of producing electrical work after the absorption of heat from the sun. There has been intense research effort into the development of high efficiency solar cells relying on emerging novel materials. Numerous studies have been carried out to control and discover the losses preventing photovoltaic devices from reaching the physical limits.  For a comprehensive review we refer to the literature~\cite{Nayak2019NRM}.  Here, we will focus on solar cells described as heat engines using  the energy-entropy analysis of photovoltaic conversion~\cite{Makrvart2008}.

Specifically, the energy conversion process is divided into the absorption and emission of the incident radiation, and conversion of the absorbed photons into useful work~\cite{Makrvart2008}. The incident photon energy $U_\mrm{in}$ is absorbed by the cells from a high temperature reservoir at temperature $T_S$. The absorbed photon is converted into energy $W$ while dumping an amount of heat $Q_w$ into a low-temperature reservoir at temperature $T_0$ of the solar cells. The energy conservation law is 
\begin{equation}
    U_\mrm{in}=W + Q_w.\label{solarEq1}
\end{equation}
The second law of thermodynamics can be expressed as
\begin{equation}
   \Sigma_i =Q_w/T_0 - S_\text{in},\label{solarEq2}
\end{equation}
where $\Sigma_i\geq 0$ is the entropy generated in the conversion process, $S_\text{in}$ is the entropy of the absorbed photon, and $Q_w/T_0$ denotes the entropy emitted into the low-temperature reservoir. Recalling that the work  performed per photon is $W=eV$, where $e$ is the electron charge and $V$ is the voltage generated by the solar cell, and combining  Eqs.~\eqref{solarEq1}-\eqref{solarEq2} gives the photogenerated voltage as
\begin{equation}
    eV=\left(1-T_0/T_s\right) U_\text{in} - T_0\Sigma_i.
\end{equation}
Therefore, the solar cell's voltage is well described by a standard thermodynamic process that generates an amount of work from energy differences between two reservoirs at temperatures $T_s$ and $T_0$. Then, considering that the solar cell emits photons as well, the energy-entropy balance for the absorption and emission processes shows that the output work is equal to the chemical potential of the emitted photons, i.e., $eV=\mu_\text{out}$. 

The thermodynamic efficiency is defined as $\eta=eV/U_\text{in}$ and the maximum efficiency for a reversible processes is simply given by the Carnot efficiency $\eta_C = 1 - T_0/T_s$.   Note that other definitions of the efficiency exist, and, for instance, it has been assumed that the solar cell is a blackbody absorbing radiation from the sun without generating entropy \cite{DeVos1993}. Moreover, studies of solar cells as endoreversible heat engine have been carried out \cite{Navarrete1997}, as well.

Over the last decade,  quantum descriptions of photovoltaic operation have attracted a great deal of attention~\cite{Scully2010PRL,Zhang2015}. In an influential work of Scully~\cite{Scully2010PRL}, it is shown that,  as in the case of the photo-Carnot quantum heat engine \cite{Scully2003},  quantum coherence can be used to enhance the performance and efficiency of a photocell. Specifically, Scully \cite{Scully2010PRL}  showed that by breaking detailed balance, the radiative recombinations are reduced, and thus we have
\begin{equation}
    eV =\hbar \omega_s\left(1-T_a/T_s\right) + \hbar \omega_0,
\end{equation}
where $V$ is the maximum induced voltage, $T_a$ is the ambient temperature, and $\hbar \omega_0$ is the photon energy of the coherent driving field. The frequency $\omega_s$ and temperature $T_s$  characterize the monochromatic beam of solar photons that  illuminate the photocell.  Scully and co-workers~\cite{Scully2011} later showed that noise-induced coherence could break detailed balance and enhance the power output of a laser or photocell quantum heat engine.

This proposal of quantum coherence enhanced photocells opened the door to various studies in light-harvesting complexes~\cite{Dorfman2013PNAS} and semiconductors~\cite{Bittner2014}.

\subsubsection{Photosynthetic reaction center}

Photosynthesis is the process of converting light energy into chemical energy suitable for living organisms. Building on the understanding of the role of quantum coherence in the photocell quantum engine, Dorfman \textit{et al.}~\cite{Dorfman2013PNAS} demonstrated that the photosynthetic reaction center can be seen as a biological quantum heat engine and possibly supports noise-induced quantum coherence.  This system transforms  high-energy thermal photon radiation into low-entropy electron flux and  is characterized by the efficiency of charge separation.

\begin{figure}
\includegraphics[width=.48\textwidth]{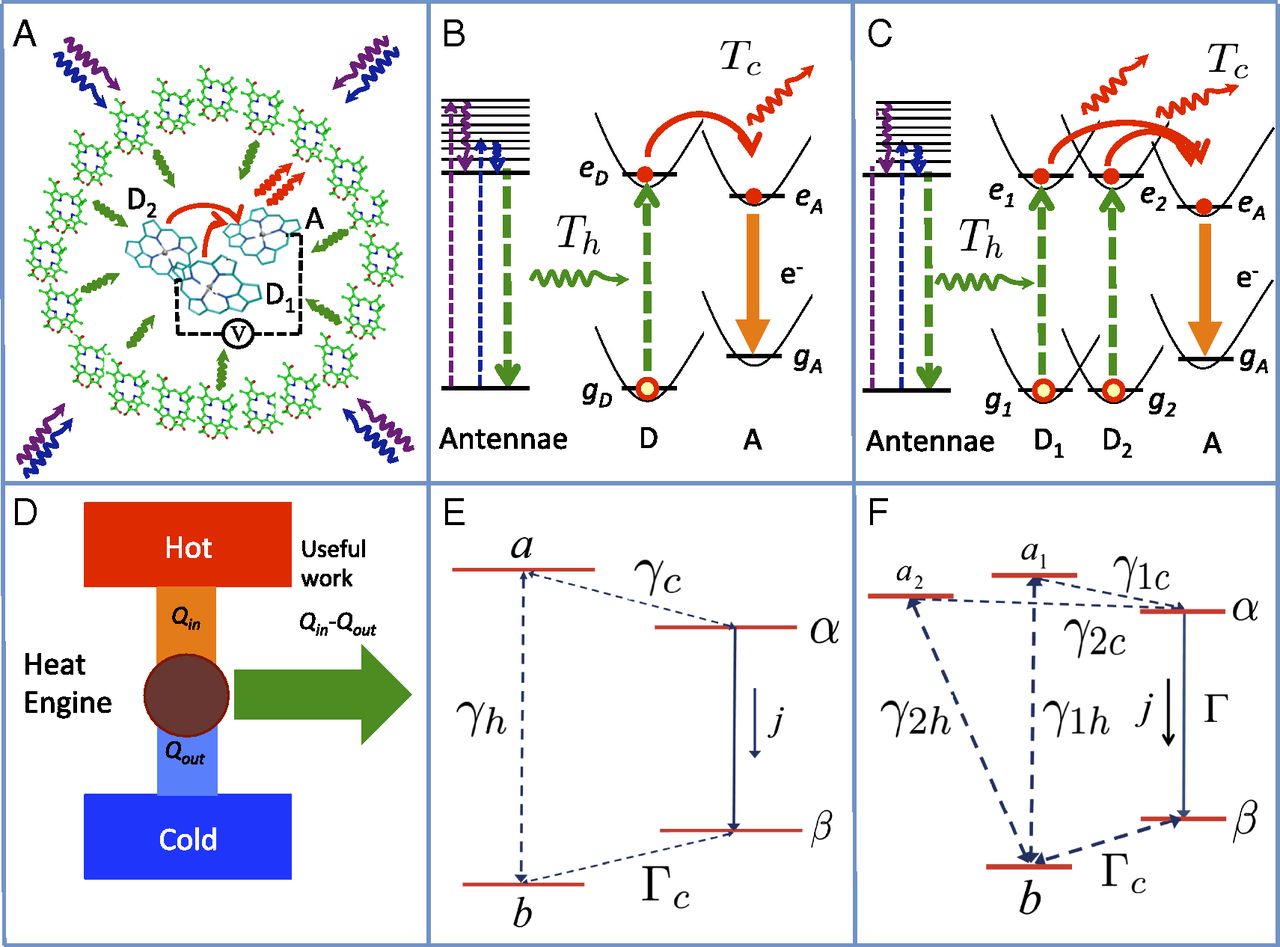}
\caption{\label{fig:photosynthesis} Schematic representation of the photosynthetic reaction center (A) charge separation between a donor D and an acceptor A molecule (B); a generic heat engine (D); (E) represents a four-level quantum heat engine; (C) and (F) are the same as (B) and (E), respectively, except that the upper level $a$ is replaced by two levels $a_1$ and $a_2$ separated by Davydov splitting.  \textit{Figure adopted from Ref.~\cite{Dorfman2013PNAS}}.}
\end{figure}

Dorfman \textit{et al.}~\cite{Dorfman2013PNAS} modeled the photoinduced charge separation between the two closely spaced identical donor $D_1$ and $D_2$, and the acceptor $A$ molecules interacting with thermal light as a four-level quantum heat engine, see Fig.~\ref{fig:photosynthesis}. The whole system is initially prepared in the ground state $b$.  The cycle starts with the absorption of solar photons that leads to the population of the donor excited states $a_1$ and $a_2$, see Fig.~\ref{fig:photosynthesis}. Both of the promoted states are only separated in energy by the small Davydov splitting, which ensures that they can be in a quantum-coherent superposition state. Then, the excited electrons can be transferred to the lower lying excited state $\alpha$ of the acceptor molecule through electronic coupling and emission of phonons. The state $\alpha$ is a charge-separated state with the electron in acceptor $A$ and the hole in the donor part. 

In the second step, the acceptor relaxes from its excited states $\alpha$ to its ground state $\alpha'$ by using the excess energy to produce useful work in form of an electric current. The  power output of the quantum heat engine is determined by the transfer rate $\Gamma$ and the steady-state ratio of populations between $\alpha$ and $\alpha'$. Assuming that the current generated flows across a load connecting the acceptor levels $\alpha$ and $\alpha'$, the voltage $V$ across the load is
\begin{equation}
eV=E_\alpha - E_{\alpha'} + k_bT_a \ln (\wp_{\alpha\alpha}/\wp_{\alpha' \alpha'})\,,
\end{equation}
where $e$ is again the elementary electric charge,  and $E_\alpha$ and $E_{\alpha'}$ are the energies of states $\alpha$ and $\alpha'$,  respectively.

Dorfman \textit{et al.}~\cite{Dorfman2013PNAS} found that a sufficiently long-lived quantum coherent phase induced between $a_1$ and $a_2$ can minimize losses by destructive interference of the loss transitions \cite{Nalbach2013PNAS}.  This proposal inspired new designs for artificial light-harvesting devices. For instance, Creatore \textit{et al.}~\cite{Creatore2013} showed that quantum interference effects of photon absorption and emission induced by the dipole-dipole interaction between molecular excited states can enhance the performance of a photocell.

\subsubsection{Enhancing light-harvesting power}

Quantum effects in biological systems are a current area of research~\cite{Cao2020SA,Kim2021QR}, and it has been demonstrated that quantum processes can contribute to the high efficiency of biological light-harvesting complexes~\cite{Harel2012}. Building on previous studies on natural and artificial  systems~\cite{Scully2011,Dorfman2013PNAS,Creatore2013}, Killoran \textit{et al.}~\cite{Killoran2015JCP} analyzed the  light-harvester as a quantum heat engine. Their work reveals the role of mixed electronic-vibrational, vibronic, coherence in biological systems.  Killoran \textit{et al.}~\cite{Killoran2015JCP} considered a situation in which the quantum heat engine connects to three external systems acting as  thermal reservoirs and facilitates all possible transitions between the exciton states. Then, by carefully taking into account both electronic and vibrational interactions, a quantitative link is found between coherent vibronic evolution and functional quantum advantage in the power output of a light-harvesting system. This proposal led to further investigations of cyclic and steady-state quantum heat engines, see Refs.~\cite{Chen2016,Mitchison2019}.

\subsubsection{Green quantum photocell}

\begin{figure}
\includegraphics[width=.48\textwidth]{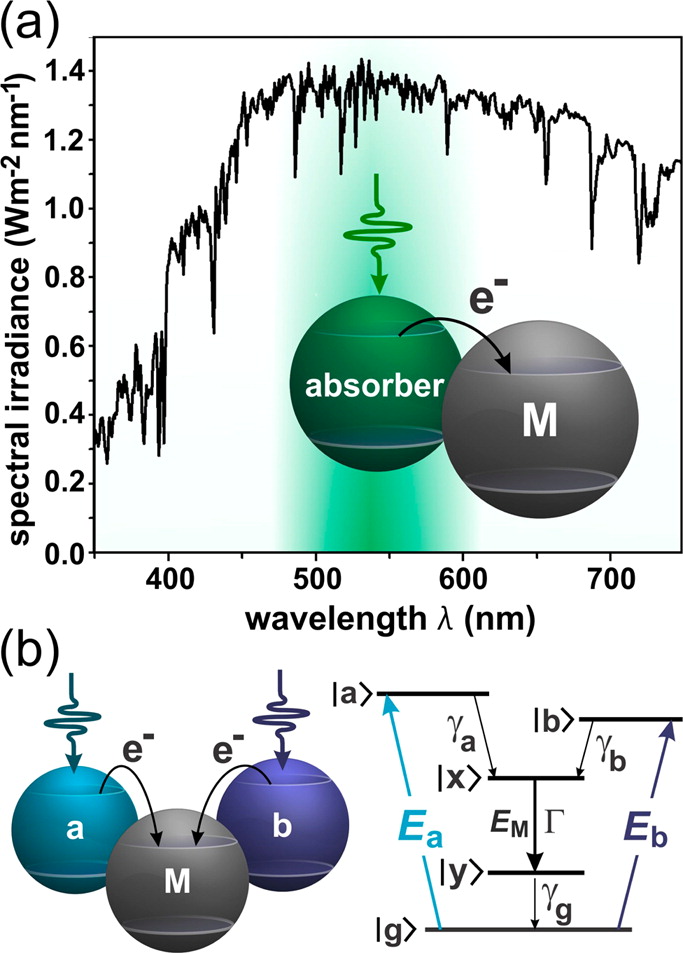}
\caption{\label{fig:green_photocell} Schematic representation of a two-channel quantum photocell.  (a) Solar spectral irradiance. Inset: One-channel quantum photocell; (b) schematic representation of a two-channel  quantum photocell. The output power is generated by the machine M.  \textit{Figure adopted from Ref.~\cite{Arp2016NL}}.}
\end{figure}

Fluctuations are ubiquitous in photovoltaic technology and photosynthesis. These fluctuations are suppressed by voltage converters and feedback controllers placed between the solar panel and battery.  Aiming to address this problem, Arp \textit{et al.}~\cite{Arp2016NL} put forward a model a of two-channel quantum photocell, see Fig.~\ref{fig:green_photocell}. 

Arp \textit{et al.}~\cite{Arp2016NL} showed that by incorporating two photon-absorbing channels, the regulation of energy flow emerges naturally within a quantum heat engine photocell. The photocell switches stochastically between the channels to convert varying incident solar power into steady-state output.  By maximizing the  absorption characteristics they found that stochastic switching suppresses power fluctuations over a broad range of the solar spectrum. Their analysis demonstrated that two-channel photocell engines regulate energy flow better than the one-channel photocell engines as well as increase power conversion efficiency.  This study  inspired a study of the photosynthetic reaction center~\cite{Li2021}.

%% file: sections/experiment_dirac.tex
\subsection{Graphene, Dirac dynamics, and other relativistic effects}

We conclude this review of physical platforms with quantum heat engines at the frontiers of physics.  In particular, we discuss a few examples of relativistic \cite{Arutyunyan2007LP} and cosmological scenarios \cite{Santos2018EPJP}.

\subsubsection{Otto engine at relativistic energies}

Relativistic quantum systems exhibit unique features not present at lower energies, such as the existence of both particles and antiparticles, and restrictions placed on the system dynamics due to the light cone. Ref. ~\cite{Myers2021NJP} examines the impact of these phenomena on the performance of an Otto engine with a working medium of a single particle in a relativistic oscillator potential.

In $(1+1)$-dimensions, the Hamiltonian for the relativistic oscillator system is \cite{Moshinsky1989},
\begin{equation}
	H_D = c \sigma_x \cdot \left(p - im \omega x \cdot \sigma_z \right)+mc^2\sigma_z
\end{equation} 
where $m$ is the particle mass, $\omega$ is the oscillator frequency, $\sigma_x$ and $\sigma_z$ are the corresponding Pauli spin matrices. The time-independent Dirac and Klein-Gordon equations for the relativistic oscillator can be exactly solved and yield the energy spectrum,
\begin{equation}
	\label{eq:DiracSpectrum}
	E_n = \pm \sqrt{2\left(n+\frac{1}{2}\right)\hbar \omega m c^2 + m^2 c^4},
\end{equation}
where $n = \{0, 1, 2, ...\}$. The positive branch corresponds to the particle energies and the negative branch to the anti-particle energies. Note that in the non-relativistic limit of $\hbar \omega / m c^2 \ll 1$ the energy spectrum in Eq. ~\eqref{eq:DiracSpectrum} reduces to that of the typical quantum harmonic oscillator with an additional term corresponding to the rest energy.

In general, the negative energy solutions present in relativistic quantum systems present difficulties in deriving the appropriate canonical ensemble, as the partition function sum will diverge. However, the relativistic oscillator belongs to a class of relativistic quantum systems that display a supersymmetric partnering of the positive and negative energy solutions referred to as the ``stability of the Dirac Sea", meaning that the positive and negative energy solutions do not mix \cite{Moreno1990MPLA, Martinez1991PRD}. 

Taking advantage of the stability of the Dirac sea, the partition function, and thus equilibrium internal energy, entropy, free energy, and heat capacity can be determined. Using the internal energy and applying the framework of endoreversible thermodynamics the efficiency in the high-temperature, relativistic regime is determined to be,    
\begin{equation}
	\label{eq:effHigh}
	\eta_{\mathrm{rel}} = 1 - \sqrt{\kappa},
\end{equation}
where $\kappa = \omega_1/\omega_2$ is the compression ratio. As $\kappa < 1$, this efficiency is notably smaller than the single particle harmonic Otto efficiency, $1 - \kappa$. This reduced efficiency can be understood as a consequence of the fact that the dynamics of the system are constrained by the light cone.

In contrast to the efficiency, the power output of the relativistic engine is found to be larger than that of a non-relativistic single particle working medium. This increased power output arises from two contributions. The first is the presence of the negative energy solutions and the second arises from the fact that, due to the linear rather than quadratic dependence on momentum, each degree of freedom for an ultra-relativistic gas contributes twice the amount to the internal energy as in the classical, non-relativistic limit. This ``relativistic equipartition" can be derived from the fact that the thermal momentum distribution of particles obeying Dirac dynamics follows the Maxwell-J{\"u}ttner distribution rather than the typical Maxwell-Boltzmann distribution found for particles obeying Schr{\"o}dinger dynamics.

Maximizing the power output with respect to the compression ratio, the efficiency at maximum power is determined in both the non-relativistic and relativistic regimes. Both cases are found to be identical to the Curzon-Ahlborn efficiency. 

While relativistic quantum systems are generally difficult to access experimentally at laboratory-scale energies, Ref. ~\cite{Myers2021NJP} provides examples of several experimental systems which display effective relativistic dynamics, including Dirac materials \cite{Wehling2014}, trapped ions \cite{Bermudez2007PRA}, and microwave resonators \cite{Franco2013PRL}, that could serve as potential platforms for a relativistic heat engine. 

\subsubsection{Otto engine in graphene}

Perhaps the most prominently studied Dirac material is graphene.  Graphene is a two-dimensional material consisting of a single layer of carbon atoms in a honeycomb lattice. As is characteristic of Dirac materials, the charge carriers in graphene display a linear dispersion relation, behaving like relativistic massless fermions with dynamics described by the Dirac-Weyl equation \cite{Wehling2014}. This makes graphene an ideal medium for probing relativistic behavior at laboratory scale energies.

Reference~\cite{Pena2020PRE} analyzes a proposed implementation of a quantum Otto cycle using a working medium of a graphene quantum dot under a perpendicular external magnetic field. Quantum dots in particular have a range of well-established experimental techniques for controlling the dot properties, including its size and shape, making them a promising candidate for practical application \cite{Pena2020PRE}. The relativistic Hamiltonian for the working medium is, 
\begin{equation}
	H = v_F (\mathbf{p}+e \mathbf{A}) \cdot \mb{\sigma} + V(r) \sigma_z\,,
\end{equation}
where $v_F$ is the Fermi velocity, $\mathbf{A}$ is the vector potential and $\sigma_i$ ($i = x,y,z$) are the Pauli spin matrices. The charge carriers are further assumed to be confined to a circular region of radius $R$, such that $V(r) = 0$ when $r<R$ and $V(r) = \infty$ when $r \geq R$. 

This potential admits two different possible boundary conditions, known as zig-zag boundary conditions (ZZBC) and infinite mass boundary conditions (IMBC) \cite{Pena2020PRE}. Under IMBC charge carriers are confined within the quantum dot while for ZZBC one component of the bispinor is required to vanish at the boundary. Reference \cite{Pena2020PRE} applies ZZBC as the continuum model under ZZBC converges well for larger radius quantum dots and low-energy states. Furthermore, unlike IMBC, ZZBC yield a non-vanishing zero-energy eigenstate.

In order to examine the thermodynamic properties of the working medium Ref. ~\cite{Pena2020PRE} numerically calculates the partition function for a range of magnetic field strengths, summing over a range of $m = -50$ to $m = 50$ for the azimuthal quantum number. Using the partition function, the equilibrium free energy, entropy, internal energy, heat capacity, and magnetization are determined.

With the equilibrium thermodynamic behavior established, Ref.~\cite{Pena2020PRE} moves on to examining the performance of two different Otto cycle implementations. For the first implementation the cycle is assumed to be completely quasistatic, such that the working medium remains in a state of thermal equilibrium at all points during the isentropic strokes. The quasistatic work can be found directly from the change in internal energy between the beginning and ending of each isentropic stroke,
\begin{equation}
	\begin{split}
		W^{qs} = &\,\, U_D(T_h, B_l) - U_A(T_A, B_l) \\
		&+ U_B (T_l, B_h) - U_C(T_C, B_h),
	\end{split} 
\end{equation}
where $A, B, C$ and $D$ indicate the four corners of the cycle, $T_l$ ($T_h$) is the cold (hot) bath temperature, and $B_l$ ($B_h$) is the smaller (larger) magnetic field strength. The temperature of the working medium at $A$ and $C$ can be found using the isentropic condition,
\begin{equation}
	S(T_l, B_h) = S(T_A, B_l) \quad \mathrm{and} \quad S(T_h, B_l) = S(T_C, B_h).
\end{equation}

For the second implementation Ref.~\cite{Pena2020PRE} considers a quantum adiabatic cycle, where the evolution during the isentropic strokes is in full accordance with the quantum adiabatic theorem. Notably, while the populations of each energy level remain fixed during the isentropic strokes, the final state is a non-thermal diagonal state. The net work extracted from the quantum adiabatic cycle is then, 
\begin{equation}
	\begin{split}
		W^{q} = &\,\, U_D(T_h, B_l) + U_B (T_l, B_h) \\
		&- \sum_{m, \tau} \left[E_{m,\tau}^l \wp_{m,\tau}(T_l, B_h) + E_{m,\tau}^h \wp_{m,\tau}(T_h, B_l)\right]
	\end{split} 
\end{equation}
where $E_{m,\tau}^{l}$ ($E_{m,\tau}^{h}$) are the eigenenergies corresponding to the smaller (larger) value of the magnetic field strength and $\wp_{m,\tau}$ are the occupation probabilities corresponding to the azimuthal and Dirac cone quantum numbers $m$ and $\tau$.   

Comparing the work extracted from the quasistatic and quantum adiabatic cycles, and noting that $\sum_{m, \tau} \left[E_{m,\tau}^l \wp_{m,\tau}(T_l, B_h)\right]$ and $\sum_{m, \tau} \left[E_{m,\tau}^h \wp_{m,\tau}(T_h, B_l)\right]$ will always be greater than or equal to the corresponding internal energies $U_A$ and $U_C$, it can be immediately seen that,  
\begin{equation}
	W^{qs} - W^{q} \geq 0.
\end{equation}
Thus the quantum adiabatic cycle will always extract the same or less work than the corresponding quasistatic cycle. This reduced work extraction arises from the fact that the quantum adiabatic cycle has additional dissipation due to ending the isentropic strokes in a non-thermal state. Furthermore, Ref. ~\cite{Pena2020PRE} finds that the quasistatic engine also has higher efficiency, and a better trade-off between efficiency and power (quantified by the simple product of efficiency and power). The behavior of the quasistatic and quantum adiabatic engines are also significantly different at large values of the magnetic field strength ratio. For the quasistatic engine the work and efficiency approach a constant value, while for the quantum adiabatic cycle both fall off sharply, with the work output quickly becoming negative, indicating that the cycle stops functioning as an engine. 

\subsubsection{Twisted bilayer graphene}

Reference \cite{Singh2021PRB} extends the study of graphene-based quantum heat engines by analyzing the performance of a magnetically driven Otto cycle with a working medium of twisted bilayer graphene (TBG). TBG, consisting of two sheets of layered graphene with one sheet rotated in relation to the other by an angle $\theta$, has seen much interest due to the fact that the system properties, including its electronic and optical behavior, can be tuned by adjusting the rotation angle \cite{Singh2021PRB}. At certain values of $\theta$, known as ``magic angles," the layers become strongly coupled and can display strongly correlated phenomena including superconductivity and zero magnetic field electronic phase transitions \cite{Singh2021PRB}.

For a single rotated graphene sheet the Hamiltonian near a Dirac point is,
\begin{equation}
	h_{\theta}(\mathbf{k}) = \mathcal{D}(\hat{z},\theta) \left[-\hbar v_F \mb{\sigma} \cdot \mathbf{k} \right] \mathcal{D}^{-1}(\hat{z},\theta), 
\end{equation}
where $\mathbf{k}$ is the crystal momentum, $v_F$ is the Fermi velocity, $\mb{\sigma} = (\sigma_x,\sigma_y)$ are the Pauli matrices, and $\mathcal{D}(\hat{z},\theta)$ is the rotation matrix. The full low-energy continuum model Hamiltonian is given by the sum of two single layer Hamiltonians, along with an interlayer tunneling term. Under an external magnetic field monolayer, bilayer, and TBG display Landau quantization. For monolayer graphene the Landau levels are,
\begin{equation}
	E_n = \pm \frac{\hbar v_F}{l_B}\sqrt{2n}, \qquad n = 0,1,2,...
\end{equation}
where the $\pm$ labels the band index (conduction and valence) and $l_B = \sqrt{\hbar/eB}$ is the Landau radius. The Landau levels of stacked bilayer graphene can also be determined analytically,
\begin{equation}
	E_n = \pm \hbar \omega_B \sqrt{n(n-1)}, \qquad n = 0,1,2,...
\end{equation}
where $\omega_B = eB/m_{\mathrm{eff}}$ is the cyclotron frequency and $m_{\mathrm{eff}}$ is the effective mass. Unlike mono- and stacked bilayer graphene, the Landau levels for TBG cannot be determined analytically, and must be found by numerically diagonalizing the Hamiltonian. For the purpose of the numerical calculations Ref. \cite{Singh2021PRB} retains the first 500 Landau levels. 

\begin{figure}
	\includegraphics[width=.48\textwidth]{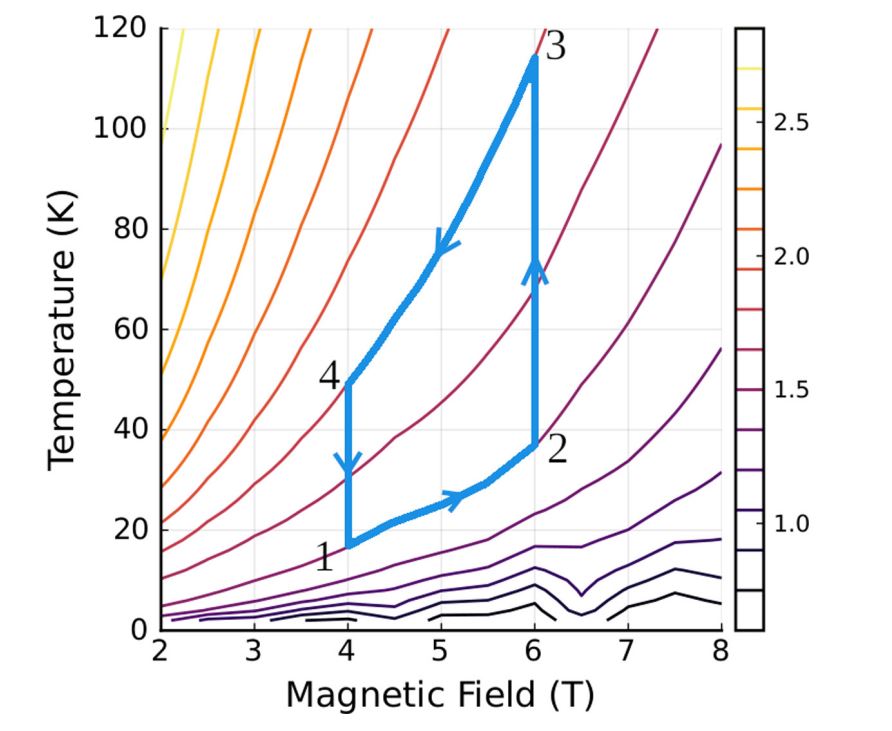}
	\caption{\label{fig:BilayerDiagram} Magnetic field - temperature diagram of a possible quantum Otto cycle implemented on magic angle twisted bilayer graphene plotted over isentropic lines. \textit{Figure adopted from Ref.~\cite{Singh2021PRB}}.}
\end{figure}

The quantum Otto cycle for TBG is depicted graphically in Fig. \ref{fig:BilayerDiagram}. The isentropic strokes are implemented by increasing or reducing the Landau radius through modulation of the external magnetic field. The isochoric strokes are implemented by bringing the working medium into contact with a classical thermal reservoir and allowing it to absorb or release heat to the reservoir while holding the Landau radius constant. Reference~\cite{Singh2021PRB} notes that the isentropic strokes can be implemented solely under the condition that the entropy remains constant, or with the stronger constraint that the occupation probabilities of each energy state remain unchanged (such that the stroke is quantumly adiabatic). For the primary portion of the analysis, Ref. ~\cite{Singh2021PRB} focuses on the ``looser" condition that $S(T_i,B_i) = S(T_f,B_f)$.                      

The heat absorbed during the hot isochore is,
\begin{equation}
	Q_{2 \rightarrow 3} = \sum_{n} E_n (B_2) \left[\wp_n(T_H,B_2)-\wp_n(T_2,B_2)\right],
\end{equation}
where $p_n$ are the occupation probabilities corresponding to Landau levels $E_n$. Similarly, the heat exchanged during the cold isochore is,
\begin{equation}
	Q_{4 \rightarrow 1} = \sum_{n} E_n (B_1) \left[\wp_n(T_C,B_1)-\wp_n(T_4,B_1)\right].
\end{equation}
The temperatures at points 2 and 4 in the cycle can be determined from the isentropic conditions,
\begin{equation}
	S(T_C, B_1) = S(T_2, B_2) \quad \mathrm{and} \quad S(T_H, B_2) = S(T_4, B_1).
\end{equation}
Using the first law, the net work can then be found from $|W| = |Q_{2 \rightarrow 3}| - |Q_{4 \rightarrow 1}|$.

For monolayer and bilayer graphene the efficiencies can be found analytically,
\begin{equation}
	\eta^{\mathrm{m}} = 1 - r^{-1} \quad \mathrm{and} \quad \eta^{\mathrm{bi}} = 1 - r^{-2},
\end{equation}
where $r = l_{b_1}/l_{B_2}$ is the compression ratio. For TBG the efficiency must be determined numerically. However, assuming that the Landau levels have the form $E_n(B) = l_B^{-\alpha}f(n)$ then the efficiency can be expressed in general as, 
\begin{equation}
	\eta = 1 - r^{-\alpha}.
\end{equation}
The numerically determined efficiencies for various angles of TBG can then be fit for the parameter $\alpha$. Reference \cite{Singh2021PRB} finds that the efficiency increases as a function of the twist angle until it reaches a maximum at $\theta^* = 0.96^{\circ}$, corresponding to the magic angle at which the renormalized Fermi velocity vanishes. As the angle continues to increase, the efficiency decreases again until it coincides with the monolayer efficiency at $\theta = 3.0^{\circ}$ at which point the large twist angle results in decoupled layers.

\subsubsection{Holographic heat engines}

A maybe less practically motivated, yet still rather active field of research is  black hole heat engines \cite{Opatrny2012AJP,Johnson2016CQG_2,Johnson2018CQG_2,Bhamidipati2017EPJC,Hendi2018PLB,Chakraborty2018IJMP,
Chakraborty2018IJMP_2,Wei2019NPB,Yerra2019MPLA,Guo2021MPLA,Feng2021CTP} and so-called holographic engines \cite{Johnson2014CQG,Johnson2016entropy,Johnson2016CQG,Hennigar2017CQG,Johnson2018CQG,Zhang2018EPJC,Mo2018JHEP,
Zhang2019JHEP,Ahmed2019CQG,Rosso2019IJMP,Ghaffarnejad2020NPB,Panah2020NPB,Sun2021CPB,Moumni2021NPB}.

The relation to quantum heat engines was made explicit by Johnson \cite{Johnson2020CQG}.  In black hole thermodynamics, a proper description of a black hole in anti-de Sitter (AdS) space time has to take into account all quantum effects \cite{Bekenstein1973PRD,Bekenstein1974PRD,Hawking1975CMP,Hawking1976PRD}. Typically, the mass $M$ plays the role of internal energy $U$, the temperature $T$ corresponds to the surface gravity divided by $2\pi$, and the entropy $S$ is given by 1/4 of the horizon area.

So-called ``holographic heat engines'' then operate with a working medium comprised of a high temperature sector of a (non-gravitational) quantum system,  i.e., the (generalized) gauge theory to which the gravitational physics in AdS is ``holographically'' dual \cite{Johnson2020CQG}.  The connection can be made even more stringent, if the cosmological constant $\Lambda$ is treated as a dynamical variable \cite{Henneaux1984PLB,Henneaux1989PLB,Sekiwa2006PRD} and one defines the pressure $P=-\Lambda/8\pi G$, where $G$ is the gravitational constant. Note that $P\geq 0$ for asymptotically AdS spacetimes. In this case, $S$ and $T$ are identified as in ``usual'' black hole thermodynamics, but the mass $M$ now plays the role of enthalpy, $H=U+PV$ \cite{Callen1985}.

Johnson \cite{Johnson2020CQG} then considered two examples of black holes, namely Kerr-AdS black holes and STU black holes, and analyzed the corresponding Otto, Brayton, and Diesel cycles. In particular, for Kerr-AdS black holes the analysis becomes pretty straight forward.  In this case, the enthalpy reads \cite{Dolan2011CQG,Caldarelli1999CQG}
\begin{equation}
H(S,P,J)=\frac{1}{2}\sqrt{\frac{\left(S+ 8PS^2/3\right)^2+4\pi^2J^2\left(1+8PS/3\right)}{\pi\, S}}\,,
\end{equation}
where $J$ is angular momentum. The thermal equation of state becomes,
\begin{equation}
T(S,P,J)=\frac{1}{8\pi H}\left[\left(1+\frac{8}{3} P S\right)\left(1+8PS\right)-4\pi^2\left(\frac{J}{S}\right)\right]\,.
\end{equation}
Hence, it becomes a simple exercise to compute the efficiency of an Otto cycle. Noting that the volume can be expressed as
\begin{equation}
V(S,P,J)=\frac{2}{3\pi H}\left[S\left(S+\frac{8}{3} P S^2\right)+2\pi^2 J^2\right]
\end{equation}
Johnson \cite{Johnson2020CQG} then showed that the efficiency becomes
\begin{equation}
\eta=1-\left(\frac{V_1}{V_2}\right)^{\gamma-1}\left(1+\mc{O}(J^2)\right)\,,
\end{equation}
where as always $\gamma=C_P/C_V$ \cite{Callen1985}.

Thus, while black holes may not become the fuel of choice for technological applications, heat engine cycles still provide the thermodynamic tool to understand the properties of complex systems.  In particular, Johnson's work \cite{Johnson2020CQG} has found attention in the study of further foundational questions in black hole physics \cite{Astefanesei2019JHEP,Debnat2020PLB,Johnson2020CQG_2,Sajadi2021NPB}.

%% file: sections/conclude.tex
\section{\label{sec:con}Final remarks}

Despite its rather slow start more about seven decades ago,  \emph{quantum thermodynamics} has recently seen a real explosion in activity.  Similarly to classical thermodynamics, an essential pillar of the theory is the understanding of thermodynamic devices, such as engines and refrigerators. The present review seeks to give an overview over the proposed and realized devices sorted according to their physical platforms. While in real-life applications of classical thermodynamics this is pretty much the ``common'' way to think about things, research in quantum devices has been predominately driven by theory.  However, at its  core thermodynamics is a theory about devices, how to describe them, and how to optimize their operation.

Given the wide applicability of thermodynamic concepts, there are many other topics we could have considered, such as, e.g.,  quantum phase transitions, quantum thermodynamics notions \cite{Goold2016JPA,LandiRMP2021}, optimal quantum control \cite{Werschnik2007JPB}, or thermodynamic uncertainty relations \cite{Agarwalla2018PRB,Liu2019PRE,Hasegawa2020PRL,Hasegawa2021PRL,Miller2021PRL}.  In addition, there has been significant amount of work on quantum versions of Maxwell's demon \cite{Vedral2009,Chapman2015PRE,Deffner2013PRE_demon,Strasberg2017PRX,Elouard2017PRL,Cottet2017PNAS,Safranek2018PRA,Engelhardt2018NJP,
Beyer2019PRL,Sanchez2019PRR,Stevens2019PRE}. However, no single review paper can cover everything,  and here we focus specifically on thermodynamic devices. As such, we hope that this review may be useful for theorists, experimentalists, as well as nanonengineers pursuing the development of quantum technologies.

We close with a quote from the seminal paper by Scully \etal\cite{Scully2003}
\begin{quote}
\emph{The deep physics behind the second law of thermodynamics is not violated; nevertheless, the quantum Carnot engine has certain features that are not possible in a classical engine.}
\end{quote}
The same holds true for any other quantum engine. We will not be able to beat the second law, but there is a host of quantum resources that can be exploited to outperform classical technologies.